\documentclass[a4paper,11pt]{article}

\pdfoutput=1

\usepackage{amsmath,amssymb,mathtools}
\usepackage{color}
\usepackage{graphicx}
\usepackage{subfigure}
\usepackage{cite}
\usepackage[colorlinks=true,linkcolor=blue, citecolor=red, urlcolor=blue, bookmarks]{hyperref}
\usepackage{multirow,makecell} 
\usepackage[figuresright]{rotating} 
\usepackage{textcomp}
\usepackage{wasysym}
\usepackage{ulem}

\usepackage[utf8]{inputenc}
\usepackage[T1]{fontenc}

\newcommand{\mr}[2]{\multirow{#1}*{#2}}
\newcommand{\mc}[3]{\multicolumn{#1}{#2}{#3}}

\usepackage[text={17.1cm,24.6cm},centering]{geometry} 

\numberwithin{equation}{section}

\def \be {\begin{equation}}
\def \ee {\end{equation}}
\def \ba {\begin{array}}
\def \ea {\end{array}}
\def \bea{\begin{eqnarray}}
\def \eea{\end{eqnarray}}
\def \nn {\nonumber}

\def \a {\alpha}
\def \b {\beta}
\def \g {\gamma}

\def \G {\Gamma}
\def \d {\delta}

\def \ve {\varepsilon}
\def \m {\mu}
\def \n {\nu}

\def \l {\lambda}
\def \lam {\lambda}
\def \Lam {\Lambda}
\def \s {\sigma}

\def \r {\rho}

\def \th {\theta}

\def \io {\iota}

\def \bs {\backslash}

\def \cA {\mathcal A}
\def \cB {\mathcal B}

\def \cF {\mathcal F}

\def \cI {\mathcal I}

\def \cN {\mathcal N}
\def \cO {\mathcal O}
\def \cP {\mathcal P}
\def \cQ {\mathcal Q}
\def \cR {\mathcal R}
\def \cS {\mathcal S}

\def \cU {\mathcal U}
\def \cV {\mathcal V}

\def \cX {\mathcal X}
\def \cY {\mathcal Y}

\def \rD {\mathrm D}

\def \rZ {\mathrm Z}

\def \f {\frac}

\def \lt {\left}
\def \rt {\right}

\def \sr {\sqrt}
\def \td {\tilde}

\def \inf {\infty}

\def \lag {\langle}
\def \rag {\rangle}

\def \ep {\mathrm{e}}
\def \ii {\mathrm{i}}

\def \arctanh {\mathop{\rm arctanh}}

\def \tr {\textrm{tr}}

\def \diag {\mathop{\textrm{diag}}}

\def \and {{~\textrm{and}~}}

\def \NS {{\textrm{NS}}}
\def \R {{\textrm{R}}}

\def \univ {{\textrm{univ}}}

\def \per {\mathop{\textrm{per}}}

\def \bos {\textrm{bos}}
\def \fer {\textrm{fer}}

\def \XXX {\textrm{XXX}}

\begin{document}

\title{
\textbf{Subsystem distances between quasiparticle excited states}
}
\author{
Jiaju Zhang$^{1}$ 
and
M. A. Rajabpour$^{2}$ 
}
\date{}
\maketitle
\vspace{-10mm}
\begin{center}
{\it
$^{1}$Center for Joint Quantum Studies and Department of Physics, School of Science,\\
      Tianjin University, 135 Yaguan Road, Tianjin 300350, China\\\vspace{1mm}
$^{2}$Instituto de Fisica, Universidade Federal Fluminense,\\
      Av. Gal. Milton Tavares de Souza s/n, Gragoat\'a, 24210-346, Niter\'oi, RJ, Brazil
}
\vspace{10mm}
\end{center}

\begin{abstract}
  We investigate the subsystem Schatten distance, trace distance and fidelity between the quasiparticle excited states of the free and the nearest-neighbor coupled fermionic and bosonic chains and the ferromagnetic phase of the spin-1/2 XXX chain. The results support the scenario that in the scaling limit when one excited quasiparticle has a large energy it decouples from the ground state and when two excited quasiparticles have a large momentum difference they decouple from each other. From the quasiparticle picture, we get the universal subsystem distances that are valid when both the large energy condition and the large momentum difference condition are satisfied, by which we mean each of the excited quasiparticles has a large energy and the momentum difference of each pair of the excited quasiparticles is large. In the free fermionic and bosonic chains, we use the subsystem mode method and get efficiently the subsystem distances, which are also valid in the coupled fermionic and bosonic chains if the large energy condition is satisfied. Moreover, under certain limit the subsystem distances from the subsystem mode method are even valid in the XXX chain. We expect that the results can be also generalized for other integrable models.
\end{abstract}

\baselineskip 18pt
\thispagestyle{empty}
\newpage


\tableofcontents

\section{Introduction}

In quantum information theory and quantum many-body systems, it is important to distinguish quantitatively two different states \cite{Nielsen:2010oan,Watrous:2018rgz,Coles:2019kdj,Cerezo:2019tuq,Chen:2020zpo,Li:2021jiv,Agarwal:2021yol}.
To differentiate two states with density matrices $\r$ and $\r'$, one may compare the expectations values of some specific local or nonlocal operator $\d\lag\cO\rag = \lag\cO\rag_\r - \lag\cO\rag_{\r'}$.
For two different states $\r\neq\r'$, there must exists some operator $\cO$ so that $\d\lag\cO\rag \neq 0$, but in practice it may be difficult to find a proper operator.
One may also calculate the differences of some nonlocal quantities such as the R\'enyi and entanglement entropies $\d S_A^{(n)} = S_{A,\r}^{(n)} - S_{A,\r'}^{(n)}$ and $\d S_A = S_{A,\r} - S_{A,\r'}$.
The R\'enyi and entanglement entropies of a subsystem $A$ in the total system in state $\r$ is defined as follows.
The Hilbert space of the total system is divided into that of the subsystem $A$ and that of its complement $B$.
One integrates out the degrees of freedom of the subsystem $B$ and obtains the reduced density matrix (RDM) $\r_{A}=\tr_B\r$ of the subsystem $A$.
The R\'enyi entropy of the RDM is defined as
\be
S_{A,\r}^{(n)}=-\f{1}{n-1}\log\tr_A\r_{A}^n,
\ee
and the entanglement entropy is  the von Neumann entropy of the RDM
\be
S_{A,\r}=-\tr_A(\r_{A}\log\r_{A}).
\ee
The entanglement entropy could be calculated as the $n\to1$ limit of the R\'enyi entropy.
The R\'enyi and entanglement entropies in various extended quantum systems have been investigated for the ground state
\cite{Bombelli:1986rw,Srednicki:1993im,Callan:1994py,Holzhey:1994we,Peschel:1998ftd,Peschel:1999pkr,Chung:2000tqg,Chung:2001oyk,%
Cheong:2002ukf,Vidal:2002rm,Peschel:2002jhw,Latorre:2003kg,Jin:2003pgk,Korepin:2004zz,Plenio:2004he,Calabrese:2004eu,Cramer:2005mx,%
Casini:2005rm,Casini:2005zv,Casini:2009sr,Calabrese:2009qy,Peschel:2009iuj,Peschel:2011jed}
and the excited states
\cite{Alba:2009th,Alcaraz:2011tn,Berganza:2011mh,Pizorn:2012aut,Essler:2012rai,Berkovits:2013mii,Taddia:2013kxu,Storms:2013wzf,%
Palmai:2014jqa,Calabrese:2014ntv,Molter:2014qsb,Taddia:2016dbm,Castro-Alvaredo:2018dja,Castro-Alvaredo:2018bij,Murciano:2018cfp,%
Castro-Alvaredo:2019irt,Castro-Alvaredo:2019lmj,Miao:2019xpp,Jafarizadeh:2019xxc,Barthel:2019zor,%
Capizzi:2020jed,You:2020osa,Haque:2020ewo,Zhang:2020ouz,Miao:2020hkj,Zhang:2020vtc,Wybo:2020fiz,Zhang:2020dtd,Zhang:2020txb,%
Chowdhury:2021qja,Zhang:2021bmy,Mussardo:2021gws}.
In this way, the compared quantities are solely defined in terms of the RDMs, but there are still potential problems.
Two states with different R\'enyi and entanglement entropies must be different, however, two different states may well have the same R\'enyi and entanglement entropies.
It is intriguing to investigate other quantities to distinguish quantitatively two different states in extended quantum many-body systems.

There are various quantities that characterize the dissimilarity, or equivalently the similarity, of two states.
For two pure states $|\psi_1\rag$ and $|\psi_2\rag$, one could just calculate the overlap $|\lag\psi_1|\psi_2\rag|^2$.
For two mixed states with density matrices $\r_1$ and $\r_2$, one may also calculate the overlap $\tr(\r_1\r_2)$.
Furthermore, one could calculate other quantities that are not simply related to the overlap.
This is especially true for the RDMs of a subsystem in the total system in various states.
When the total system is in a pure state, the RDM of a subsystem is often in a mixed state.
For example, the quantities could be the Schatten distance, trace distance, fidelity, relative entropy and other information metrics.
These quantities have been investigated in various extended systems in for example
\cite{Fagotti:2013jzu,Cardy:2014rqa,Lashkari:2014yva,Lashkari:2015dia,Arias:2016nip,Sarosi:2016atx,Sarosi:2016oks,He:2017vyf,Basu:2017kzo,%
Arias:2017dda,He:2017txy,Suzuki:2019xdq,Zhang:2019kwu,Kusuki:2019hcg,Zhang:2019wqo,Mendes-Santos:2019tmf,Zhang:2019itb,Zhang:2020mjv,%
Arias:2020sgz,deBoer:2020snb,Kudler-Flam:2021rpr,Yang:2021enf,Chen:2021pls,Capizzi:2021zga,Kudler-Flam:2021alo}.
In this paper we will investigate the subsystem Schatten distance, trace distance and fidelity in the fermionic, bosonic and spin-1/2 XXX chains.
For two RDMs $\r_{A}$ and $\s_{A}$, the subsystem Schatten distance with index $n\geq1$ is defined as
\be
D_n(\r_{A},\s_{A}) = \Big( \f{\tr_A|\r_A-\s_{A}|^n}{2} \Big)^{1/n}.
\ee
For convenience in this paper, we introduce a normalization state $\l_{A}$ and write the normalized subsystem Schatten distance as
\be
D_n(\r_{A},\s_{A};\l_A) = \Big( \f{\tr_A|\r_A-\s_{A}|^n}{2\tr \l_A^n} \Big)^{1/n}.
\ee
For an even integer $n$, it is just
\be
D_n(\r_{A},\s_{A};\l_A) = \Big( \f{\tr_A(\r_A-\s_{A})^n}{2\tr \l_A^n} \Big)^{1/n}.
\ee
The special $n=1$ case of the Schatten distance is just the trace distance
\be
D_1(\r_{A},\s_{A}) =\f12 \tr_A|\r_{A}-\s_{A}|,
\ee
which is independent of the normalization state, i.e.\ that $D_1(\r_{A},\s_{A};\l_A)=D_1(\r_{A},\s_{A})$.
The trace distance could be calculated from the replica trick proposed in \cite{Zhang:2019wqo,Zhang:2019itb}.
One first calculates the Schatten distance with the index $n$ being a general even integer and then takes the analytical continuation $n\to1$. The fidelity of two RDMs $\r_{A}$ and $\s_{A}$ is
\be
F(\r_{A},\s_{A}) = \tr_A\sr{\sr{\r_{A}}\s_{A}\sr{\r_{A}}}.
\ee
Though it is not apparent by definition, the fidelity is symmetric to its two arguments $F(\r_{A},\s_{A})=F(\s_{A},\r_{A})$.
As the case of the trace distance, we do not need to introduce a normalization state for the fidelity.
Note that the Schatten and trace distances denote the dissimilarity of two configurations, while the fidelity denotes the two configurations' similarity.

In extended quantum many-body systems, it is interesting to investigate universal behaviors of the R\'enyi and entanglement entropies.
Recently, a new universal behavior of the R\'enyi entropy and the entanglement entropy in quasiparticle excited states of integrable models was discovered in \cite{Castro-Alvaredo:2018dja,Castro-Alvaredo:2018bij,Castro-Alvaredo:2019irt,Castro-Alvaredo:2019lmj} (one could also see earlier partial results in \cite{Pizorn:2012aut,Berkovits:2013mii,Molter:2014qsb}).
The universal differences of the quasiparticle excited state R\'enyi and entanglement entropies with those in the ground state are independent of the models and the values of the quasiparticle momenta.
To obtain the universal R\'enyi and entanglement entropies, one has to take the limit that each of the relevant quasiparticle is highly excited above the ground state and each pair of the excited quasiparticle has a large momentum difference, which we will call respectively the large energy condition and the large momentum difference condition.
The universal excess R\'enyi and entanglement entropies could be written out by a simple semiclassical quasiparticle picture with the quantum effects of distinguishability and indistinguishability of the excited quasiparticles.
The same universal formulas could be obtained in the classical limit of a one-dimensional quantum gas in presence of an external potential \cite{Mussardo:2021gws}.
By relaxing the limit that quasiparticle momentum differences are large, we have obtained additional contributions to the R\'enyi and entanglement entropies in \cite{Zhang:2020vtc,Zhang:2020dtd}.
The results were further formulated into three conjectures for the R\'enyi and entanglement entropies in \cite{Zhang:2021bmy}, and these conjectures were also checked extensively therein.

In this paper, we generalize the results of quasiparticle excited state R\'enyi and entanglement entropies \cite{Castro-Alvaredo:2018dja,Castro-Alvaredo:2018bij,Zhang:2020vtc,Zhang:2020dtd,Zhang:2021bmy} to the subsystem Schatten and trace distances and fidelity.
Some preliminary results in the two-dimensional non-compact bosonic theory have been presented in \cite{Zhang:2020ouz,Zhang:2020txb}, and in this paper we will show more systematic details.
From the quasiparticle picture, we obtain universal Schatten and trace distances and fidelity that are independent of the models and the explicit values of the quasiparticle momenta.
The universal Schatten and trace distances and fidelity are valid when both the large energy condition and large momentum difference condition are satisfied.
By relaxing the large momentum difference condition, we obtain additional corrections to the universal results that are different in different models and dependent on the momentum differences of the excited quasiparticles.
We formulate the results of the Schatten and trace distances and fidelity into three conjectures, check these conjectures extensively in the fermionic and bosonic chains and spin-1/2 XXX chains, and obtain consistent results.

The universal R\'enyi and entanglement entropies in \cite{Castro-Alvaredo:2018dja,Castro-Alvaredo:2018bij,Castro-Alvaredo:2019irt,Castro-Alvaredo:2019lmj} and the universal Schatten and trace distances and fidelity in this paper are just special cases of the three conjectures for the R\'enyi and entanglement entropies in \cite{Zhang:2021bmy} and the three corresponding conjectures for the subsystem distances in this paper.
The universal formulas for the R\'enyi and entanglement entropies and the Schatten and trace distances and fidelity are obtained from the assumption that when the momentum of each pair of different excited quasiparticles is large all the different excited quasiparticles decouple from each other.
The three conjectures for the R\'enyi and entanglement entropies and the Schatten and trace distances and fidelity are based on the scenario that in the scaling limit when one excited quasiparticle has a large energy it decouples from the ground state and when two excited quasiparticles have a large momentum difference they decouple from each other.
We consider the subsystem with $\ell$ successive sites $A$ on a circular chain with $L$ sites in the scaling limit $L\to+\infty$, $\ell\to+\infty$ with fixed ratio $x\equiv\f{\ell}{L}$.
We take the ground state $|G\rag$, single-particle state $|k\rag$, and double-particle state $|kk'\rag$ as examples.
In the condition that the energy $\ve_k$ of the excited quasiparticle with momentum $k$ is large, the quasiparticle decouples from the ground state, and there are universal excess entanglement entropy and trace distance
\be \label{introee}
S_{A,k} - S_{A,G} = -x \log x - (1-x) \log(1-x),
\ee
\be \label{introd1}
D_1(\r_{A,k},\r_{A,G}) = x.
\ee
The RHS of (\ref{introee}) is nothing but the Shannon entropy of the probability distribution $\{x,1-x\}$.
The RHS of (\ref{introd1}) is just the classical trace distance between the probability distributions $\{x,1-x\}$ and $\{0,1\}$.
In the condition that the energy $\ve_k$ of the excited quasiparticle with momentum $k$ is large and the momentum difference $|k-k'|$ of the two excited quasiparticles with momentum $k$ and $k'$ is large, the quasiparticle with momentum $k$ decouples from not only the ground state but also the quasiparticle with momentum $k'$, and there are universal excess entanglement entropy and trace distance
\be
S_{A,kk'} - S_{A,k'} = -x \log x - (1-x) \log(1-x),
\ee
\be
D_1(\r_{A,kk'},\r_{A,k'}) = x.
\ee
We will give more examples and details in the main text of the paper.

The remaining part of the paper is arranged as follows:
In section~\ref{SectionRevSum} we review the three conjectures for the R\'enyi and entanglement entropies in \cite{Zhang:2021bmy} and formulate the corresponding three conjectures for the Schatten and trace distances and fidelity.
In section~\ref{SectionFreeFermion} we calculate the Schatten and trace distances and fidelity in the free fermionic chain from the subsystem mode method and check the results from various variations of the correlation matrix method.
In section~\ref{SectionInteractingFermion} we check the three conjectures for the subsystem distances in the nearest-neighbor coupled fermionic chains using the correlation matrix method.
In section~\ref{SectionFreeBoson} we calculate the Schatten and trace distances and fidelity in the free bosonic chain from the subsystem mode method and check the Schatten distance with an even index from the wave function method.
In section~\ref{SectionInteractingBoson} we check the three conjectures for the Schatten distance with an even index in the nearest-neighbor coupled bosonic chains using the wave function method.
In section~\ref{SectionXXX} we formulate the three conjectures for the trace distance and fidelity among the ground state and magnon excited states of the ferromagnetic phase of the spin-1/2 XXX chain and check these conjectures from the local mode method.
We conclude with discussions in section~\ref{SectionConclusion}.
In appendix~\ref{appNOB} we present an efficient procedure to calculate the subsystem distances for density matrices in a nonorthonormal basis.
In appendix~\ref{appRec} we give the derivation of a formula that is useful for the recursive correlation matrix method in the free fermionic chain.

\section{Three conjectures for the entropies and distances} \label{SectionRevSum}

In integrable models, one may use the set of the momenta $K=\{k_1,\cdots,k_r\}$ of the excited quasiparticles to denote the state of the total system as $|K\rag$.
In this paper, we consider circular quantum chains of $L$ sites.
The quantity $k$ that we call momentum is actually the number of waves of the total system, and the actual momentum $p$ is related to $k$ as $p=\f{2\pi k}{L}$.
In the free fermionic and bosonic chains the momenta  may be integers or half-integers depending on whether the boundary conditions are periodic or anti-periodic.
In the spin-1/2 XXX chain the momenta $k$ may not necessarily be integers or half-integers and could be real and even complex numbers.

In \cite{Zhang:2021bmy}, we have made three conjectures for the R\'enyi and entanglement entropies of the subsystem $A=[1,\ell]$ in quasiparticle excited states.
The first conjecture is that in the large energy condition
\be \label{condition1}
\f{1}{\ve_k} \ll \min(\ell,L-\ell), ~\forall k \in K,
\ee
one may define the finite-dimensional effective RDM $\td\r_{A,K}$ for the quasiparticle excited state $|K\rag$ to account for the difference of the R\'enyi and entanglement entropies in the excited state $|K\rag$ and the ground state $|G\rag$
\bea
&& S_{A,K}^{(n)} - S_{A,G}^{(n)} = - \f{1}{n-1} \log \tr \td\r_{A,K}^n, \label{conjecture1RE}\\
&& S_{A,K} - S_{A,G} = - \tr ( \td\r_{A,K} \log \td\r_{A,K} ). \label{conjecture1EE}
\eea
In other words, one could write the RDM as
\be \label{conjecture1RDM}
\r_{A,K} \cong \td\r_{A,K} \otimes \r_{A,G}.
\ee
In (\ref{conjecture1RDM}) we use `$\cong$', instead of `$=$', which just means that for the subsystem the excited quasiparticles decouple from the ground state as the background and the LHS and RHS lead to the same R\'enyi entropy.
Actually, we cannot use `$=$', as the LHS and RHS of (\ref{conjecture1RDM}) have different dimensions.
As we will see later, in the scaling limit the actual nontrivial dimension of the effective RDM only depends on the excited quasiparticles $K$, not on the real Hilbert space of the subsystem $A$.
For the same reason, in (\ref{conjecture1RE}) and (\ref{conjecture1EE}) we have used the trace `$\tr$' instead of `$\tr_A$'.
For the free fermionic chain in section~\ref{SectionFreeFermion}, the free bosonic chain in section~\ref{SectionFreeBoson} and the ferromagnetic phase XXX chain in section~\ref{SectionXXX}, one may view the effective RDMs $\td\r_{A,K}$ as the actual RDMs $\r_{A,K}$. For the coupled fermionic chain in section~\ref{SectionInteractingFermion} and the coupled bosonic chain in section~\ref{SectionInteractingBoson} one may view the effective RDMs $\td\r_{A,K}$ as the corresponding RDM $\r_{A,K}$ in respectively the free fermionic and bosonic chains.
In this viewpoint one could also use $\tr_A$ in (\ref{conjecture1RE}) and (\ref{conjecture1EE}).
In the ferromagnetic phase of the spin-1/2 XXX chain, one may also view the effective RDM $\td\r_{A,K}$ as a directive product of the RDM in the free fermionic and bosonic chains.

The second conjecture is that for the set of momenta $K$ satisfying the large energy condition and the sets $K$ and $K'$ satisfying the large momentum difference condition
\bea
&& \f{1}{\ve_k} \ll \min(\ell,L-\ell), ~ \forall k \in K, \label{condition21}\\
&& |k-k'| \gg 1, ~ \forall k\in K, ~ \forall k'\in K',  \label{condition22}
\eea
there are the differences of the R\'enyi and entanglement entropies
\bea
&& S_{A,K\cup K'}^{(n)} - S_{A,K'}^{(n)} = - \f{1}{n-1} \log \tr \td\r_{A,K}^n, \label{conjecture2RE}\\
&& S_{A,K\cup K'} - S_{A,K'} = - \tr ( \td\r_{A,K} \log \td\r_{A,K} ), \label{conjecture2EE}
\eea
and the effective RDM
\be \label{conjecture2RDM}
\r_{A,K\cup K'} \cong \td\r_{A,K} \otimes \r_{A,K'},
\ee
with the same effective RDM $\td\r_{A,K}$ as the one in (\ref{conjecture1RDM}).
In certain limit, the excited quasiparticles decouple from the background.
In the effective RDM (\ref{conjecture1RDM}) the ground state RDM $\r_{A,G}$ is viewed as the background, while in (\ref{conjecture2RDM}) the RDM $\r_{A,K'}$ is viewed as the background.

The third conjecture is that for the two sets of momenta $K$ and $K'$ satisfying the large momentum difference condition
\be \label{condition3}
|k-k'| \gg 1, ~ \forall k\in K, ~ \forall k' \in K',
\ee
there are relations
\bea
&& S_{A,K\cup K'}^{(n)} - S_{A,G}^{(n)} = S_{A,K}^{(n)} + S_{A,K'}^{(n)} - 2 S_{A,K'}^{(n)}, \label{conjecture3RE} \\
&& S_{A,K\cup K'} - S_{A,G} = S_{A,K} + S_{A,K'} - 2 S_{A,K'}. \label{conjecture3EE}
\eea
For the third conjecture, we do not necessarily have general analytical expressions on the RHS of (\ref{conjecture3RE}) and (\ref{conjecture3EE}).

The essence of the above three conjectures for the R\'enyi and entanglement entropy is the scenario that a set of quasiparticles satisfying the large energy condition decouple from the ground state and two sets of quasiparticles satisfying the large momentum difference condition decouple from each other.
Based on this scenario, we formulate the corresponding three conjectures for the Schatten and trace distances and fidelity as follows.
\begin{itemize}
  \item The first conjecture is that for two states $|K_1\rag$ and $|K_2\rag$ both satisfying the large energy condition (\ref{condition1}), i.e.\ that
\be
\f{1}{\ve_k} \ll \min(\ell,L-\ell), ~\forall k \in K_1\cup K_2,
\ee
there are the normalized Schatten distance, and trace distance and fidelity
\bea \label{Dn1rAK1rAK2}
&& D_n(\r_{A,K_1},\r_{A,K_2};\r_{A,G}) = D_n(\td\r_{A,K_1},\td\r_{A,K_2}), \label{conjecture1SD} \\
&& D_1(\r_{A,K_1},\r_{A,K_2}) = D_1(\td\r_{A,K_1},\td\r_{A,K_2}), \label{conjecture1TD} \\
&& F(\r_{A,K_1},\r_{A,K_2}) = F(\td\r_{A,K_1},\td\r_{A,K_2}). \label{conjecture1F}
\eea
  \item The second conjecture is that for two momentum sets $K_1$ and $K_2$ both satisfying the large energy condition (\ref{condition21}) and the large momentum difference condition with respect to the momentum set $K'$ (\ref{condition22}), i.e.\ that
\bea
&& \f{1}{\ve_k} \ll \min(\ell,L-\ell), ~ \forall k \in K_1\cup K_2, \\
&& |k-k'| \gg 1, ~ \forall k\in K_1\cup K_2, ~ \forall k'\in K',
\eea
there are the Schatten and trace distances and fidelity
\bea
&& D_n(\r_{A,K_1 \cup K'},\r_{A,K_2 \cup K'};\r_{A,K'}) = D_n(\td\r_{A,K_1},\td\r_{A,K_2}), \label{conjecture2SD}\\
&& D_1(\r_{A,K_1 \cup K'},\r_{A,K_2 \cup K'}) = D_1(\td\r_{A,K_1},\td\r_{A,K_2}), \label{conjecture2TD}\\
&& F(\r_{A,K_1 \cup K'},\r_{A,K_2 \cup K'}) = F(\td\r_{A,K_1},\td\r_{A,K_2}). \label{conjecture2F}
\eea
  \item
The third conjecture is that for two sets $K_1$ and $K_2$ both satisfying the large momentum difference condition with the momentum set $K'$ (\ref{condition3}), i.e.\ that
\be
|k-k'| \gg 1, ~ \forall k\in K_1\cup K_2, ~ \forall k'\in K',
\ee
there are
\bea
&& D_n(\r_{A,K_1 \cup K'},\r_{A,K_2 \cup K'};\r_{A,K'}) = D_n(\r_{A,K_1},\r_{A,K_2};\r_{A,G}), \label{conjecture3SD} \\
&& D_1(\r_{A,K_1 \cup K'},\r_{A,K_2 \cup K'}) = D_1(\r_{A,K_1},\r_{A,K_2}),  \label{conjecture3TD}\\
&& F(\r_{A,K_1 \cup K'},\r_{A,K_2 \cup K'}) = F(\r_{A,K_1},\r_{A,K_2}). \label{conjecture3F}
\eea
\end{itemize}

The three conjectures for the R\'enyi and entanglement entropies have been checked extensively in \cite{Zhang:2020vtc,Zhang:2020dtd,Zhang:2021bmy}.
In this paper we will check the corresponding three conjectures for the Schatten and trace distances and fidelity in the fermionic, bosonic and XXX chains.
Some preliminary results in the two-dimensional non-compact bosonic theory have been presented in \cite{Zhang:2020ouz,Zhang:2020txb}.

\section{Free fermionic chain} \label{SectionFreeFermion}

In this section, we consider the free fermionic chain.
We calculate the Schatten and trace distances and fidelity from the subsystem mode method.
We also check the results from various variations of the correlation matrix method, among which the diagonalized truncated correlation matrix method is the most efficient one.

\subsection{Quasiparticle excited states}

The translation invariant free fermionic chain of $L$ sites has the Hamiltonian
\be
H = \sum_{j=1}^L \Big( a_j^\dag a_j - \f12 \Big),
\ee
with the spinless fermions $a_j,a_j^\dag$.
The quasiparticle modes are Fourier transformations of the local modes
\be \label{fermionmodesbkbkdag}
b_k = \f{1}{\sr{L}}\sum_{j=1}^L \ep^{-\ii j p_k}a_j, ~~
b_k^\dag = \f{1}{\sr{L}}\sum_{j=1}^L \ep^{\ii j p_k}a_j^\dag, ~~
p_k = \f{2\pi k}{L}.
\ee
Here $p_k$ is the actual momentum and $k$ is the total number of waves, which is an integer and a half-integer depending on the boundary conditions of the spinless fermions $a_j,a_j^\dag$.
Note that $p_k\cong p_k+2\pi$ and $k\cong k+L$.
As we mentioned in section~\ref{SectionRevSum}, we just call $k$ momentum.
We only consider the case that $L$ is an even integer.
For the states in the Neveu-Schwarz (NS) sector, i.e.\ antiperiodic boundary conditions for the spinless fermions $a_{L+1}=-a_1$, $a_{L+1}^\dag=-a_{1}^\dag$, we have the half-integer momenta
\be \label{FermionNS}
\NS ~ {\textrm{sector:}}~k = \f{1-L}{2}, \cdots,-\f12,\f12,\cdots,\f{L-1}{2}.
\ee
For the states in the Ramond (R) sector, i.e.\ periodic boundary conditions for the spinless fermions $a_{L+1}=a_1$, $a_{L+1}^\dag=a_{1}^\dag$, we have the integer momenta
\be \label{FermionR}
\R ~ {\textrm{sector:}}~k = 1-\f{L}{2}, \cdots,-1,0,1,\cdots,\f{L}{2}-1,\f{L}{2}.
\ee

The ground state of the Hamiltonian is annihilated by all the local and global lowering modes
\be
a_j |G\rag = b_k |G\rag = 0, ~ \forall j, \forall k.
\ee
The ground state $|G\rag$ in the free fermionic chain is the ground state of both the NS sector and the R sector, i.e.\ that $|G\rag=|G_\NS\rag=|G_\R\rag$.
The general excited state in the NS sector is generated by applying the raising operators $b_k^\dag \in \NS$ on the ground state
\be
|K\rag=|k_1\cdots k_r\rag = b_{k_1}^\dag\cdots b_{k_r}^\dag|G\rag, ~ k_1,\cdots, k_r \in \NS.
\ee
The general excited state in the R sector is generated by applying the raising operators $b_k^\dag \in \R$ on the ground state
\be
|K\rag=|k_1\cdots k_r\rag = b_{k_1}^\dag\cdots b_{k_r}^\dag|G\rag, ~ k_1,\cdots, k_r \in \R.
\ee
One may consider the subsystem difference of two states in the same sector or two states in different sectors.
Note that one state in the NS sector and another state in the R sector with the same energy may not necessary be orthogonal.

\subsection{Subsystem mode method}

The subsystem mode method was used in \cite{Zhang:2020vtc,Zhang:2020dtd,Zhang:2021bmy} to calculate the R\'enyi and entanglement entropies in the quasiparticle excited states of the free fermionic and bosonic chains.
Especially the subsystem mode method was formulated systematically in \cite{Zhang:2021bmy}, and one could see details therein.
In this subsection, we give a brief and self-consistent review of the subsystem mode method and further adapt the method for the calculation of the subsystem distances.

We choose the subsystem $A=[1,\ell]$ and its complement $B=[\ell+1,L]$.
We focus on the scaling limit that $L\to+\infty$ and $\ell\to+\infty$ with fixed ratio $x\equiv\f{\ell}{L}$.
The ground state $|G\rag$ could be written as a direct product form $|G\rag=|G_A\rag\otimes|G_B\rag$ with $a_j|G_A\rag=0$ for all $j\in A$ and $a_j|G_B\rag=0$ for all $j\in B$.
We divide the quasiparticle modes as sums $b_k=b_{A,k}+b_{B,k}$, $b_k^\dag=b_{A,k}^\dag+b_{B,k}^\dag$ with the subsystem modes
\bea
&& b_{A,k} = \f{1}{\sr{L}}\sum_{j=1}^\ell \ep^{-\ii j p_k}a_j, ~~
   b_{A,k}^\dag = \f{1}{\sr{L}}\sum_{j=1}^\ell \ep^{\ii j p_k}a_j^\dag, \nn\\
&& b_{B,k} = \f{1}{\sr{L}}\sum_{j=\ell+1}^L \ep^{-\ii j p_k}a_j, ~~
   b_{B,k}^\dag = \f{1}{\sr{L}}\sum_{j=\ell+1}^L \ep^{\ii j p_k}a_j^\dag.
\eea
The subsystem modes satisfy the nontrivial anti-commutation relations
\bea
&& \{ b_{A,k_1}, b_{A,k_2} \} = \{ b_{A,k_1}^\dag, b_{A,k_2}^\dag \} = 0, \\
&& \{ b_{A,k_1}, b_{A,k_2}^\dag \} = \a_{k_1-k_2}, ~~
   \{ b_{B,k_1}, b_{B,k_2}^\dag \} = \b_{k_1-k_2}, \nn
\eea
with the definitions of the factors
\bea
&& \a_k \equiv
 \f{1}{L} \sum_{j=1}^\ell \ep^{- \ii j p_k}
=\lt\{
\ba{cc}
\f{\ell}{L} & k=0\\
\ep^{-\f{\pi\ii k(\ell+1)}{L}} \f{\sin\f{\pi k\ell}{L}}{L\sin\f{\pi k}{L}} & k\neq0
\ea
\rt.\!\!\!, \label{alphak} \\
&& \b_k \equiv
 \f{1}{L} \sum_{j=\ell+1}^L \ep^{- \ii j p_k}
=\lt\{
\ba{cc}
1-\f{\ell}{L} & k=0\\
\ep^{-\f{\pi\ii k(L+\ell+1)}{L}} \f{\sin\f{\pi k(L-\ell)}{L}}{L\sin\f{\pi k}{L}} & k\neq0
\ea
\rt.\!\!\!. \label{betak}
\eea
There is $\b_0=1-\a_0$, and for $k\in\rZ$ and $k\neq0$ there is $\b_k=-\a_k$.
In this paper, we will also mention a bit the case that the momentum differences are not integers.

For an arbitrary ordered set of momenta $K=\{k_1,\cdots,k_r\}$ with $k_1<\cdots<k_r$, we define the products of subsystem modes
\bea
&& b^\dag_{A,K} = b^\dag_{A,k_1}\cdots b^\dag_{A,k_r}, ~~
   b_{A,K} = (b^\dag_{A,K})^\dag = b_{A,k_r} \cdots b_{A,k_1}, \nn\\
&& b^\dag_{B,K} = c^\dag_{B,k_1}\cdots c^\dag_{B,k_r}, ~~
   b_{B,K} = (b^\dag_{B,K})^\dag = b_{B,k_r} \cdots b_{B,k_1}.
\eea
The excited state of the total system could be written as
\be
|K\rag=\sum_{K' \subseteq K} s_{K,K'} b^\dag_{A,K'} b^\dag_{B,K \bs K'}|G\rag,
\ee
with $K \bs K'$ being the complement set of $K'$ contained in $K$.
We have defined the factor
\be
s_{K,K'} = \lt\{
\ba{cl}
0                             & K' \nsubseteq K \\
{\rm sig}[K',K\bs K']   & K' \subseteq K
\ea
\rt.\!\!\!,
\ee
with ${\rm sig}[K',K\bs K']$ denoting the signature of the two ordered sets $K'$ and $K\bs K'$ joining together without changing the orders of the momenta in each of them.

We get the RDM in the nonorthonormal basis $b^\dag_{A,K'}|G_A\rag$ with $K'\subseteq K$ written as
\be \label{rhoAK}
\r_{A,K}=
\sum_{K_1,K_2\subseteq K}
[\cP_{A,K}]_{K_1K_2}
b^\dag_{A,K_1}
|G_A\rag\lag G_A|
b_{A,K_2},
\ee
with the entries of the $2^{|K|} \times 2^{|K|}$ matrix $\cP_{A,K}$
\be
[\cP_{A,K}]_{K_1K_2} = s_{K,K_1} s_{K,K_2} \lag b_{B,K\bs K_2} b^\dag_{B,K\bs K_1} \rag_G.
\ee
We have used $|K|$ to denote the number of quasiparticles in the set $K$.
We need to evaluate the expectation values $\lag b_{A,K_1} b^\dag_{A,K_2} \rag_G$ and $\lag b_{B,K_1} b^\dag_{B,K_2} \rag_G$ in the ground state, which are just the determinants
\bea
&& \lag b_{A,K_1} b^\dag_{A,K_2} \rag_G =
\lt\{
\ba{cl}
0                  & |K_1| \neq |K_2| \\
\det \cA_{K_1 K_2} & |K_1| = |K_2|
\ea
\rt.\!\!\!, \nn\\
&& \lag b_{B,K_1} b^\dag_{B,K_2} \rag_G =
\lt\{
\ba{cl}
0                  & |K_1| \neq |K_2| \\
\det \cB_{K_1 K_2} & |K_1| = |K_2|
\ea
\rt.\!\!\!.
\eea
The $|K_1|\times|K_2|$ matrices $\cA_{K_1 K_2}$ and $\cB_{K_1 K_2}$ have the entries
\be \label{cAK1K2cBK1K2}
[\cA_{K_1 K_2}]_{k_1k_2} = \a_{k_1 - k_2}, ~~
[\cB_{K_1 K_2}]_{k_1k_2} = \b_{k_1-k_2}, ~~
k_1 \in K_1, k_2 \in K_2.
\ee
For later convenience, we also define the $|K|\times|K|$ matrices $\cA_K\equiv\cA_{KK}$ and $\cB_K\equiv\cB_{KK}$.

For two sets of momenta $K_1$ and $K_2$, we have the union set $K_1\cup K_2$ in which each of the repeated momenta appears only once.
For example, for $K_1=\{\f12\}$ and $K_2=\{\f32\}$ we have $K_1\cup K_2=\{\f12,\f32\}$ and for $K_1=\{\f12\}$ and $K_2=\{\f12,\f32\}$ we have $K_1\cup K_2=\{\f12,\f32\}$.
We get the RDMs in the nonorthonormal basis $c^\dag_{A,K'}|G_A\rag$ with $K' \subseteq K_1 \cup K_2$
\bea
&& \r_{A,K_1}=
\sum_{K'_1,K'_2\subseteq K_1\cup K_2}
[\cP_{A,K_1}]_{K'_1K'_2}
b^\dag_{A,K'_1}
|G_A\rag\lag G_A|
b_{A,K'_2}, \nn\\
&& \r_{A,K_2}=
\sum_{K'_1,K'_2\subseteq K_1\cup K_2}
[\cP_{A,K_2}]_{K'_1K'_2}
b^\dag_{A,K'_1}
|G_A\rag\lag G_A|
b_{A,K'_2},
\eea
with the entries of the $2^{|K_1\cup K_2|} \times 2^{|K_1\cup K_2|}$ matrices $\cP_{A,K_1}$ and $\cP_{A,K_2}$
\bea
&& [\cP_{A,K_1}]_{K'_1K'_2} = \lt\{
\ba{cl}
s_{K_1,K'_1} s_{K_1,K'_2} \lag b_{B,K_1\bs K'_2} b^\dag_{B,K_1\bs K'_1} \rag_G & K'_1,K'_2 \subseteq K_1 \\
0 & \rm{otherwise}
\ea
\rt.\!\!\!, \nn\\
&& [\cP_{A,K_2}]_{K'_1K'_2} =  \lt\{
\ba{cl}
s_{K_2,K'_1} s_{K_2,K'_2} \lag b_{B,K_2\bs K'_2} b^\dag_{B,K_2\bs K'_1} \rag_G.
 & K'_1,K'_2 \subseteq K_2 \\
0 & \rm{otherwise}
\ea
\rt.\!\!\!.
\eea
We also define the $2^{|K_1\cup K_2|} \times 2^{|K_1\cup K_2|}$ matrix $\cQ_{A,K_1\cup K_2}$ with entries
\be
[\cQ_{A,K_1\cup K_2}]_{K'_1K'_2} = \lag b_{A,K'_1} b^\dag_{A,K'_2} \rag_G, ~ K'_1,K'_2\subseteq K_1\cup K_2.
\ee
With the $2^{|K_1\cup K_2|} \times 2^{|K_1\cup K_2|}$ matrices $\cP_{A,K_1}$, $\cP_{A,K_2}$ and $\cQ_{A,K_1\cup K_2}$, we may follow the procedure in appendix~\ref{appNOB} and calculate the Schatten and trace distances and fidelity.
Note that the matrices $\cP_{A,K_1}$, $\cP_{A,K_2}$ and $\cQ_{A,K_1\cup K_2}$ are block diagonal with $|K_1\cup K_2|+1$ blocks.

In the above strategy, we need to use the matrices with sizes that grow exponentially with the number of the excited quasiparticles, while the calculation complexity does not depend on the Schatten index $n$.
There is another strategy to calculate the Schatten distance with an even index $n=2,4,\cdots$.
The quantity $\tr_A(\r_{A,K_1}-\r_{A,K_2})^n$ could be evaluated by binomial expansion.
For example, to calculate the second Schatten distance $D_2(\r_{A,K_1},\r_{A,K_2})$ we need to evaluate
\be
\tr_A(\r_{A,K_1} - \r_{A,K_2})^2 = \tr_A\r_{A,K_1}^2 - 2 \tr_A( \r_{A,K_1}\r_{A,K_2} ) + \tr_A\r_{A,K_2}^2,
\ee
and to calculate the fourth Schatten distance $D_4(\r_{A,K_1},\r_{A,K_2})$ we need to evaluate
\bea
&& \tr_A(\r_{A,K_1} - \r_{A,K_2})^4 =
   \tr_A(\r_{A,K_1}^4)
 - 4 \tr_A( \r_{A,K_1}^3\r_{A,K_2} )
 + 4\tr_A(\r_{A,K_1}^2\r_{A,K_2}^2) \\
&& \phantom{\tr_A(\r_{A,K_1} - \r_{A,K_2})^4 =}
 + 2\tr_A(\r_{A,K_1}\r_{A,K_2}\r_{A,K_1}\r_{A,K_2})
 - 4 \tr_A( \r_{A,K_1}\r_{A,K_2}^3 )
 + \tr_A(\r_{A,K_2}^4). \nn
\eea
We evaluate each term in the binomial expansion following
\bea
&& \tr_A( \r_{A,K_1} \r_{A,K_2} \cdots \r_{A,K_n}) =
\det\lt(\ba{cccc}
\cB_{K_1}    & \cA_{K_1K_2} &        &                  \\
             & \cB_{K_2}    & \ddots &                  \\
             &              & \ddots & \cA_{K_{n-1}K_n} \\
(-)^{n-1}\cA_{K_nK_1} &              &        & \cB_{K_n}        \\
\ea\rt) \\
&& \phantom{\tr_A( \r_{A,K_1} \r_{A,K_2} \cdots \r_{A,K_n})} =
\Big( \prod_{a=1}^n \det \cB_{K_a} \Big) \det ( 1
+ \cA_{K_1K_2} \cB_{K_2}^{-1}
  \cA_{K_2K_3} \cB_{K_3}^{-1}
  \cdots
  \cA_{K_nK_1} \cB_{K_1}^{-1} ). \nn
\eea
In the second strategy, the sizes of the relevant matrices grow lineally with the number of the excited quasiparticles, the calculation complexity also grows with the Schatten index $n$.

When the numbers of the excited particles are small, the subsystem mode method is efficient for analytical calculations. When the numbers of the excited particles are not so large, the subsystem mode method is still efficient for numerical evaluations.

\subsection{Recursive correlation matrix method}

To verify the results from the subsystem mode method, we calculate numerically the subsystem distances using the correlation matrix method \cite{Chung:2001oyk,Vidal:2002rm,Peschel:2002jhw,Latorre:2003kg}.
We use the $\ell\times\ell$ correlation matrix $C_{A,K}$ with the entries
\be
[C_{A,K}]_{j_1j_2} = \lag a^\dag_{j_1} a_{j_2} \rag_K = h_{j_1-j_2}^K, ~~
j_1,j_2 = 1,2,\cdots,\ell.
\ee
There is the function
\be
h_j^K = \f1L \sum_{k\in K} \ep^{-\f{2\pi\ii j k}{L}}.
\ee
We use $\r_C$ to denote the RDM corresponding to the $\ell\times\ell$ correlation matrix $C$. For example, we have $\r_{C_{A,K}}=\r_{A,K}$.

To calculate the Schatten distance with an even integer index from the correlation matrices, we use the recursive formula
\bea \label{RecursiveC}
&& \r_{C_1}\r_{C_2} = \tr(\r_{C_1}\r_{C_2}) \r_{C_3},\\
&& \tr(\r_{C_1}\r_{C_2}) = \det(1-C_1-C_2+2C_1C_2), \nn\\
&& C_3=C_1(1-C_1-C_2+2C_2C_1)^{-1}C_2,\nn
\eea
We show the derivation of recursive formula (\ref{RecursiveC}) in appendix~\ref{appRec}.

To calculate the fidelity from the correlation matrices, we use the formula \cite{Casini:2018cxg}
\be
F(\r_{C_1},\r_{C_2})=
\sr{\det(1-C_1)}
\sr{\det(1-C_2)}
\det\Big(1+\sr{\sr{\f{C_1}{1-C_1}}\f{C_2}{1-C_2}\sr{\f{C_1}{1-C_1}}}\Big).
\ee

In the recursive correlation matrix method, we only need to use the matrices with sizes increasing algebraically with respect to the size of the subsystem $\ell$, and so it is very efficient.
The drawback is that from this method one could only calculate the Schatten distance with an even integer index and the fidelity.

\subsection{Contracted correlation matrix method}

From correlation matrix, one may construct the numerical RDM explicitly \cite{Vidal:2002rm,Latorre:2003kg}.
In the subsystem $A=[1,\ell]$, we use the complete basis of the operators
\be
\cO_{i_1i_2\cdots i_\ell} \equiv  \cO_{i_1}^1 \cO_{i_2}^2 \cdots \cO_{i_\ell}^\ell, ~~
\cO_{i=0,1,2,3}^j \equiv \{ a_j, a_j^\dag, a_j a_j^\dag, a_j^\dag a_j \}.
\ee
It is easy to check
\be
\tr( \cO_{i_1i_2\cdots i_\ell}^\dag \cO_{i'_1i'_2\cdots i'_\ell} )
= \d_{i_1i'_1} \d_{i_2 i'_2} \cdots \d_{i_\ell i'_\ell}.
\ee
We get the RDM written as
\be
\r_{A,K} = \sum_{i_1,i_2,\cdots,i_\ell \in \{0,1,2,3\}}
           \lag \cO_{i_1i_2\cdots i_\ell}^\dag \rag_K \cO_{i_1i_2\cdots i_\ell}.
\ee
The expectation value $\lag \cO_{i_1i_2\cdots i_\ell}^\dag \rag_K$ could be evaluated using the anticommutation relations of the local modes $a_j,a_j^\dag$ and the determinant formula from the Wick contractions
\be
\lag a_{j_1}^\dag a_{j_2}^\dag \cdots a_{j_i}^\dag a_{j'_{i'}} \cdots a_{j'_2} a_{j'_1} \rag_K
= \d_{ii'} \det_{j \in J,j' \in J'} h_{j-j'}^K, ~~
J = \{j_1,j_2,\cdots,j_i\}, ~
J' = \{j'_1,j'_2,\cdots,j'_{i'}\}.
\ee
Note that the orders of the sets $J$ and $J'$ are important, otherwise there would appear possible minus sign.
With the explicit numerical RDMs, in principle we could calculate everything defined from the RDMs.

In the contracted correlation matrix method, we need to process the matrices with sizes increase exponentially with respect to the size of the subsystem $\ell$, and so it is usually not so efficient. We could only consider the subsystem with a rather small size, say $\ell \lesssim 8$.

\subsection{Diagonalized correlation matrix method}

The RDM could be written in terms of the modular Hamiltonian \cite{Cheong:2002ukf,Peschel:2002jhw}
\be
\r_C = \det(1-C) \exp
           \Big( {-\sum_{j_1,j_2=1}^\ell H_{j_1j_2}a_{j_1}^\dag a_{j_2}} \Big),
\ee
with the matrix
\be
H=\log(C^{-1} - 1).
\ee
The correlation matrix $C$ is Hermitian and could be diagonalized as
\be
U^\dag C U = \td C,
\ee
with the diagonal matrix $\td C=\diag(\m_1,\m_2,\cdots,\m_\ell)$ and the unitary matrix $U=(u_1,u_2,\cdots,u_\ell)$ constructed with the eigenvectors and eigenvalues of the matrix $C$
\be
C u_j = \m_j u_j, ~ j=1,2,\cdots,\ell.
\ee
The matrix $H$ is also diagonal under the same basis
\be
U^\dag H U = \diag(\n_1,\n_2,\cdots,\n_\ell), ~~ \n_j = \log(\m_j^{-1}-1), ~ j=1,2,\cdots,\ell.
\ee

We define the new modes
\be
\td a_j = \sum_{j'=1}^\ell U^\dag_{jj'} a_{j'}, ~~
\td a_j^\dag = \sum_{j'=1}^\ell U_{j'j} a_{j'}^\dag,
\ee
in terms of which the RDM takes the form
\be
\r_C = \Big[ \prod_{j=1}^\ell(1-\m_j) \Big] \exp
           \Big( - \sum_{j=1}^\ell \n_j \td a_j^\dag \td a_j \Big)
         = \prod_{j=1}^\ell [ (1-\m_j) + (2\m_j-1) \td a_j^\dag \td a_j ].
\ee
In this way, we construct the explicit numerical RDMs and in principle could calculate everything defined from the RDMs.
To calculate the fidelity, it is convenient to use the square root of the RDM
\be
\sqrt{\r_C} = \prod_{j=1}^\ell [ \sqrt{1-\m_j} + ( \sqrt{\m_j} - \sqrt{1-\m_j} ) \td a_j^\dag \td a_j ].
\ee

The diagonalized correlation matrix method is more efficient than contracted correlation matrix method, but it is not efficient enough, as we still need to construct the explicit RDM with size increasing exponentially with the subsystem size $\ell$.
Explicitly, from this method we could consider the subsystem with size $\ell \lesssim 12$.

\subsection{Diagonalized truncated correlation matrix method}

We generalize the diagonalized correlation matrix method to the cases of a much larger subsystem size.
Note that the correlation matrix $C_{A,K}$ in state $|K\rag$ has rank $\min(\ell,|K|)$, and when $\ell>|K|$ we may truncate it into a $|K|\times |K|$ matrix. The subspace after truncation is nothing but the subspace generated by the subsystem modes $c_{A,k}^\dag$ with $k\in K$.

We consider two states $|K_1\rag$ and $|K_2\rag$ with the correlation matrices $C_{A,K_1}$ and $C_{A,K_2}$, and we denote $r=|K_1\cup K_2|$. When $\ell>|K_1\cup K_2|$, we may truncate the correlation matrices $C_{A,K_1}$ and $C_{A,K_2}$ into $r\times r$ matrices.
Firstly, we collect all the $r_1=|K_1|$ eigenvectors of $C_{A,K_1}$ with nonvanishing eigenvalues $u_1,u_2,\cdots,u_{r_1}$ and all the $r_2=|K_2|$ eigenvectors of $C_{A,K_2}$ with nonvanishing eigenvalues $u_{r_1+1},u_{r_1+2},\cdots,u_{r_1+r_2}$.
All the $r_1+r_2$ vectors $u_1,u_2,\cdots,u_{r_1+r_2}$ form a $r$-dimensional complex linear space, in which we find $r$ orthonormal basis $v_1,v_2,\cdots,v_r$.
Note that there is $r<r_1+r_2$ if $K_1\cap K_2 \neq \varnothing$.
Then we construct the $\ell \times r$ matrix
\be
V = ( v_1, v_2,\cdots, v_r ),
\ee
with $v_1,v_2,\cdots,v_r$ viewed as $\ell$-component column vectors.
We define the new truncated correlation matrices of size $r\times r$
\be
\td C_{A,K_1} = V^\dag C_{A,K_1} V, ~~
\td C_{A,K_2} = V^\dag C_{A,K_2} V.
\ee
Finally, we construct the $2^r\times 2^r$ truncated RDMs $\td \r_{A,K_1}$ and $\td \r_{A,K_2}$ from the $r\times r$ truncated correlation matrices $\td C_{A,K_1}$ and $\td C_{A,K_2}$ using the diagonalized correlation matrix method in the above section.
With the truncated RDMs $\td \r_{A,K_1}$ and $\td \r_{A,K_2}$, we calculate the Schatten and trace distances and fidelity.

The diagonalized truncated correlation matrix method in the free fermionic chain is an exact method, and no approximation has been used.
For a large subsystem, the size of the truncated RDMs only depends on the number of excited quasiparticles and the method is very efficient when the number of excited quasiparticles is not so large, say $|K_1\cup K_2| \lesssim 12$.

\subsection{Schatten and trace distances}

We first give the universal Schatten and trace distances from the quasiparticle picture, which are valid when the momentum differences of all the excited quasiparticles are large. Then we give several examples of analytical exact results of the Schatten and trace distances from the subsystem mode method. We also check the analytical results using various numerical methods if applicable, however, we only show several examples of the numerical checks.

\subsubsection{Universal Schatten and trace distances}

The effective RDM of the subsystem $A$ in the ground state is
\be
\r_{A,G} = |0][0|,
\ee
where $|0]=|G_A\rag$ denoting the ground state of the subsystem $A$, i.e.\ that state with no quasiparticle in it.
The single-particle state effective RDM takes the form
\be
\r_{A,k} = (1-x) |0][0| + x |k][k|,
\ee
where $|k]$ denotes the state of the subsystem $A$ with one quasiparticle of momentum $k$.
In the limit that the momentum difference of each pair of the excited quasiparticle is large, all the excited quasiparticles are independent and the RDM in a general state $|k_1\cdots k_r\rag$ takes a universal form
\be \label{FermionSURDM}
\r_{A,k_1\cdots k_r}^\univ=\bigotimes_{i=1}^r \r_{A,k_i}.
\ee
We consider two general states $|k_1\cdots k_r k'_1\cdots k'_{r'}\rag$ and $|k_1\cdots k_r k''_1\cdots k''_{r''}\rag$ with $r$ overlapping excited quasiparticles.
From the universal RDM (\ref{FermionSURDM}) and assumption that different quasiparticles are independent, we get the universal Schatten and trace distances
\bea
&& D_n^\univ( \r_{A,k_1\cdots k_r k'_1\cdots k'_{r'}}, \r_{A,k_1\cdots k_r k''_1\cdots k''_{r''}} ) =
\f{1}{2^{1/n}} [ x^n+(1-x)^n ]^{r/n} \{
  | (1-x)^{r'} - (1-x)^{r''} |^n \nn\\
&& \phantom{D_n^\univ( \r_{A,k_1\cdots k_r k'_1\cdots k'_{r'}}, \r_{A,k_1\cdots k_r k''_1\cdots k''_{r''}} ) =}
+ [x^n+(1-x)^n]^{r'}
+ [x^n+(1-x)^n]^{r''} \nn\\
&& \phantom{D_n^\univ( \r_{A,k_1\cdots k_r k'_1\cdots k'_{r'}}, \r_{A,k_1\cdots k_r k''_1\cdots k''_{r''}} ) =}
- (1-x)^{n r'} - (1-x)^{n r''}
\}^{1/n},  \label{FermionDnuniv}\\
&& D^\univ_1( \r_{A,k_1\cdots k_r k'_1\cdots k'_{r'}}, \r_{A,k_1\cdots k_r k''_1\cdots k''_{r''}} ) = 1 - (1-x)^{\max(r',r'')}.  \label{FermionD1univ}
\eea
The overlapping excited quasiparticles could be viewed as the background, and we obtain the normalized Schatten distance
\bea
&& D_n^\univ( \r_{A,k_1\cdots k_r k'_1\cdots k'_{r'}}, \r_{A,k_1\cdots k_r k''_1\cdots k''_{r''}} ; \r_{A,k_1\cdots k_r} ) = \f{1}{2^{1/n}} \{
  | (1-x)^{r'} - (1-x)^{r''} |^n \nn\\
&& \phantom{D_n^\univ( \r_{A,k_1\cdots k_r k'_1\cdots k'_{r'}}, \r_{A,k_1\cdots k_r k''_1\cdots k''_{r''}} ; \r_{A,k_1\cdots k_r} ) = }
+ [x^n+(1-x)^n]^{r'}
+ [x^n+(1-x)^n]^{r''} \nn\\
&& \phantom{D_n^\univ( \r_{A,k_1\cdots k_r k'_1\cdots k'_{r'}}, \r_{A,k_1\cdots k_r k''_1\cdots k''_{r''}} ; \r_{A,k_1\cdots k_r} ) = }
- (1-x)^{n r'} - (1-x)^{n r''}
\}^{1/n}. \label{FermionNormalizedDnuniv}
\eea
Remember that there is no need to normalize the trace distance.
Both the normalized Schatten distance and the trace distance is independent of the background $\r_{A,k_1\cdots k_r}$.
A special case of the universal Schatten and trace distances (\ref{FermionDnuniv}) and (\ref{FermionD1univ}) are
\bea
&& D^\univ_n( \r_{A,G}, \r_{A,k_1\dots k_r} ) = \f{1}{2^{1/n}} \{ [1-(1-x)^r]^n + [x^n+(1-x)^n]^r - (1-x)^{n r}  \}^{1/n}, \label{FermionDnunivGK} \\
&& D^\univ_1( \r_{A,G}, \r_{A,k_1\dots k_r} ) = 1-(1-x)^r. \label{FermionD1univGK}
\eea
We emphasize that the validity of the universal Schatten and trace distances (\ref{FermionDnuniv}), (\ref{FermionD1univ}), (\ref{FermionNormalizedDnuniv}), (\ref{FermionDnunivGK}) and (\ref{FermionD1univGK}) requires that all the momentum differences among the excited quasiparticles are large.

\begin{figure}[p]
  \centering
  \includegraphics[height=1.24\textwidth]{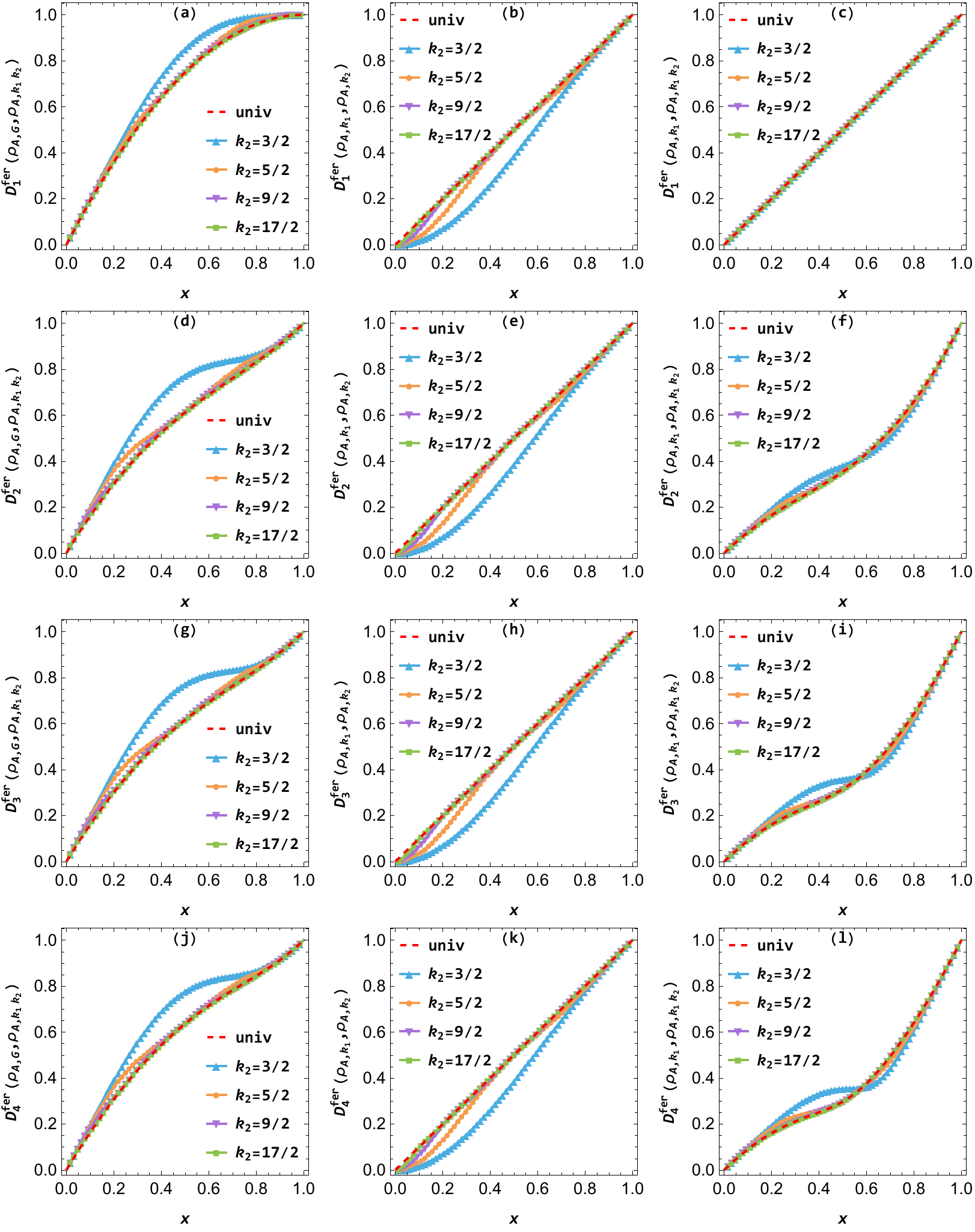}\\
  \caption{The universal Schatten and trace distances from the semiclassical quasiparticle picture (\ref{UniversalDnGk1k2}), (\ref{UniversalD1Gk1k2}), (\ref{UniversalDnD1k1k2}), (\ref{UniversalDnk1k1k2}) and (\ref{UniversalD1k1k1k2}) (dashed red lines), the analytical Schatten and trace distances from the subsystem mode method (\ref{FermionDnGk1k2}), (\ref{FermionD1Gk1k2}), (\ref{FermionDnD1k1k2}), (\ref{FermionDnk1k1k2}) and (\ref{FermionD1k1k1k2}) (solid lines), and the numerical results from the diagonalized truncated correlation matrix method (symbols) in the free fermionic chain.
  We have fixed $L=64$, $k_1=1/2$.
  We use different colors for different values of the momenta $k_2$.}
  \label{FigureFermionDnD1}
\end{figure}

In figure~\ref{FigureFermionDnD1}, we see that the Schatten and trace distances approach the universal Schatten and trace distances in the large momentum difference condition.

\subsubsection{$\r_{A,G}$ VS $\r_{A,K}$}

The RDM $\r_{A,G}=|G_A\rag\lag G_A|$ is a pure state and the two RDMs $\r_{A,G}$ and $\r_{A,K}$ commutes.
We get the exact Schatten and trace distances from the subsystem mode method
\bea
&& D_n^\fer(\r_{A,G},\r_{A,K}) = \f{1}{2^{1/n}} [ (1-\cF_{A,G,K}^\fer)^n - (\cF_{A,G,K}^\fer)^n + \cF_{A,K}^{(n),\fer} ]^{1/n}, \label{FermionDnrAGrAk1cdotskr} \\
&& D_1^\fer(\r_{A,G},\r_{A,K}) = 1-\cF_{A,G,K}^\fer, \label{FermionD1rAGrAk1cdotskr}
\eea
where we have
\bea
&& \cF_{A,G,K}^\fer \equiv \tr_A(\r_{A,G}\r_{A,K}) = \det ( 1- \cA_{K} ), \label{FermionFAGK} \\
&& \cF_{A,K}^{(n),\fer} \equiv \tr_A \r_{A,K}^n = \det [ ( 1- \cA_{K} )^n + \cA_{K}^n ]. \label{FermionFAKn}
\eea
Remember that the $|K|\times|K|$ matrix $\cA_{K} \equiv \cA_{KK}$ is defined following (\ref{cAK1K2cBK1K2}).
In the limit that all the momentum differences are large, we have $\cA_K=\diag(x,\cdots,x)$ and $\cF_{A,G,K}^\fer=(1-x)^r$, $\cF_{A,K}^{(n),\fer}=[x^n+(1-x)^n]^r$. The Schatten and trace distances (\ref{FermionDnrAGrAk1cdotskr}) and (\ref{FermionD1rAGrAk1cdotskr}) become the universal Schatten and trace distances (\ref{FermionDnunivGK}) and (\ref{FermionD1univGK}).

Explicitly, we get for the special case $r=1$
\be \label{FermionDnD1Gk}
D_n^\fer(\r_{A,G},\r_{A,k}) = D_1^\fer(\r_{A,G},\r_{A,k}) = x,
\ee
and the special case $r=2$
\bea
&& \hspace{-12mm}
   D_n^\fer(\r_{A,G},\r_{A,k_1k_2})=\f{1}{2^{1/n}} \{
    [x(2-x)+|\a_{12}|^2]^n
  - [(1-x)^2-|\a_{12}|^2]^n \label{FermionDnGk1k2}  \\
&& \hspace{-12mm}\phantom{D_n^\fer(\r_{A,G},\r_{A,k_1k_2})=}
  + [ (x + |\a_{12}|)^n + (1 - x - |\a_{12}|)^n ]
    [ (x - |\a_{12}|)^n + (1 - x + |\a_{12}|)^n ]
\}^{1/n}, \nn \\
&& \hspace{-12mm}D_1^\fer(\r_{A,G},\r_{A,k_1k_2})= x(2-x)+|\a_{12}|^2, \label{FermionD1Gk1k2}
\eea
with the shorthand $\a_{12} \equiv \a_{k_1-k_2}$ and the definition of $\a_k$ (\ref{alphak}).
The universal version of these results are
\be \label{UniversalDnD1Gk}
D_n^\univ(\r_{A,G},\r_{A,k})=D_1^\univ(\r_{A,G},\r_{A,k})=x,
\ee
\bea
&& D_n^\univ(\r_{A,G},\r_{A,k_1k_2}) = \f{1}{2^{1/n}} [ (x(2-x))^n - (1-x)^{2n} + (x^n+(1-x)^n)^2 ]^{1/n}, \label{UniversalDnGk1k2} \\
&& D_1^\univ(\r_{A,G},\r_{A,k_1k_2}) = x(2-x). \label{UniversalD1Gk1k2}
\eea
Note that there is $D_n^\fer(\r_{A,G},\r_{A,k}) = D_n^\univ(\r_{A,G},\r_{A,k})$, while generally there is $D_n^\fer(\r_{A,G},\r_{A,k_1k_2})\neq D_n^\univ(\r_{A,G},\r_{A,k_1k_2})$.
In fact, there is $D_n^\fer(\r_{A,G},\r_{A,k_1k_2}) = D_n^\univ(\r_{A,G},\r_{A,k_1k_2})$ in the large momentum difference condition $|k_1-k_2|\to+\infty$.

We compare the analytical results (\ref{FermionDnGk1k2}) and (\ref{FermionD1Gk1k2}) with the corresponding numerical ones from the diagonalized truncated correlation matrix method in the first column of figure~\ref{FigureFermionDnD1}.
There are perfect matches between the analytical and the numerical results.

\subsubsection{$\r_{A,k_1}$ VS $\r_{A,k_2}$}

From the subsystem mode method, we get the Schatten and trace distances
\be \label{FermionDnD1k1k2}
D_n^\fer(\r_{A,k_1},\r_{A,k_2})=D_1^\fer(\r_{A,k_1},\r_{A,k_2})=\sr{x^2-|\a_{12}|^2}.
\ee
Note that it is independent of the index $n$.
In the large momentum difference condition, there is the universal Schatten and trace distances
\be
D_n^\univ(\r_{A,k_1},\r_{A,k_2})=D_1^\univ(\r_{A,k_1},\r_{A,k_2})=x. \label{UniversalDnD1k1k2}
\ee
We compare these analytical results of the Schatten and trace distances with the corresponding numerical ones as well as the universal Schatten and trace distances in the second column of figure~\ref{FigureFermionDnD1}.

The results (\ref{FermionDnD1k1k2}) also apply to the case that one state $|k_1\rag$ is in the NS sector and another state $|k_2\rag$ is in the R sector, i.e.\ that $k_1$ is a half integer, $k_2$ is an integer, and so $k_1-k_2$ is a half integer.
For finite half-integer $k_1-k_2$ in the scaling limit, the two states $|k_1\rag$ and $|k_2\rag$ are not orthogonal and we have $|\lag k_1|k_2\rag|=\f{1}{\pi|k_1-k_2|}\neq0$. It is easy to check that $\lim_{x\to1}D_n(\r_{A,k_1},\r_{A,k_2})= \sqrt{1-|\lag k_1|k_2\rag|^2}\neq1$.

\subsubsection{$\r_{A,k_1}$ VS $\r_{A,k_1k_2}$}

From the subsystem mode method, we get
\bea
&& D_n^\fer(\r_{A,k_1},\r_{A,k_1k_2}) = [ (x^2-|\a_{12}|^2)^n+( x(1-x)+|\a_{12}|^2)^n ]^{1/n}, \label{FermionDnk1k1k2} \\
&& D_1^\fer(\r_{A,k_1},\r_{A,k_1k_2}) = x. \label{FermionD1k1k1k2}
\eea
The corresponding universal version are
\bea
&& D_n^\univ(\r_{A,k_1},\r_{A,k_1k_2}) = x [ x^n + (1-x)^n ]^{1/n}, \label{UniversalDnk1k1k2} \\
&& D_1^\univ(\r_{A,k_1},\r_{A,k_1k_2}) = x. \label{UniversalD1k1k1k2}
\eea
We compare these analytical results of the Schatten and trace distances with the numerical ones and the universal ones in the third column of figure~\ref{FigureFermionDnD1}.

\subsubsection{$\r_{A,K_1}$ VS $\r_{A,K_2}$}

For general states with more quasiparticles, it is difficult to obtain the analytical results, but still we may get the numerical results efficiently from the subsystem mode method and the diagonalized truncated correlation matrix method. There are perfect matches between the results obtained from different methods. We will not show the results here.

\subsubsection{A conjecture for trace distance}

From the trace distances (\ref{FermionDnD1Gk}) and (\ref{FermionD1k1k1k2}), it is tempting to conjecture the trace distance in the free fermionic chain%
\footnote{A more general conjecture is unfortunately not true. We check numerically that there is generally
\[
D_1^\fer(\r_{A,k_1\cdots k_r k'_1 \cdots k'_{r'}},\r_{A,k_1\cdots k_r k''_1 \cdots k''_{r''}}) \neq
D_1^\fer(\r_{A,k'_1 \cdots k'_{r'}},\r_{A,k''_1 \cdots k''_{r''}}).
\]
In fact, to make in the scaling limit
\[
D_1^\fer(\r_{A,k_1\cdots k_r k'_1 \cdots k'_{r'}},\r_{A,k_1\cdots k_r k''_1 \cdots k''_{r''}}) =
D_1^\fer(\r_{A,k'_1 \cdots k'_{r'}},\r_{A,k''_1 \cdots k''_{r''}}),
\]
it is required that
\[ |k_i - k'_{i'}| \to + \infty, ~ \forall i \in \{ 1,\cdots,r \}, \forall i' \in \{ 1,\cdots,r' \}, \]
\[ |k_i - k''_{i''}| \to + \infty, ~ \forall i \in \{ 1,\cdots,r \}, \forall i'' \in \{ 1,\cdots,r'' \}. \]}
\be \label{FermionUniversalD1}
D_1^\fer(\r_{A,k_1\cdots k_r},\r_{A,k_1\cdots k_r k_{r+1}}) = x.
\ee
We have checked it numerically for extensive examples, which we will not show here.
It would be interesting to derive it rigorously.

\subsubsection{Universal short interval expansion}

It is interesting to look into the behavior of the Schatten and trace distances in short interval expansion.
From the short interval expansion of the results (\ref{FermionDnD1Gk}), (\ref{FermionDnGk1k2}), (\ref{FermionDnD1k1k2}), (\ref{FermionDnk1k1k2}) and (\ref{FermionD1k1k1k2}), it is tempting to conjecture the universal leading order behavior of the general Schatten distance for the cases that the momentum differences of all the relevant quasiparticles are finite in the scaling limit
\be \label{FermionDnsie}
D_n^\fer(\r_{A,k_1\cdots k_r k'_1 \cdots k'_{r'}},\r_{A,k_1\cdots k_r k''_1 \cdots k''_{r''}}) = | r' - r'' | x + O(x^2).
\ee
The leading order of the result is independent of the Schatten index $n$.
Note that $|r' - r''|$ is just the difference of the excited quasiparticle numbers of the two states.
We do not know how to derive it for general states, but we have checked it extensively using the numerical realization of the subsystem mode method, which we will not show here.

Note that the universal Schatten and trace distances (\ref{FermionDnuniv}) and (\ref{FermionD1univ}) do not satisfy the short interval expansion (\ref{FermionDnsie}).
We stress that the validity of the universal Schatten and trace distances (\ref{FermionDnuniv}) and (\ref{FermionD1univ}) requires large momentum differences while the validity of the universal short interval expansion of Schatten and trace distances (\ref{FermionDnsie}) requires finite momentum differences.

\subsection{Fidelity}

We present the universal subsystem fidelity from the semiclassical quasiparticle picture and examples of the analytical fidelity from the subsystem mode methods. We also check the analytical fidelity numerically using the diagonalized truncated correlation matrix method.

\subsubsection{Universal fidelity}

\begin{figure}[t]
  \centering
  \includegraphics[height=0.31\textwidth]{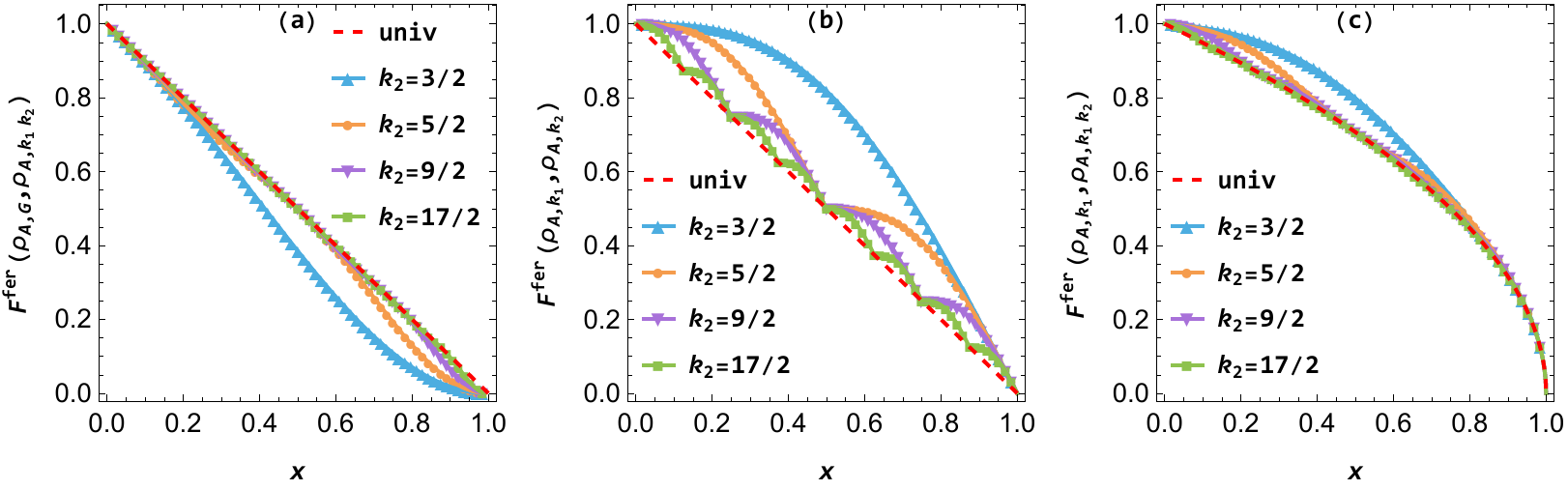}\\
  \caption{The universal fidelities from the semiclassical quasiparticle picture (\ref{UniversalFrAGrAk1k2}), (\ref{UniversalFrAk1rAk2}) and (\ref{UniversalFrAk1rAk1k2}) (dashed red lines), the analytical fidelities from the subsystem mode method (\ref{FermionFrAGrAk1k2}), (\ref{FermionFrAk1rAk2}) and (\ref{FermionFrAk1rAk1k2}) (solid lines), and the corresponding numerical results from the diagonalized truncated correlation matrix method (symbols) in the free fermionic chain. We have set $L=64$, $k_1=1/2$.}
  \label{FigureFermionF}
\end{figure}

In the limit that the momentum differences of the excited quasiparticles are large, we get the universal fidelity from the semiclassical quasiparticle picture
\be \label{FermionFidelityuniv}
F^\univ( \r_{A,k_1\cdots k_r k'_1\cdots k'_{r'}}, \r_{A,k_1\cdots k_r k''_1\cdots k''_{r''}} ) = (1-x)^{(r'+r'')/2}.
\ee
There is the special case
\be
F^\univ(\r_{A,G},\r_{A,k_1\cdots k_r})=(1-x)^{r/2}.
\ee
In figure~\ref{FigureFermionF}, we see the fidelity approach the universal fidelity in the large momentum difference condition.

\subsubsection{$\r_{A,G}$ VS $\r_{A,K}$}

In the free fermionic chain, the RDM of the ground state is a pure state, and there is a simpler result for the fidelity
\be
F^\fer(\r_{A,G},\r_{A,K})=\sr{\cF_{A,G,K}^\fer}.
\ee
with $\cF_{A,G,K}^\fer$ (\ref{FermionFAGK}).
With $D_1^\fer(\r_{A,G},\r_{A,K})$ (\ref{FermionD1rAGrAk1cdotskr}) and the fact that $0 \leq \cF_{A,G,K}^\fer \leq 1$, it is easy to see the expected inequality \cite{Nielsen:2010oan,Watrous:2018rgz}
\be
1-F^\fer(\r_{A,G},\r_{A,K}) \leq D_1^\fer(\r_{A,G},\r_{A,K}) \leq \sr{1-[F^\fer(\r_{A,G},\r_{A,K})]^2}.
\ee
For the special case $r=1$ there is
\be \label{FermionFrAGrAk}
F^\fer(\r_{A,G},\r_{A,k})=\sr{1-x},
\ee
and for $r=2$ there is
\be \label{FermionFrAGrAk1k2}
F^\fer(\r_{A,G},\r_{A,k_1k_2})=\sr{(1-x)^2-|\a_{12}|^2}.
\ee
Note the universal fidelities
\bea
&& F^\univ(\r_{A,G},\r_{A,k}) = \sr{1-x}, \label{UniversalFrAGrAk} \\
&& F^\univ(\r_{A,G},\r_{A,k_1k_2}) = 1-x. \label{UniversalFrAGrAk1k2}
\eea
We compare the analytical results with the numerical ones and the universal ones in the first panel of figure~\ref{FigureFermionF}.

\subsubsection{$\r_{A,k_1}$ VS $\r_{A,k_2}$}

We get the fidelity from the subsystem mode method
\be \label{FermionFrAk1rAk2}
F^\fer(\r_{A,k_1},\r_{A,k_2})=1-x+|\a_{12}|.
\ee
The corresponding universal fidelity is
\be
F^\univ(\r_{A,k_1},\r_{A,k_2}) = 1-x. \label{UniversalFrAk1rAk2}
\ee
We compare the analytical results with the numerical and universal ones in the second panel of figure~\ref{FigureFermionF}.

\subsubsection{$\r_{A,k_1}$ VS $\r_{A,k_1k_2}$}

From the subsystem mode method we get
\be \label{FermionFrAk1rAk1k2}
F^\fer(\r_{A,k_1},\r_{A,k_1k_2})=\sr{(1-x)[(1-x)^2-|\a_{12}|^2]}+\sr{x^2(1-x)+(1+x)|\a_{12}|^2},
\ee
with corresponding universal fidelity
\be
F^\univ(\r_{A,k_1},\r_{A,k_1k_2}) = \sqrt{1-x}. \label{UniversalFrAk1rAk1k2}
\ee
We show the results in the third panel of figure~\ref{FigureFermionF}.

\subsubsection{$\r_{A,K_1}$ VS $\r_{A,K_2}$}

For more general cases we calculate the fidelity numerically from the subsystem mode method and the diagonalized truncated correlation matrix method.
Different methods lead to the same results.
We will not show details here.

\section{Nearest-neighbor coupled fermionic chain} \label{SectionInteractingFermion}

We use the correlation matrix method and check the three conjectures for the subsystem distances between the quasiparticle excited states in the nearest-neighbor coupled fermionic chain.

\subsection{Quasiparticle excited states}

We consider the chain of $L$ spinless fermions $a_j,a_j^\dag$ with the Hamiltonian
\be
H = \sum_{j=1}^L \Big[ \lam \Big( a_j^\dag a_j - \f12 \Big)
                     - \f12 ( a_j^\dag a_{j+1} + a_{j+1}^\dag a_j )
                     - \f{\g}{2} ( a_j^\dag a_{j+1}^\dag + a_{j+1} a_j ) \Big].
\ee
It could be diagonalized following \cite{Lieb:1961fr,Katsura:1962hqz,Pfeuty:1970ayt}
\be \label{DiagonalFermionicChain}
H = \sum_k \ve_k \Big( c_k^\dag c_k - \f12 \Big), ~~
\ve_k = \sr{(\l - \cos p_k)^2 + \g^2 \sin^2 p_k }, ~~
p_k=\f{2\pi k}{L}.
\ee
Here $p_k$ is the actual momentum and $k$ is the total number of waves, which is an integer and a half-integer depending on the boundary conditions.
The quasiparticle modes $c_k,c_k^\dag$ are Bogoliubov transformation of the modes $b_k,b_k^\dag$ (\ref{fermionmodesbkbkdag})
\be
c_k = b_k \cos\f{\th_k}{2} - \ii b_{-k}^\dag \sin\f{\th_k}{2}, ~~
c_k^\dag = b_k^\dag \cos\f{\th_k}{2} + \ii b_{-k} \sin\f{\th_k}{2},
\ee
where the angle $\th_k$ is determined by
\be \label{fermionthetak}
\ep^{\ii\th_k} = \f{\l-\cos p_k+\ii\g\sin p_k}{\ve_k}.
\ee
As we have mentioned before, we just call $k$ momentum in this paper.
In this paper, we only consider the case that $L$ is an even integer.
For the states in the NS sector we have the half-integer momenta (\ref{FermionNS}), and for the states in the R sector we have the integer momenta (\ref{FermionR}).

The ground states $|G_\NS\rag$ and $|G_\R\rag$ in NS and R sectors are annihilated respectively by all the corresponding lowering operators $c_k$
\bea
&& c_k|G_\NS\rag=0, ~\forall k \in \NS, \nn\\
&& c_k|G_\R\rag=0, ~\forall k \in \R.
\eea
The general excited state in the NS sector is generated by applying the raising operators $c_k^\dag \in \NS$ on the NS sector ground state
\be
|K\rag=|k_1\cdots k_r\rag=c_{k_1}^\dag\cdots c_{k_r}^\dag|G_\NS\rag, ~ k_1,\cdots, k_r \in \NS.
\ee
The general excited state in the R sector is generated similarly
\be
|K\rag=|k_1\cdots k_r\rag=c_{k_1}^\dag\cdots c_{k_r}^\dag|G_\R\rag, ~ k_1,\cdots, k_r \in \R.
\ee

\subsection{Recursive correlation matrix method}

We calculate numerically the results using the correlation matrix method \cite{Chung:2001oyk,Vidal:2002rm,Peschel:2002jhw,Latorre:2003kg}.
In the nearest-neighbor coupled fermionic chain, one could define the Majorana modes
\be
d_{2j-1} = a_j + a_j^\dag, ~~
d_{2j} = \ii ( a_j - a_j^\dag ), ~
j=1,2,\cdots,\ell.
\ee
In the general excited state $|K\rag=|k_1k_2\cdots k_r\rag$, one defines the $2\ell \times 2\ell$ correlation matrix $\G^K$ with entries
\be
\G^K_{m_1m_2} = \lag d_{m_1} d_{m_2} \rag_K - \d_{m_1m_2}, ~~
m_1,m_2 = 1,2,\cdots,2\ell.
\ee
Explicitly, there are
\be
\G^K_{2j_1-1,2j_2-1}=\G^K_{2j_1,2j_2}=f_{j_2-j_1}^K, ~~
\G^K_{2j_1-1,2j_2}=-\G^K_{2j_2,2j_1-1}=g_{j_2-j_1}^K, ~~
j_1,j_2 = 1,2,\cdots,\ell,
\ee
with the definitions
\be
f^K_j \equiv  \f{2\ii}{L} \sum_{k \in K}\sin(j  p_k), ~~
g^K_j \equiv -\f{\ii}{L} \sum_{k \notin K} \cos( j p_k-\th_k ) + \f{\ii}{L} \sum_{k \in K} \cos( j p_k-\th_k ).
\ee
See the definitions of $\ve_k$ and $ p_k$ in (\ref{DiagonalFermionicChain}) and the definition of $\th_k$ in (\ref{fermionthetak}).
The RDM is fully determined by the correlation matrix, and so one may use $\r_\G$ to denote the RDM corresponding to the correlation matrix $\G$.

To evaluate the Schatten distance with index $n$ being an even integer, we use the recursive formula \cite{Balian:1969tb,Fagotti:2010yr}%
\footnote{Generally there would be sign ambiguity in $\tr(\r_{\G_1}\r_{\G_2})$, but there is no such problem for all the examples we consider in this paper. One could see details in \cite{Fagotti:2010yr}.}
\bea \label{RecursiveGamma}
&& \r_{\G_1}\r_{\G_2} = \tr(\r_{\G_1}\r_{\G_2}) \r_{\G_3}, \nn\\
&& \tr(\r_{\G_1}\r_{\G_2}) = \sr{\det\f{1+\G_2\G_1}{2}}, \nn\\
&& \G_3=1-(1-\G_1)(1+\G_2\G_1)^{-1}(1-\G_2).
\eea

We calculate the fidelity from the correlation matrices using the formula \cite{Banchi:2014uht}
\be \label{FrG1rG2}
F^\fer(\r_{\G_1},\r_{\G_2}) =
\Big(\det\f{1-\G_1}{2}\Big)^{1/4}
\Big(\det\f{1-\G_2}{2}\Big)^{1/4}
\Big[\det\Big(1+\sr{\sr{\f{1+\G_1}{1-\G_1}}\f{1+\G_2}{1-\G_2}\sr{\f{1+\G_1}{1-\G_1}}}\Big)\Big]^{1/2}.
\ee
The correlation matrix $\G$ often has a lot of eigenvalues equaling or close to one, and we have to introduce a cutoff to regularize the artificial divergence in the above formula (\ref{FrG1rG2}).

\subsection{Contracted correlation matrix method}

To evaluate the trace distance and other Schatten distances with odd integer indices $n$, we need to construct numerically the explicit RDMs from the correlation functions \cite{Vidal:2002rm,Latorre:2003kg}
\be
\r_{A,K} = \f{1}{2^\ell} \sum_{s_1,\cdots,s_{2\ell}\in\{0,1\}}
             \lag d_{2\ell}^{s_{2\ell}} \cdots d_1^{s_1} \rag_K
             d_1^{s_1} \cdots d_{2\ell}^{s_{2\ell}},
\ee
where the multi-point correlation functions are evaluated from the two-point correlation functions by Wick contractions. In the contracted correlation matrix method, we could only consider the subsystem with a rather small size, say $\ell \lesssim 6$.

\subsection{Canonicalized correlation matrix method}

The RDM could be written in terms of the modular Hamiltonian as \cite{Fagotti:2010yr}
\be
\r_\G = \sqrt{\det\f{1-\G}{2}} \exp\Big( -\f12 \sum_{m_1,m_2=1}^{2\ell} W_{m_1m_2} d_{m_1} d_{m_2} \Big),
\ee
with the matrix
\be
W = \arctanh \G.
\ee
Both the correlation matrix $\G$ and the matrix $W$ are purely imaginary skew-symmetric matrices, and they could be transformed into canonical form as \cite{Becker:1973qv}
\be
Q^T \G Q= \bigoplus_{j=1}^\ell \lt( \ba{cc} 0 & \ii \g_j \\ - \ii \g_j & 0 \ea \rt), ~~
Q^T W Q= \bigoplus_{j=1}^\ell \lt( \ba{cc} 0 & \ii \d_j \\ - \ii \d_j & 0 \ea \rt), ~~
\d_j=\arctanh \g_j,
\ee
with the $2\ell\times2\ell$ real orthogonal matrix $Q$ satisfying $Q^TQ=QQ^T=1$ and the real numbers $\g_j\in[-1,1]$, $j=1,2,\cdots,\ell$.
We define the new Majorana modes
\be
\td d_{m_1} \equiv \sum_{m_2=1}^{2\ell} Q_{m_2m_1} d_{m_2} , ~ m_1=1,2,\cdots,2\ell,
\ee
and write the explicit RDM as
\be
\r_\G = \prod_{j=1}^\ell \f{1-\ii\g_j\td d_{2j-1}\td d_{2j}}{2}.
\ee
From the explicit RDM, in principle we could calculate everything.
To calculate the fidelity, it is convenient to use the formula
\be
\sqrt{\r_\G} = \prod_{j=1}^\ell \f{(\sqrt{1+\g_j}+\sqrt{1-\g_j})-\ii(\sqrt{1+\g_j}-\sqrt{1-\g_j})\td d_{2j-1}\td d_{2j}}{2\sqrt{2}}.
\ee

The canonicalized correlation matrix method is a little more efficient than the contracted correlation matrix method in the previous subsection.
With the canonicalized correlation matrix method we could consider the subsystem with size $\ell \lesssim 12$.

\subsection{Checks of the three conjectures}

\begin{figure}[p]
  \centering
  \includegraphics[height=\textwidth]{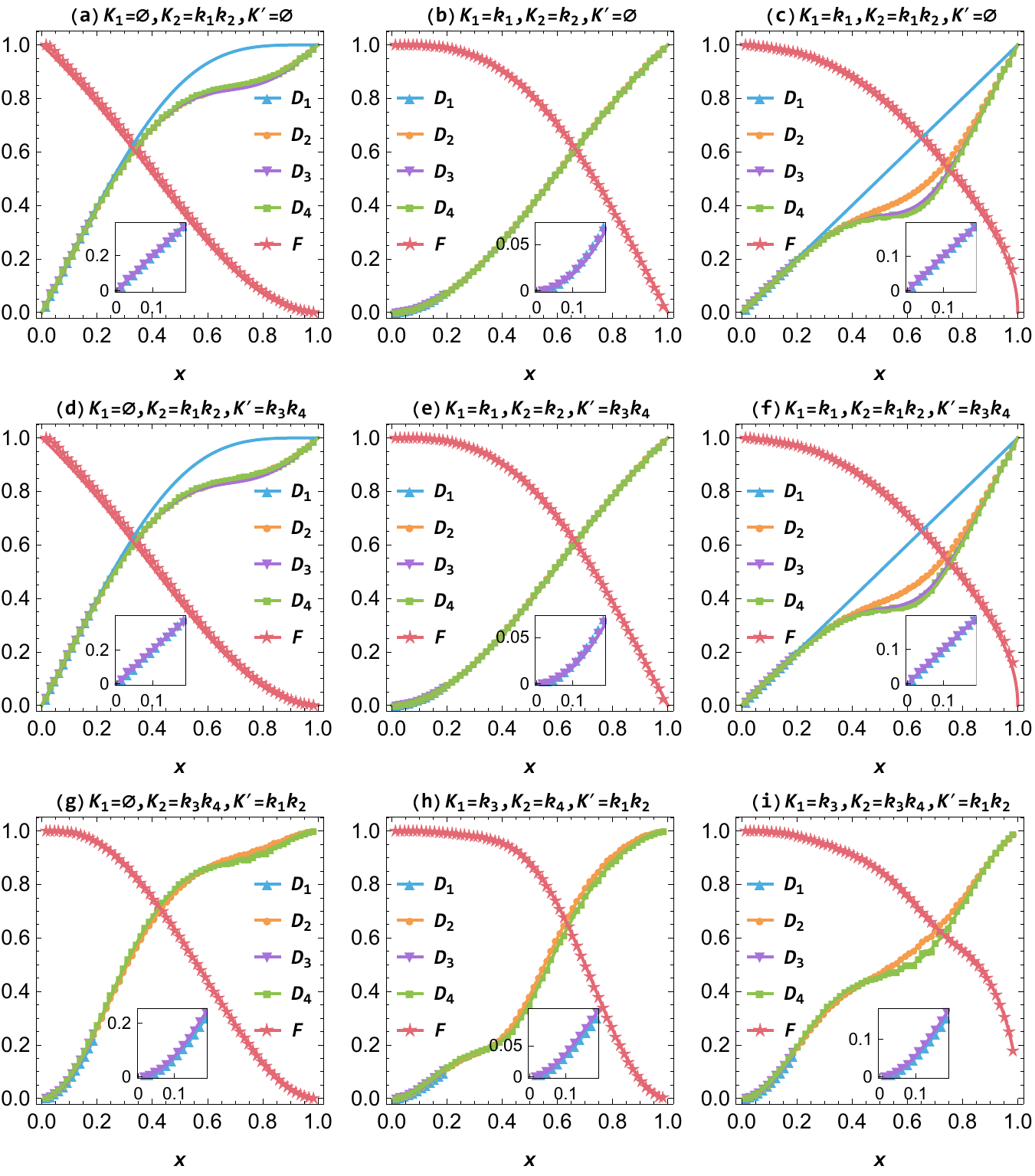}\\
  \caption{Checks of the first conjecture (\ref{conjecture1SD}), (\ref{conjecture1TD}) and (\ref{conjecture1F}) (the first row), the second conjecture (\ref{conjecture2SD}), (\ref{conjecture2TD}) and (\ref{conjecture2F}) (the second row), and the third conjecture (\ref{conjecture3SD}), (\ref{conjecture3TD}) and (\ref{conjecture3F}) (the third row) in the nearest-neighbor coupled fermionic chain.
  In each panel, the symbols are numerical results for
  the trace distance $D_1\equiv D_1^\fer(\r_{A,K_1\cup K'},\r_{A,K_2\cup K'})$,
  the normalized Schatten distance $D_n \equiv D_n^\fer(\r_{A,K_1\cup K'},\r_{A,K_2\cup K'};\r_{A,K'})$ with $n=2,3,4$, and
  the fidelity $F\equiv F^\fer(\r_{A,K_1\cup K'},\r_{A,K_2\cup K'})$, among which $D_1$ and $D_3$ are from the canonicalized correlation matrix method and $D_2$, $D_4$ and $F$ are from the recursive correlation matrix method.
  In the first and the second rows, the solid lines are the analytical conjectured results from the subsystem mode method in the free fermionic chain.
  In the third row, the solid lines are the numerical conjectured results in the nearest-neighbor coupled fermionic chain.
  In each panel, we give the inset with the results of $D_1$ and $D_3$.
  We have set $\g=\l=1$, $(k_1,k_2)=(\f12,\f32)+\f{L}{4}$ and $(k_3,k_4)=(\f12,\f32)$.
  For the analytical results in the first and second rows we have set $L=+\infty$, and for all numerical results we have set $L=64$.}
  \label{FigureFermionConjectures}
\end{figure}

We have introduced the three conjectures for the subsystem distances in section~\ref{SectionRevSum}.
We check the first conjecture (\ref{conjecture1SD}), (\ref{conjecture1TD}) and (\ref{conjecture1F}), the second conjecture (\ref{conjecture2SD}), (\ref{conjecture2TD}) and (\ref{conjecture2F}), and the third conjecture (\ref{conjecture3SD}), (\ref{conjecture3TD}) and (\ref{conjecture3F}) in respectively the first row, the second row and the third row of figure~\ref{FigureFermionConjectures}.

\section{Free bosonic chain} \label{SectionFreeBoson}

We calculate the Schatten and trace distances and fidelity in the free bosonic chain from the subsystem mode method.
We also obtain the same Schatten distances with even integer indices from the wave function method.

\subsection{Quasiparticle excited states}

We consider the translational invariant chain of $L$ independent harmonic oscillators
\be
H = \f{1}{2} \sum_{j=1}^L \big( p_j^2 + m^2 q_j^2 \big).
\ee
In terms of the local bosonic modes
\be
a_j =\sr{\f{m}{2}}\Big( q_j + \f{\ii}{m} p_j \Big), ~~
a_j^\dag =\sr{\f{m}{2}}\Big( q_j - \f{\ii}{m} p_j \Big),
\ee
the Hamiltonian becomes
\be
H = \sum_{j=1}^{L} \Big( a_j^\dag a_j +\f12 \Big).
\ee

The quasiparticle modes are
\be
b_k = \f{1}{\sr{L}}\sum_{j=1}^L\ep^{-\ii j p_k}a_j, ~~
b_k^\dag = \f{1}{\sr{L}}\sum_{j=1}^L\ep^{\ii j p_k}a_j^\dag.
\ee
We only consider the periodic boundary conditions $a_{L+1}=a_1$, $a_{L+1}^\dag=a_1^\dag$ with $L$ being an even integer, and so there are integer momenta
\be
k=1-\f{L}{2},\cdots,-1,0,1,\cdots,\f{L}{2}-1,\f{L}{2}.
\ee
The ground state $|G\rag$ is defined as
\be
a_j | G \rag = b_k | G \rag = 0, ~ \forall j, \forall k.
\ee
A general quasiparticle excited state takes the form
\be
|K\rag=|k_1^{r_1}\cdots k_s^{r_s}\rag =
\f{( b^\dag_{k_1} )^{r_1} \cdots ( b^\dag_{k_s} )^{r_s}}{\sr{N_K}} | G \rag,
\ee
with the normalization factor
\be
N_K = r_1!\cdots r_s!.
\ee

\subsection{Subsystem mode method}

The subsystem is $A=[1,\ell]$ and its complement is $B=[\ell+1,L]$.
We divide the quasiparticle modes into the subsystem modes as $b_k=b_{A,k}+b_{B,k}$ and $b_k^\dag=b_{A,k}^\dag+b_{B,k}^\dag$, which satisfy the commutation relations
\bea
&& [ b_{A,k_1}, b_{A,k_2}] = [ b_{B,k_1}^\dag , b_{B,k_2}^\dag ] = 0, \nn\\
&& [ b_{A,k_1}, b_{A,k_2}^\dag ] = \a_{k_1-k_2}, ~~
   [ b_{B,k_1}, b_{B,k_2}^\dag ] = \b_{k_1-k_2},
\eea
with $\a_k$ and $\b_k$ defined the same as those in (\ref{alphak}) and (\ref{betak}).

For an arbitrary set $K=\{k_1^{r_1},\cdots,k_s^{r_s}\}$, which we may write for short $K=k_1^{r_1}\cdots k_s^{r_s}$ when there is no ambiguity, we have the number of excited quasiparticles
\be \label{Rdefinition}
R = |K| = \sum_{i=1}^s r_i.
\ee
With the subsystem modes $b_{A,k}$, $b_{B,k}$, $b_{A,k}^\dag$ and $b_{B,k}^\dag$, we define the products
\bea
&& b^\dag_{A,K} = (b^\dag_{A,k_1})^{r_1}\cdots (b^\dag_{A,k_s})^{r_s}, ~~
   b_{A,K} = (b^\dag_{A,K})^\dag = (b_{A,k_s})^{r_s} \cdots (b_{A,k_1})^{r_1}, \nn\\
&& b^\dag_{B,K} = (b^\dag_{B,k_1})^{r_1}\cdots (b^\dag_{B,k_s})^{r_s}, ~~
   b_{B,K} = (b^\dag_{B,K})^\dag = (b_{B,k_s})^{r_s} \cdots (b_{B,k_1})^{r_1}.
\eea
Then there is the excited state
\be
|K\rag=\sum_{K' \subseteq K} s_{K,K'} b^\dag_{A,K'} b^\dag_{B,K \bs K'}|G\rag,
\ee
with $K \bs K'$ being the complement of $K'$ contained in $K$ and the factor $s_{K,K'}$ defined as
\be
s_{K,K'} = \lt\{
\ba{cl}
0                                            & K' \nsubseteq K \\
\prod_{i=1}^s \f{\sr{r_i!}}{r'_i!(r_i-r'_i)!} & K' \subseteq K
\ea
\rt.\!\!\!,
\ee
Then we get the RDM
\be
\r_{A,K}=
\sum_{K_1,K_2\subseteq K}
s_{K,K_1} s_{K,K_2}
\lag b_{B,K\bs K_2} b^\dag_{B,K\bs K_1} \rag_G
b^\dag_{A,K_1}
|G_A\rag\lag G_A|
b_{A,K_2}.
\ee
Note the possible momenta repetitions of the sets and subsets used in the bosonic chain.

Then we have the RDM in the form of (\ref{rcP})
\be
\r_{A,K}=
\sum_{K_1,K_2\subseteq K}
[\cP_{A,K}]_{K_1K_2}
b^\dag_{A,K_1}
|G_A\rag\lag G_A|
b_{A,K_2},
\ee
with the entries of the $|K|\times|K|$ matrix $\cP_{A,K}$
\be
[\cP_{A,K}]_{K_1K_2} = s_{K,K_1} s_{K,K_2} \lag b_{B,K\bs K_2} b^\dag_{B,K\bs K_1} \rag_G,
\ee
We need to evaluate the expectation values $\lag c_{A,K_1} c^\dag_{A,K_2} \rag_G$ and $\lag c_{B,K_1} c^\dag_{B,K_2} \rag_G$, which are just the permanents
\bea
&& \lag b_{A,K_1} b^\dag_{A,K_2} \rag_G =
\lt\{
\ba{cl}
0                  & |K_1| \neq |K_2| \\
\per \cA_{K_1 K_2} & |K_1| = |K_2|
\ea
\rt.\!\!\!, \nn\\
&& \lag b_{B,K_1} b^\dag_{B,K_2} \rag_G =
\lt\{
\ba{cl}
0                  & |K_1| \neq |K_2| \\
\per \cB_{K_1 K_2} & |K_1| = |K_2|
\ea
\rt.\!\!\!.
\eea
where the $|K_1|\times|K_2|$ matrices $\cA_{K_1 K_2}$ and $\cB_{K_1 K_2}$ have the entries
\be
[\cA_{K_1 K_2}]_{k_1k_2} = \a_{k_1-k_2}, ~~
[\cB_{K_1 K_2}]_{k_1k_2} = \b_{k_1-k_2}, ~~
k_1\in K_1, k_2\in K_2,
\ee
with the definitions of $\a_k$ and $\b_k$ in (\ref{alphak}) and (\ref{betak}).
We also define the $|K|\times|K|$ matrices $\cA_K\equiv\cA_{KK}$ and $\cB_K\equiv\cB_{KK}$ for later convenience.

For two sets of momenta $K_1$ and $K_2$, we define the specific union set $K_1\cup K_2$ as follows.
Firstly, we write $K_1=k_1^{r_1}\cdots k_s^{r_s}$ and $K_2=k_1^{r'_1}\cdots k_s^{r'_s}$ with the $s$ momenta $k_i$, $i=1,\cdots,s$ appearing at least once in $K_1$ or $K_2$ and some of the $2s$ integers $r_i$, $r'_i$, $i=1,\cdots,s$ being possibly zero.
Then, we define $K_1\cup K_2 \equiv k_1^{r''_1}\cdots k_s^{r''_s}$ with $r''_i=\max(r_i,r'_i)$, $i=1,\cdots,s$.
For example, from $K_1=1^223=1^2234^0$, $K_2=12^44=12^43^04$ we get the union set $K_1\cup K_2=1^22^434$, and there are also $|K_1|=4$, $|K_2|=6$, $|K_1\cup K_2|=8$.

We obtain the RDMs in the nonorthonormal basis $c^\dag_{A,K'}|G_A\rag$ with $K'\in K_1\cup K_2$
\bea
&& \r_{A,K_1}=
\sum_{K'_1,K'_2\subseteq K_1\cup K_2}
[\cP_{A,K_1}]_{K'_1K'_2}
b^\dag_{A,K'_1}
|G_A\rag\lag G_A|
b_{A,K'_2}, \nn\\
&& \r_{A,K_2}=
\sum_{K'_1,K'_2\subseteq K_1\cup K_2}
[\cP_{A,K_2}]_{K'_1K'_2}
b^\dag_{A,K'_1}
|G_A\rag\lag G_A|
b_{A,K'_2},
\eea
with the entries of the $2^{|K_1\cup K_2|} \times 2^{|K_1\cup K_2|}$ matrices $\cP_{A,K_1}$ and $\cP_{A,K_2}$
\bea
&& [\cP_{A,K_1}]_{K'_1K'_2} = \lt\{
\ba{cl}
s_{K_1,K'_1} s_{K_1,K'_2} \lag b_{B,K_1\bs K'_2} b^\dag_{B,K_1\bs K'_1} \rag_G & K'_1,K'_2 \subseteq K_1 \\
0 & \rm{otherwise}
\ea
\rt.\!\!\!, \nn\\
&& [\cP_{A,K_2}]_{K'_1K'_2} =  \lt\{
\ba{cl}
s_{K_2,K'_1} s_{K_2,K'_2} \lag b_{B,K_2\bs K'_2} b^\dag_{B,K_2\bs K'_1} \rag_G.
 & K'_1,K'_2 \subseteq K_2 \\
0 & \rm{otherwise}
\ea
\rt.\!\!\!.
\eea
We also define the $2^{|K_1\cup K_2|} \times 2^{|K_1\cup K_2|}$ matrix $\cQ_{A,K_1\cup K_2}$ with entries
\be
[\cQ_{A,K_1\cup K_2}]_{K'_1K'_2} = \lag b_{A,K'_1} b^\dag_{A,K'_2} \rag_G, ~ K'_1,K'_2\subseteq K_1\cup K_2.
\ee
With the $2^{|K_1\cup K_2|} \times 2^{|K_1\cup K_2|}$ matrices $\cP_{A,K_1}$, $\cP_{A,K_2}$ and $\cQ_{A,K_1\cup K_2}$, we follow the procedure in appendix~\ref{appNOB} and calculate the Schatten and trace distances and fidelity.

Besides the above strategy, we have another strategy to calculate the Schatten distance with an even index $n=2,4,\cdots$.
The quantity $\tr_A(\r_{A,K_1}-\r_{A,K_2})^n$ could be evaluated by binomial expansion, and we evaluate each term in the expansion following
\be \label{GenPerFor}
\tr_A ( {\r_{A,K_1}\r_{A,K_2}\cdots\r_{A,K_n}} ) =
\f{1}{N_{K_1}N_{K_2}\cdots N_{K_n}} \per
\lt(\ba{cccc}
\cB_{K_1}    & \cA_{K_1K_2} &        &                  \\
             & \cB_{K_2}    & \ddots &                  \\
             &              & \ddots & \cA_{K_{n-1}K_n} \\
\cA_{K_nK_1} &              &        & \cB_{K_n}        \\
\ea\rt).
\ee

\subsection{Wave function method}

We also calculate the Schatten distances with even integer indices from the wave function method \cite{Castro-Alvaredo:2018dja,Castro-Alvaredo:2018bij}.
One could also see the wave function method in \cite{Zhang:2020txb,Zhang:2020dtd}.
From the wave function method it is easy to get the same permanent formula (\ref{GenPerFor}) in the free bosonic chain.
We will not give details of the derivation of the permanent formula (\ref{GenPerFor}) from the wave function method.
We will review briefly the wave function method in the nearest-neighbor coupled bosonic chain in subsection~\ref{wavefunctionmethod}.

\subsection{Schatten and trace distances}

We give examples of the Schatten and trace distances in the free bosonic chain from the semiclassical quasiparticle picture and the subsystem mode method.

\subsubsection{$\r_{A,k^r}$ VS $\r_{A,k^s}$}

Without loss of generality we require $r<s$.
From the quasiparticle picture and the subsystem mode method, we get the same Schatten and trace distances
\bea
&& \hspace{-15mm}
   D_n^\bos( \r_{A,k^r}, \r_{A,k^s} ) = \f{1}{2^{1/n}} \Big\{
    \sum_{i=0}^{r} | C_r^i(1-x)^{r-i} - C_s^i(1-x)^{s-i} |^n x^{n i}
  + \sum_{i=r+1}^s [ C_s^i (1-x)^{s-i} x^i ]^n
   \Big\}^{1/n},  \label{BosonDnkrks} \\
&& \hspace{-15mm} D_1^\bos( \r_{A,k^r}, \r_{A,k^s} ) = \sum_{i=0}^{i_0} [ C_r^i(1-x)^{r-i} - C_s^i(1-x)^{s-i} ] x^{i}, \label{BosonD1krks}
\eea
where $i_0$ is the largest integer in the range $[0,r]$ that satisfies
\be
C_r^{i_0} \geq C_s^{i_0} (1-x)^{s-r}.
\ee
Note that $D_n^\bos( \r_{A,k^r}, \r_{A,k^s} )=D_n^\univ( \r_{A,k^r}, \r_{A,k^s} )$ and $D_1^\bos( \r_{A,k^r}, \r_{A,k^s} )=D_1^\univ( \r_{A,k^r}, \r_{A,k^s} )$.

\begin{figure}[t]
  \centering
  \includegraphics[height=0.31\textwidth]{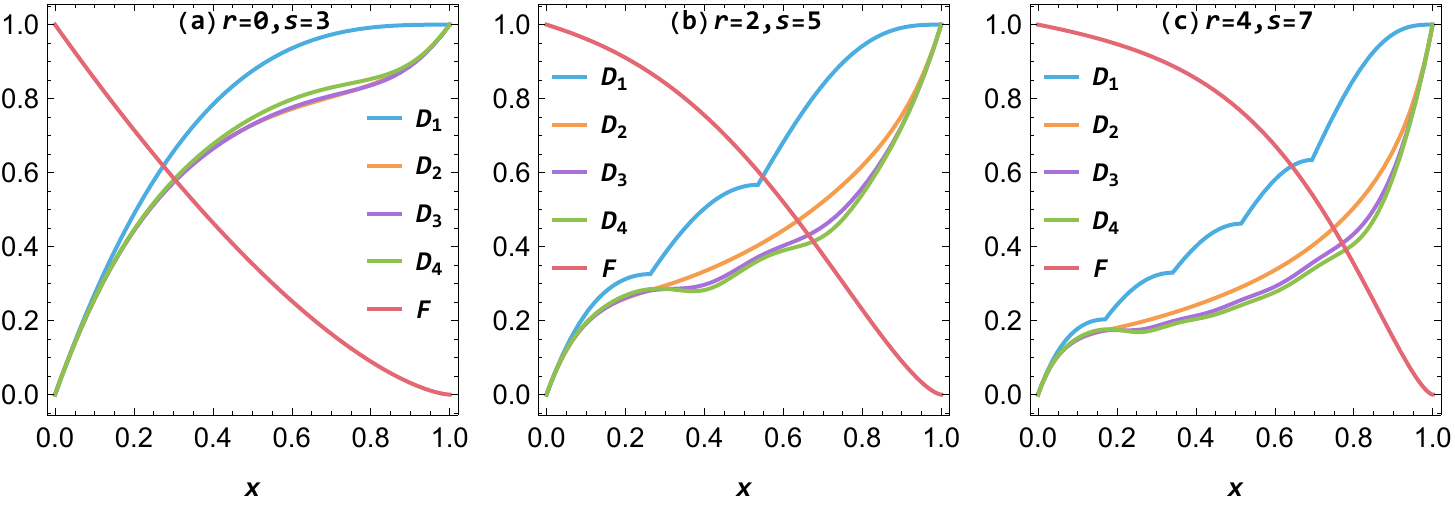}\\
  \caption{Examples of the Schatten and trace distances $D_n\equiv D_n^\bos( \r_{A,k^r}, \r_{A,k^s} )$ with $n=1,2,3,4$ (\ref{BosonDnkrks}) and (\ref{BosonD1krks}) and the fidelity $F\equiv F^\bos( \r_{A,k^r}, \r_{A,k^s} )$ (\ref{BosonFkrks}) in the free bosonic chain.}
  \label{FigureBosonDnD1Fkrks}
\end{figure}

We show examples of the results in figure~\ref{FigureBosonDnD1Fkrks}.
It is interesting to note that the derivative of the trace distance $D_1^\bos( \r_{A,k^r}, \r_{A,k^s} )$ with respect to $x$ is not continuous.
In the range $x\in[0,1]$, the derivative of the trace distance has $\min(r,s)$ discontinuous points.

\subsubsection{$\r_{A,G}$ VS $\r_{A,K}$}

The universal Schatten and trace distances (\ref{FermionDnuniv}) and (\ref{FermionD1univ}) still apply to the RDMs in the bosonic chain, but in the bosonic chain there are more general cases.
We just consider the universal distance between the RDMs in the ground state $|G\rag$ and the most general quasiparticle excited state $|K\rag=|k_1^{r_1}\cdots k_s^{r_s}\rag$.
From the quasiparticle picture, we get the universal Schatten and trace distances
\bea
&& \hspace{-10mm}
D_n^\univ(\r_{A,G},\r_{A,K}) =
\f{1}{2^{1/n}} \Big\{ [1-(1-x)^R]^n
                - (1-x)^{n R}
                + \prod_{i=1}^s \sum_{p=0}^{r_i}[ C_{r_i}^p x^p (1-x)^{r_i-p} ]^n \Big\}^{1/n}, \label{BosonDnuniv}\\
&& \hspace{-10mm}
D_1^\univ(\r_{A,G},\r_{A,K}) = 1 - (1-x)^R. \label{BosonD1univ}
\eea
Remember the total number of excited quasiparticles $R = |K| = \sum_{i=1}^s r_i$ (\ref{Rdefinition}).
In the free bosonic chain, the universal Schatten and trace distances are valid in the condition that all the momentum differences among the excited quasiparticles are large.

In the free bosonic chain, we get the exact Schatten and trace distances from the subsystem mode method
\bea
&& \hspace{-5mm} D_n^\bos(\r_{A,G},\r_{A,K}) = \f{1}{2^{1/n}} \big[ (1-\cF_{A,G,K}^\bos)^n - (\cF_{A,G,K}^\bos)^n + \cF_{A,K}^{(n),\bos} \big]^{1/n},  \label{BosonDnGk1r1cdotsksrs}\\
&& \hspace{-5mm} D_1^\bos(\r_{A,G},\r_{A,K}) = 1-\cF_{A,G,K}^\bos, \label{BosonD1Gk1r1cdotsksrs}
\eea
where we have
\bea
&& \cF_{A,G,K}^\bos
 \equiv \tr_A(\r_{A,G}\r_{A,K})
 = \f{1}{N_K}\per\cB_{K},   \label{BosonFAGK}\\
&& \cF_{A,K}^{(n),\bos}
 \equiv \tr_A \r_{A,K}^n
 = \f{1}{N_K^n}\per\lt(\ba{cccc}
\cB_{K} & \cA_{K} &        & \\
        & \cB_{K} & \ddots & \\
                  &        & \ddots & \cA_{K} \\
\cA_{K} &         &        & \cB_{K} \\
\ea\rt). \label{BosonFAKn}
\eea

\begin{figure}[p]
  \centering
  \includegraphics[height=1.24\textwidth]{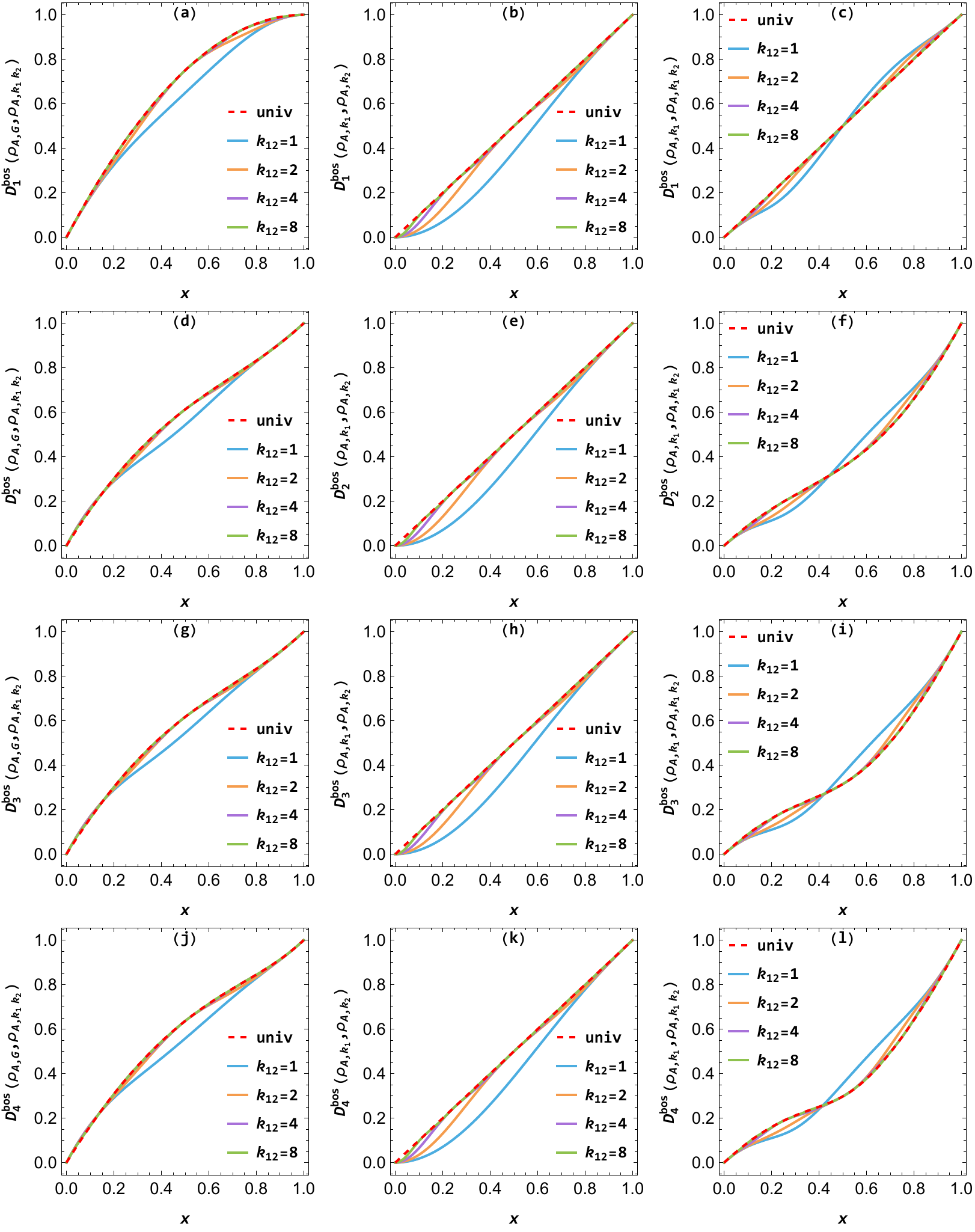}\\
  \caption{The universal Schatten and trace distances from the semiclassical quasiparticle picture (\ref{UniversalDnGk1k2}), (\ref{UniversalD1Gk1k2}), (\ref{UniversalDnD1k1k2}), (\ref{UniversalDnk1k1k2}) and (\ref{UniversalD1k1k1k2}) (dashed red lines) and the analytical Schatten and trace distances from the subsystem mode method (\ref{BosonDnGk1k2}), (\ref{BosonD1Gk1k2}), (\ref{BosonDnD1k1k2}), (\ref{BosonDnk1k1k2}) and (\ref{FermionD1k1k1k2}) (solid lines) in the free bosonic chain.
  We use different colors to denote different quasiparticle momentum differences $k_{12}\equiv k_1-k_2$.
  We have fixed $L=+\inf$.}
  \label{FigureBosonDnD1}
\end{figure}

One special case of the universal Schatten and trace distances (\ref{BosonDnuniv}) and (\ref{BosonD1univ}) are
\bea
&& D_n^\bos(\r_{A,G},\r_{A,k^r})
                = \f{1}{2^{1/n}} \Big\{ [1-(1-x)^r]^n
                - (1-x)^{n r}
                + \sum_{p=0}^{r}[ C_{r}^p x^p (1-x)^{r-p} ]^n \Big\}^{1/n}, \label{BosonDnGkr} \\
&& D_1^\bos(\r_{A,G},\r_{A,k^r}) = 1 - (1-x)^r. \label{BosonD1Gkr}
\eea
For more general cases, there are corrections to the universal Schatten and trace distances.
For example, we obtain the Schatten and trace distances
\bea
&& \hspace{-5mm} D_n^\bos(\r_{A,G},\r_{A,k_1k_2})=\f{1}{2^{1/n}} \{
  [x(2-x)-|\a_{12}|^2]^n
+ [ x^2+|\a_{12}|^2 ]^n  \label{BosonDnGk1k2} \\
&& \hspace{-5mm} \phantom{D_n^\bos(\r_{A,G},\r_{A,k_1k_2})=}
+ (x+|\a_{12}|)^n(1-x-|\a_{12}|)^n
+ (x-|\a_{12}|)^n(1-x+|\a_{12}|)^n
\}^{1/n}, \nn \\
&& \hspace{-5mm} D_1^\bos(\r_{A,G},\r_{A,k_1k_2})= x(2-x)-|\a_{12}|^2. \label{BosonD1Gk1k2}
\eea
Remember the shorthand $\a_{12} \equiv \a_{k_1-k_2}$ with the definition of $\a_k$ in (\ref{alphak}).
We show the results in figure~\ref{FigureBosonDnD1}.

\subsubsection{$\r_{A,k_1}$ VS $\r_{A,k_2}$}

From the subsystem mode method, we get the Schatten and trace distances
\be \label{BosonDnD1k1k2}
D_n^\bos(\r_{A,k_1},\r_{A,k_2})=D_1^\bos(\r_{A,k_1},\r_{A,k_2})=\sr{x^2-|\a_{12}|^2}.
\ee
We show the results in figure~\ref{FigureBosonDnD1}.

\subsubsection{$\r_{A,k_1}$ VS $\r_{A,k_1k_2}$} \label{subsectionBosonDnD1k1k1k2}

From the subsystem mode method, we get the Schatten and trace distances
\bea
&& D_n^\bos(\r_{A,k_1},\r_{A,k_1k_2})=\f{1}{2^{1/n}} \Big\{
  [x(1-x)-|\a_{12}|^2]^n
 + (x^2+|\a_{12}|^2)^n \nn\\
&& \phantom{D_n^\bos(\r_{A,k_1},\r_{A,k_1k_2})=}
 + \Big[ \f12 \sr{x^2-8x(1-2x)|\a_{12}|^2} - \f12 x(1-2x) + |\a_{12}|^2 \Big]^n \nn\\
&& \phantom{D_n^\bos(\r_{A,k_1},\r_{A,k_1k_2})=}
 + \Big[ \f12 \sr{x^2-8x(1-2x)|\a_{12}|^2} + \f12 x(1-2x) - |\a_{12}|^2 \Big]^n
\Big\}^{1/n},  \label{BosonDnk1k1k2} \\
&& D_1^\bos(\r_{A,k_1},\r_{A,k_1k_2})=\f{1}{2}( x + \sr{x^2-8x(1-2x)|\a_{12}|^2} ), \label{BosonD1k1k1k2}
\eea
which are shown in figure~\ref{FigureBosonDnD1}.

\subsubsection{$\r_{A,K_1}$ VS $\r_{A,K_2}$}

For more general cases, we calculate the Schatten and trace distances numerically, which we will not show here.

\subsubsection{Universal short interval expansion}

With the above examples, we conjecture that there is universal short interval expansion of the Schatten and trace distances
\be \label{BosonDnsie}
D_n^\bos(\r_{A,K_1},\r_{A,K_2}) = | R_1 - R_2 | x + O(x^2),
\ee
with the definition (\ref{Rdefinition}), generalizing the result (\ref{FermionDnsie}) in the fermionic chain.
In \cite{Zhang:2020txb} there have been extensive numerical checks for the special case
\be
D_n^\bos(\r_{A,G},\r_{A,K}) = R x + O(x^2).
\ee
We check extensive examples to support the conjecture (\ref{BosonDnsie}), which we will not show in this paper.

\subsection{Fidelity}

We calculate the fidelity from the subsystem mode method.
As the density matrices of the total system and RDMs in the excited states in the bosonic chain are not Gaussian, it is difficult to evaluate the square root of the RDMs, and we could not calculate the general fidelity from the wave function method.

\subsubsection{$\r_{A,k^r}$ VS $\r_{A,k^s}$}

From the quasiparticle picture and subsystem mode method, we get the same result of the fidelity
\be
F^\bos(\r_{A,k^r},\r_{A,k^s}) = \sum_{i=0}^{\min(r,s)} \sqrt{C_r^i C_s^i} x^i (1-x)^{\f{r+s}{2}-i}. \label{BosonFkrks}
\ee
Note that $F^\bos(\r_{A,k^r},\r_{A,k^s})=F^\univ(\r_{A,k^r},\r_{A,k^s})$.
We show examples of the results in the figure~\ref{FigureBosonDnD1Fkrks}.

\subsubsection{$\r_{A,G}$ VS $\r_{A,K}$}

From the quasiparticle picture, we get the universal fidelity
\be
F^\univ(\r_{A,G},\r_{A,K})=(1-x)^{R/2},
\ee
which is valid in the free bosonic chain when the large momentum difference condition is satisfied.
More generally, from the subsystem mode method we get the exact fidelity in the free bosonic chain
\be
F^\bos(\r_{A,G},\r_{A,K})=\sr{\cF_{A,G,K}^\bos},
\ee
with the definition of $\cF_{A,G,K}^\bos$ (\ref{BosonFAGK}).

For the single-particle state, there is
\be
F^\bos(\r_{A,G},\r_{A,k})=\sr{1-x},
\ee
which is the same as the fidelity (\ref{FermionFrAGrAk}) in the free fermionic chain and the universal fidelity (\ref{UniversalFrAGrAk}).

\begin{figure}[t]
  \centering
  \includegraphics[height=0.3\textwidth]{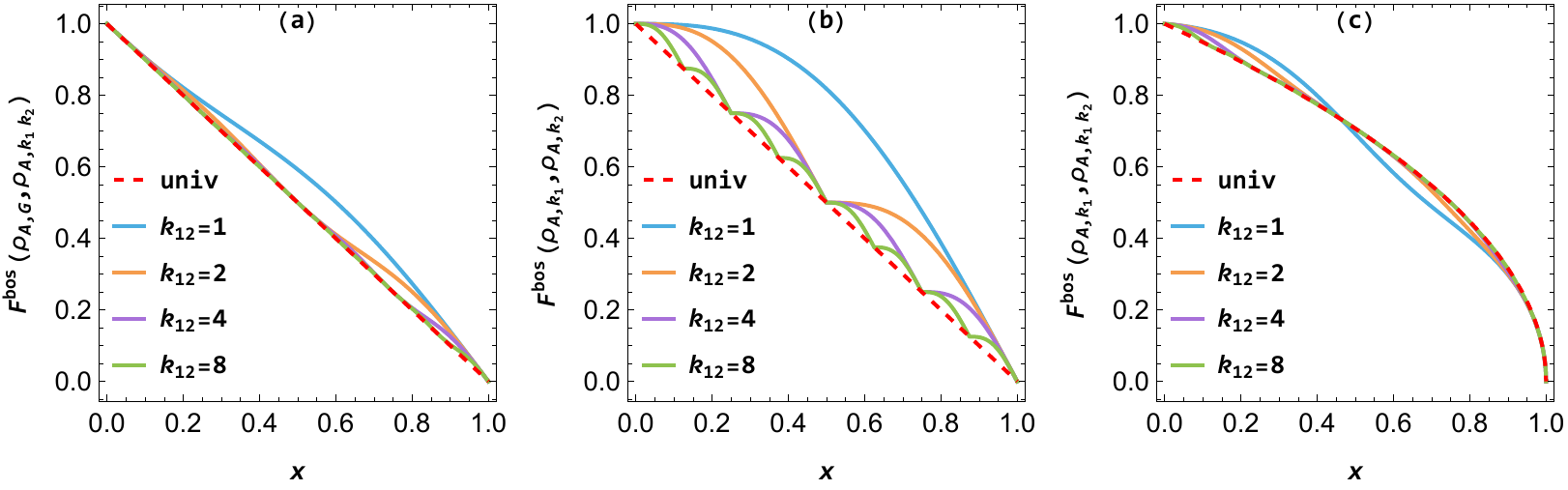}\\
  \caption{The fidelities (\ref{BosonFrAGrAk1k2}), (\ref{BosonFrAk1rAk2}) and (\ref{BosonFrAk1rAk1k2}) (solid lines) and their corresponding universal fidelities (\ref{UniversalFrAGrAk1k2}), (\ref{UniversalFrAk1rAk2}) and (\ref{UniversalFrAk1rAk1k2}) (red dotted lines) in the free bosonic chain.
  We have set $L=+\inf$.}
  \label{FigureBosonF}
\end{figure}

For the double-particle state, there is
\be \label{BosonFrAGrAk1k2}
F^\bos(\r_{A,G},\r_{A,k_1k_2})=\sr{(1-x)^2+|\a_{12}|^2},
\ee
which is different from the fidelity $F^\fer(\r_{A,G},\r_{A,k_1k_2})$ (\ref{FermionFrAGrAk1k2}) in the free fermionic chain.
We show it in figure~\ref{FigureBosonF}.

\subsubsection{$\r_{A,k_1}$ VS $\r_{A,k_2}$}

We get the fidelity from the subsystem mode method
\be \label{BosonFrAk1rAk2}
F^\bos(\r_{A,k_1},\r_{A,k_2})=1-x+|\a_{12}|,
\ee
which is the same as the fidelity $F^\fer(\r_{A,k_1},\r_{A,k_2})$ (\ref{FermionFrAk1rAk2}) in the fermionic chain.
We show it in figure~\ref{FigureBosonF}.

\subsubsection{$\r_{A,k_1}$ VS $\r_{A,k_1k_2}$}

We also get
\be \label{BosonFrAk1rAk1k2}
F^\bos(\r_{A,k_1},\r_{A,k_1k_2})=\sr{(1-x)[(1-x)^2+|\a_{12}|^2]}+\sr{x^2(1-x)+(1-3x)|\a_{12}|^2},
\ee
which is different from the fidelity $F^\fer(\r_{A,k_1},\r_{A,k_1k_2})$ (\ref{FermionFrAk1rAk1k2}) in the fermionic chain.
The result is shown in figure~\ref{FigureBosonF}.

\subsubsection{$\r_{A,K_1}$ VS $\r_{A,K_2}$}

For more general cases, we calculate the fidelity numerically and will not show the results in this paper.

\section{Nearest-neighbor coupled bosonic chain} \label{SectionInteractingBoson}

We use the correlation matrix method and check the three conjectures for the Schatten distances with even integer indices in the quasiparticle excited states of the nearest-neighbor coupled bosonic chain.

\subsection{Quasiparticle excited states}

We consider the chain of nearest-neighbor coupled harmonic oscillators
\be
H = \f{1}{2} \sum_{j=1}^L \big[ p_j^2 + m^2 q_j^2 + (q_j-q_{j+1})^2 \big],
\ee
with periodic boundary condition $q_{L+1}=q_1$.
It could be diagonalized as
\be
H = \sum_k \ve_k \Big( c_k^\dag c_k +\f12 \Big), ~~
\ve_k = \sr{m^2+4\sin^2\f{p_k}{2}},~~
p_k=\f{2\pi k}{L}.
\ee
The ground state $|G\rag$ is defined as
\be
c_k | G \rag = 0, ~ \forall k = 1-\f{L}{2}, \cdots,-1,0,1,\cdots,\f{L}{2}-1,\f{L}{2}.
\ee
A general excited state takes the form
\be
|K\rag=|k_1^{r_1}\cdots k_s^{r_s}\rag =
\f{( c^\dag_{k_1} )^{r_1} \cdots ( c^\dag_{k_s} )^{r_s}}{\sr{N_K}} | G \rag,
\ee
with the normalization factor $N_K = r_1!\cdots r_s!$.

\subsection{Wave function method} \label{wavefunctionmethod}

We denote the canonical coordinates of $A$ as $\cR=(q_1,\cdots,q_{\ell})$ and the canonical coordinates of $B$ as $\cS=(q_{\ell+1},\cdots,q_L)$. For each quasiparticle state $|K\rag$, we have the wave function $\lag\cR,\cS|K\rag$, which could be found for example in \cite{Zhang:2020txb}.
In the replica trick, there are $n$ copies of the system, and we have the canonical coordinates $\cQ=( \cR_1,\cS_1,\cdots,\cR_n,\cS_n )$ with $\cR_a=(q_{a,1},\cdots,q_{a,\ell})$ and $\cS_a=(q_{a,\ell+1},\cdots,q_{a,L})$, $a=1,\cdots,n$.
We get the trace of the product
\bea
&& \tr_A ( {\r_{A,K_1}\r_{A,K_2}\cdots\r_{A,K_n}} ) =
                   \int \rD \cQ
                   \lag \cR_1, \cS_1 | K_1 \rag \lag K_1 | \cR_2, \cS_1 \rag \\
&& \phantom{\tr_A ( {\r_{A,K_1}\r_{A,K_2}\cdots\r_{A,K_n}} ) =}
                   \times
                   \lag \cR_2, \cS_2 | K_2 \rag \lag K_2 | \cR_3, \cS_2 \rag
                   \cdots
                   \lag \cR_n, \cS_n | K_n \rag \lag K_n | \cR_1, \cS_n \rag, \nn
\eea
from which we calculate the Schatten distance with an even integer.

\subsection{Checks of the three conjectures}

\begin{figure}[p]
  \centering
  \includegraphics[height=\textwidth]{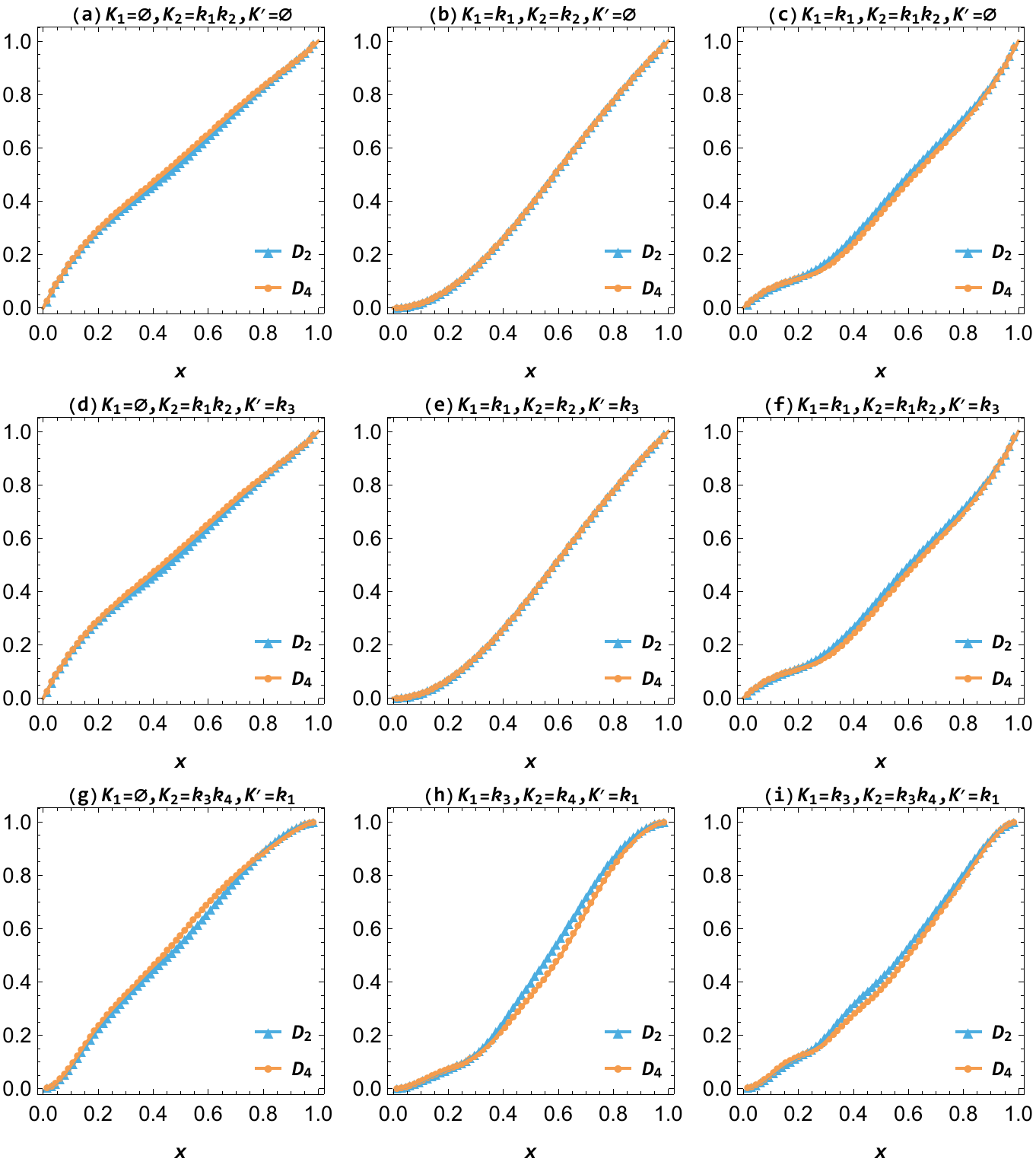}\\
  \caption{Checks of the first conjecture (\ref{conjecture1SD}) (the first row), the second conjecture (\ref{conjecture2SD}) (the second row), and the third conjecture (\ref{conjecture3SD}) (the third row) in the nearest-neighbor coupled bosonic chain. The symbols in each panel are numerical results for the normalized Schatten distance $D_n \equiv D_n^\bos(\r_{A,K_1\cup K'},\r_{A,K_2\cup K'};\r_{A,K'})$ with $n=2,4$, which are from the wave function method.
  In the first and the second rows, the solid lines are the analytical conjectured results from the subsystem mode method in the free bosonic chain.
  In the third row, the solid lines are the numerical conjectured results in the nearest-neighbor coupled bosonic chain.
  We have set $m=10^{-6}$, $(k_1,k_2)=(1,2)+\f{L}{4}$ and $(k_3,k_4)=(1,2)$.
  For the analytical results in the first and second rows we have set $L=+\infty$, and for all other cases we have set $L=64$.}
  \label{FigureBosonConjectures}
\end{figure}

We have introduced the three conjectures for subsystem distances in section~\ref{SectionRevSum}.
We check the first conjecture (\ref{conjecture1SD}), the second conjecture (\ref{conjecture2SD}), and the third conjecture (\ref{conjecture3SD}) in respectively the first row, second and third rows of figure~\ref{FigureBosonConjectures}.

\section{XXX chain} \label{SectionXXX}

In the spin-1/2 XXX chain, we focus on the trace distance and fidelity among the ferromagnetic ground state and the magnon excited states.

\subsection{Magnon excited states}

We consider the spin-1/2 XXX chain in positive transverse field $h>0$ with the Hamiltonian
\be \label{XXXHamiltonian}
H = -\f14 \sum_{j=1}^L ( \s_j^x\s_{j+1}^x + \s_j^y\s_{j+1}^y + \s_j^z\s_{j+1}^z ) - \f{h}{2} \sum_{j=1}^L \s_j^z,
\ee
and periodic boundary conditions $\s_{L+1}^{x,y,z}=\s_1^{x,y,z}$. We focus on the case with the total number of sites $L$ being four times of an integer.

The XXX chain is in the ferromagnetic phase, and the unique ground state is
\be
|G\rag = |\!\uparrow\uparrow\cdots\uparrow\rag.
\ee
The low-lying excited states are magnon excited states and can be obtained from the coordinate Bethe ansatz \cite{Gaudin:1983kpuCaux:2014uuq,Karbach:1998abi}.
We use the Bethe quantum numbers of the excited magnons $\cI=\{I_1,\cdots,I_m\}$, which are integers in the range $[0,I-1]$, to denote the magnon excited states.

A general magnon excited state $|\cI\rag=|I_1\cdots I_m\rag$ takes the form
\be \label{statepsi}
|\cI\rag = \f{1}{\sr{\cN}} \sum_{1\leq j_1 < \cdots <j_m\leq L} \cU_{j_1\cdots j_m} |j_1\cdots j_m\rag.
\ee
The normalization factor is
\be \label{statepsicN}
\cN = \sum_{1\leq j_1 < \cdots <j_m\leq L} | \cU_{j_1\cdots j_m} |^2.
\ee
We use $|j_1\cdots j_m\rag$ to denote configuration that the spins on the sites $j_1,\cdots,j_m$ are spin downward and all the other $L-m$ sites are spin upward.
The ansatz for the wave function is
\be \label{statepsicUj1cdotsjR}
\cU_{j_1\cdots j_m} = \sum_{\cP \in \cS_m}
\exp\Big( \ii \sum_{i=1}^R j_i p_{\cP i}
        + \f{\ii}{2} \sum_{1\leq i < i' \leq m} \th_{\cP i \cP i'} \Big),
\ee
where $\cS_m$ is the permutation group.
The phase $\th_{ii'}$ is determined by the equation
\be
\ep^{\ii\th_{ii'}} = - \f{1+\ep^{\ii(p_i+p_{i'})}-2\ep^{\ii p_{i}}}
                         {1+\ep^{\ii(p_{i}+p_{i'})}-2\ep^{\ii p_{i'}}}.
\ee
We always use the convention $p_i=\f{2\pi k_i}{L}$ with the actual momenta
\be
p_i = \f{2 \pi I_i}{L} + \f1L \sum_{i'\neq i} \th_{ii'},
\ee
and the momenta
\be \label{XXXkifromIi}
k_i = I_i + \f{1}{2\pi}\sum_{i'\neq i} \th_{ii'}.
\ee

When there is no ambiguity, we will also use the momenta of the excited magnons $K=\{k_1,\cdots,k_m\}$ to denote the same state. Note that nontrivial relation between the Bethe numbers and the momenta (\ref{XXXkifromIi}).

\subsection{Local mode method}

We have the subsystem $A=[1,\ell]$ and its complement $B=[\ell+1,L]$ in the state $|\cI\rag$ (\ref{statepsi})  with $m$ magnons.
We define the indices $\cX_i=(x_1,\cdots,x_i)$ to denote the configuration of the subsystem $A$ that the sites at $(x_1,\cdots,x_i)$ are flipped.
Similarly, we define $\cY_i=(y_{i+1},\cdots,y_m)$ to characterize the configurations of the subsystems $B$ that the sites at $(y_{i+1},\cdots,y_m)$ are flipped.
The tensor $\cU$ could be written as a $C_\ell^i \times C_{L-\ell}^{m-i}$ matrix $\cU_i$ with entries
\be
[\cU_i]_{\cX_i\cY_i} = \cU_{x_1 \cdots x_i y_{i+1} \cdots y_m}.
\ee
We write the magnon excited state $|\cI\rag$ in the orthonormal basis $|\cX_i \cY_i\rag$
\be
|\cI\rag = \f{1}{\sr{\cN}} \sum_{i=0}^m\sum_{\cX_i,\cY_i}[\cU_i]_{\cX_i \cY_i}|\cX_i \cY_i\rag.
\ee
Then we get the RDM
\be \label{rA}
\r_{A,\cI} =
\sum_{i=\max(0,m+\ell-L)}^{\min(\ell,m)}
\sum_{\cX_i,\cX'_i} [\cV_i]_{\cX_i\cX'_i} |\cX_i\rag \lag \cX'_i|,
\ee
with the $C_\ell^i \times C_\ell^i$ matrix
\be
\cV_i \equiv \f{\cU_i \cU_i^\dag}{\cN}.
\ee
The matrix $\cV_i$ is well-defined only for $i$ in the range
\be
\max(0,m+\ell-L) \leq i \leq \min(\ell,m),
\ee
and we also define $\cV_i=0$ for other values of $i$.

For another general state $|\cI'\rag$ with $m'$ particle, we get the RDM similar to (\ref{rA})
\be
\r_{A,\cI'} =
\sum_{i=\max(0,m'+\ell-L)}^{\min(\ell,m')}
\sum_{\cX_i,\cX'_i} [\cV'_i]_{\cX_i\cX'_i} |\cX_i\rag \lag \cX'_i|,
\ee
with $\cV'_i$ defined in the same way as above.
We get the Schatten distance and trace distance
\be
D_n(\r_{A,\cI},\r_{A,\cI'})= \Big[ \f12
\sum_{i=\min( \max(0,m+\ell-L), \max(0,m'+\ell-L) )}^{\max(\min(\ell,m),\min(\ell,m'))}
\tr ( | \cV_i - \cV'_i |^n) \Big]^{1/n},
\ee
\be \label{XXXD1formulaLMM}
D_1(\r_{A,\cI},\r_{A,\cI'}) =  \f12
\sum_{i=\min( \max(0,m+\ell-L), \max(0,m'+\ell-L) )}^{\max(\min(\ell,m),\min(\ell,m'))}
 \tr | {\cV_i} - {\cV'_i} |,
\ee
\be
F(\r_{A,\cI},\r_{A,\cI'}) = \sum_{i=\max(0,m+\ell-L,m'+\ell-L)}^{\min(\ell,m,m')}
\tr  [ ( \cV_{i}^{1/2}\cV'_{i}\cV_{i}^{1/2} )^{1/2} ].
\ee

For two states $|\cI\rag$ and $|\cI'\rag$ with $m\leq m'$, we have the formula of trace distance
\be
D_1(\r_{A,\cI},\r_{A,\cI'}) = \f12 \Big( 1
 + \sum_{i=\max(0,m+\ell-L)}^{\min(\ell,m)} \tr | {\cV_i} - {\cV'_i} |
 - \sum_{i=\max(0,m'+\ell-L)}^{\min(\ell,m)} \tr \cV'_i \Big),
\ee
which is a little more efficient for numerical evaluations than formula (\ref{XXXD1formulaLMM}).

\subsection{$\r_{A,G}$ VS $\r_{A,\cI}$}

For the subsystem $A=[1,\ell]$, we get the trace distance and fidelity between the ground state RDM $\r_{A,G}$ and the RDM in the general magnon excited state $|\cI\rag=|I_1\cdots I_m\rag$
\be \label{XXXD1GcI}
D_1^\XXX(\r_{A,G},\r_{A,\cI}) = 1 - \cF_{A,G,\cI}^\XXX,
\ee
\be \label{XXXFGcI}
F^\XXX(\r_{A,G},\r_{A,\cI}) = \sqrt{\cF_{A,G,\cI}^\XXX},
\ee
with the coefficient
\be
\cF_{A,G,\cI}^\XXX \equiv \tr_A(\r_{A,G}\r_{A,\cI}) = \f1\cN \sum_{\ell+1\leq j_1 < \cdots <j_m\leq L} | \cU_{j_1\cdots j_m} |^2.
\ee
As the fidelity $F^\XXX(\r_{A,G},\r_{A,\cI})$ (\ref{XXXFGcI}) is simply related to the trace distance $D_1^\XXX(\r_{A,G},\r_{A,\cI})$ (\ref{XXXD1GcI}), in the following subsections we will only show several examples of the trace distance when one of the two states is the ground state.

\subsection{$\r_{A,G}$ VS $\r_{A,I}$}

The single-magnon state is
\be
|I\rag = \f{1}{\sqrt{L}} \sum_{j=1}^L \ep^{\f{2\pi\ii j I }{L}} |j\rag,
\ee
with the Bethe quantum number $I\in[0,L-1]$.
There are physical momentum $p$ and momentum $k$ with the relation $p=\f{2\pi k}{L}=\f{2\pi I}{L}$.
We obtain the trace distance
\be
D_1^\XXX(\r_{A,G},\r_{A,I}) = x,
\ee
which is the same as the trace distances in the free fermionic chain $D_1^\fer(\r_{A,G},\r_{A,k})$ (\ref{FermionDnD1Gk}), and free bosonic chain $D_1^\bos(\r_{A,G},\r_{A,k^r})$ (\ref{BosonD1Gkr}) with $r=1$, and the universal trace distance  $D_1^\univ(\r_{A,G},\r_{A,k})$ (\ref{UniversalDnD1Gk}).

\subsection{$\r_{A,G}$ VS $\r_{A,I_1I_2}$}

The double-magnon states could be scattering states or bound states and take the form
\be
|I_1I_2\rag = \f{1}{\sr{\cN}} \sum_{1\leq j_1 < j_2 \leq L} \cU_{j_1j_2}|j_1j_2\rag,
\ee
with Bethe numbers $I_1,I_2$ satisfying $0\leq I_1\leq I_2\leq L-1$ and
\be
\cU_{j_1j_2} = \ep^{\ii(j_1p_1+j_2p_2+\f12\th)} + \ep^{\ii(j_1p_2+j_2p_1-\f12\th)}.
\ee
The normalization factor is
\be
\cN=\sum_{1\leq j_1 < j_2 \leq L} |\cU_{j_1j_2}|^2.
\ee
The two magnons have physical momenta $p_1,p_2$ and momenta $k_1,k_2$ being related as
\be
p_1 = \f{2\pi k_1}{L}, ~~
p_2 = \f{2\pi k_2}{L}.
\ee
Note that $k_1,k_2$ may not necessarily be integers or half-integers and may be possibly complex numbers for bound states.
The total physical momentum, total momentum, and total Bethe number of the state are
\be
p=p_1+p_2, ~~ k=k_1+k_2, ~~ I=I_1+I_2,
\ee
with
\be
p = \f{2\pi k}{L}, ~~ k = I.
\ee
The total Bethe number $I$ is an integer in the range $[0,2L-2]$.
The angle $\th$ is determined by the equation
\be \label{Betheequation}
\ep^{\ii\th} = - \f{1+\ep^{\ii(p_1+p_2)}-2\ep^{\ii p_1}}{1+\ep^{\ii(p_1+p_2)}-2\ep^{\ii p_2}}.
\ee
To the equation (\ref{Betheequation}), there are three classes of solutions \cite{Karbach:1998abi}, which we reorganize into three cases following \cite{Zhang:2021bmy}.

\subsubsection{Case I state}

For the case I state, there are
\be
p_1=I_1=k_1=p_2=I_2=k_2=\th=0,
\ee
and the state is
\be
|00\rag = \sr{\f{2}{L(L-1)}} \sum_{1\leq j_1<j_2\leq L} |j_1j_2\rag.
\ee
We get the trace distance
\be
D_1^\XXX(\r_{A,G},\r_{A,00}) = 1-\frac{(L-\ell)(L-\ell-1)}{L(L-1)}.
\ee
In the scaling limit, it is just
\be
D_1^\XXX(\r_{A,G},\r_{A,00}) = x(2-x) = D_1^\bos(\r_{A,G},\r_{A,k^2}),
\ee
where $D_1^\bos(\r_{A,G},\r_{A,k^2})$ is the trace distance in the bosonic chain, i.e.\ the $r=2$ case of $D_1^\bos(\r_{A,G},\r_{A,k^r})$ (\ref{BosonD1Gkr}).

\subsubsection{Case II states}

For case II states, there are
\be
p_1=\f{2\pi I_1+\th}{L}, ~~ p_2=\f{2\pi I_2-\th}{L},
\ee
and
\be
k_1=I_1+\f{\th}{2\pi}, ~~ k_2=I_2-\f{\th}{2\pi}.
\ee
The angle $\th \in [-\pi,\pi]$ is a real solution to the equation (\ref{Betheequation}).
In the scaling limit, there is
\be \label{XXXBetheAngle}
\lim_{L\to+\inf} \theta = f(\io_1,\io_2),
\ee
with the scaled Bethe numbers
\be \label{ratiosi1i2}
\io_1 = \lim_{L\to+\inf} \f{I_1}{L}, ~~ \io_2 = \lim_{L\to+\inf} \f{I_2}{L}.
\ee
There are properties that
\be \label{f0i2fi110}
f(0,\io_2)=f(\io_1,1)=0,
\ee
and
\be \label{fiipi}
f(\io,\io)=\pi, ~ 0<\io<1.
\ee

We get the trace distance
\bea
&& \hspace{-9mm}
   D_1^\XXX(\r_{A,G},\r_{A,I_1I_2}) = 1 \label{XXXD1GI1I2caseII} \\
&& \hspace{-9mm} ~~~
   - \f{(L-\ell)(L-\ell-1)(1-\cos p_{12})+(L-\ell)\cos(p_{12}-\th)-(L-\ell-1)\cos\th-\cos[(L-\ell)p_{12}-\th]}
       {L(L-1)(1-\cos p_{12})+L\cos(p_{12}-\th)-(L-1)\cos\th-\cos(L p_{12}-\th)}, \nn
\eea
which in the scaling limit interpolates between the results in the fermionic and bosonic chain.
We have $p_{12} = p_1 - p_2 = \f{2\pi k_{12}}{L}$ and $k_{12}=k_1-k_2=I_1-I_2+\f{\th}{\pi}$.
There are three cases in the scaling limit.
\begin{itemize}
  \item For the cases with $\io_1=\io_2=0$, or $\io_1=\io_2=1$, or $\io_1=0$ and $\io_2=1$, there is $\th=0$. The trace distance (\ref{XXXD1GI1I2caseII}) becomes
    \be
    D_1^\XXX(\r_{A,G},\r_{A,I_1I_2}) = x(2-x) - \Big[ \f{\sin(\pi k_{12} x)}{\pi k_{12}} \Big]^2 = D_1^\bos(\r_{A,G},\r_{A,k_1k_2}),
    \ee
    where $D_1^\bos(\r_{A,G},\r_{A,k_1k_2})$ is the trace distance in the free bosonic chain (\ref{BosonD1Gk1k2}).
  \item For the cases with $\io_1=\io_2\in(0,1)$, here is $\th=\pi$. The trace distance (\ref{XXXD1GI1I2caseII}) becomes
    \be
    D_1^\XXX(\r_{A,G},\r_{A,I_1I_2}) = x(2-x) + \Big[ \f{\sin(\pi k_{12} x)}{\pi k_{12}} \Big]^2 = D_1^\fer(\r_{A,G},\r_{A,k_1k_2}),
    \ee
    where $D_1^\fer(\r_{A,G},\r_{A,k_1k_2})$ is the trace distance in the free fermionic chain (\ref{FermionD1Gk1k2}).
  \item For all the other cases, there is the large momentum difference
     \be
     \lim_{L\to+\inf}|k_{12}| = +\inf.
     \ee
     The trace distance (\ref{XXXD1GI1I2caseII}) becomes
     \be
     D_1^\XXX(\r_{A,G},\r_{A,I_1I_2}) = x(2-x) = D_1^\univ(\r_{A,G},\r_{A,k_1k_2}),
     \ee
     where $D_1^\univ(\r_{A,G},\r_{A,k_1k_2})$ (\ref{UniversalD1Gk1k2}) is the universal trace distance from the quasiparticle picture, i.e. $r=2$ case of $D^\univ_1( \r_{A,G}, \r_{A,k_1\dots k_r} )$ (\ref{FermionD1univGK}) and $R=2$ case of $D_1^\univ(\r_{A,G},\r_{A,K})$ (\ref{BosonD1univ}).
\end{itemize}

\subsubsection{Case IIIa states}

One case III state is a bound state of two magnons with the length scale $1/v$.
When $v\to0$ the two magnons are loosely bounded, and when $v\to+\inf$ the two magnons are tightly bounded.
In the case III cases, there are two subclasses depending on the Bethe numbers, which we call the case IIIa cases and the case IIIb cases.

For case IIIa solutions there are
\be
p_1=\f{\pi I}{L}+\ii v, ~~
p_2=\f{\pi I}{L}-\ii v, ~~
\th=\pi+\ii L v,
\ee
with the possible values of the total Bethe number
\be
I=\td I,\td I+2,\cdots,\f{L}{2}-1,\f{3L}{2}+1,\f{3L}{2}+3,\cdots,2L-\td I.
\ee
There is the odd integer $\td I \approx \sr{L}/\pi$.
The Bethe numbers of the two magnons are
\be
I_1=\f{I-1}{2}, ~~
I_2=\f{I+1}{2}.
\ee

We get the trace distance
\bea
&& \hspace{-9mm}
D_1^\XXX(\r_{A,G},\r_{A,I_1I_2}) = 1  \label{XXXD1GI1I2caseIIIa} \\
&& \hspace{-9mm} ~~~
 - \f{2(L-\ell)\sinh v\sinh[(L-1)v] - \cosh(Lv) +\cosh[(L-2\ell)v] - 2(L-\ell)(L-\ell-1)\sinh^2v}
     {2L [ \sinh((L-1)v) - (L-1)\sinh v ] \sinh v}. \nn
\eea
In the scaling limit, the parameter $v$ is in the range
\be
\f{1}{2L} \lesssim v \leq +\inf.
\ee
For $v=\f{u}{L}$ with fixed $u$ in the scaling limit, the trace distance (\ref{XXXD1GI1I2caseIIIa}) becomes
\be \label{XXXD1GI1I2caseIIIap}
D_1^\XXX(\r_{A,G},\r_{A,I_1I_2}) = 1 - \f{2(1-x)u[\sinh u-(1-x)u]-\cosh u+\cosh[(1-2x)u]}{2u(\sinh u - u)}.
\ee

\begin{figure}[t]
  \centering
  \includegraphics[height=0.31\textwidth]{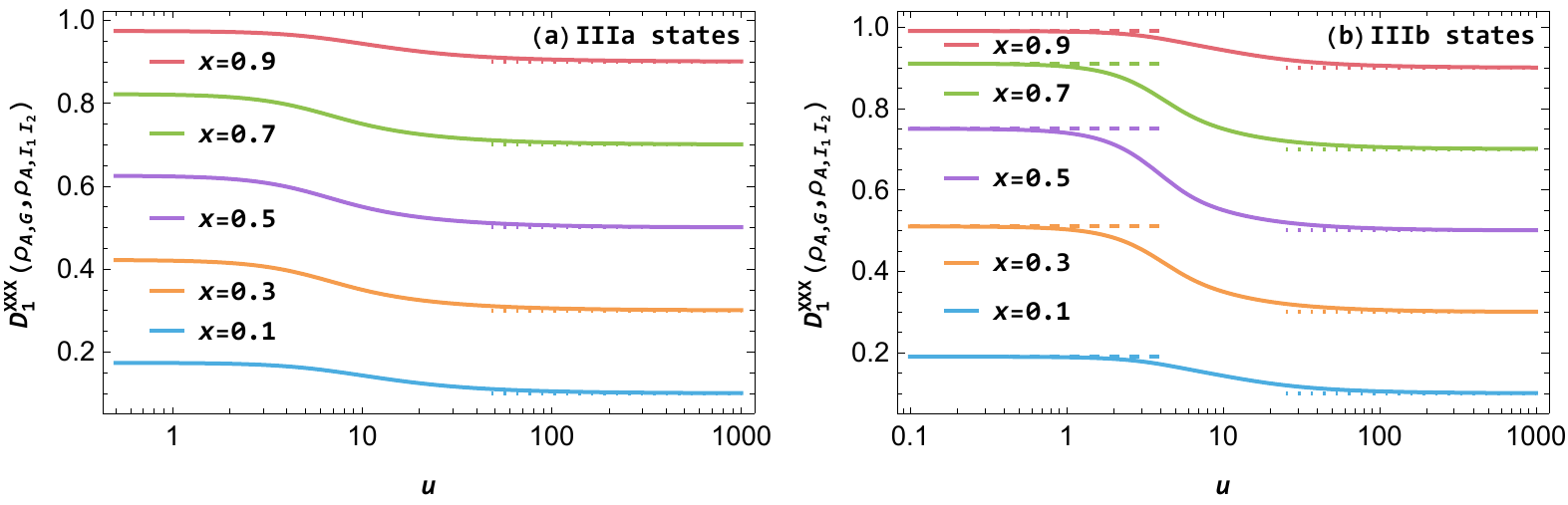}\\
  \caption{The subsystem trace distances between the ground state and the case IIIa double-particle bound states $D_1^\XXX(\r_{A,G},\r_{A,I_1I_2})$ (\ref{XXXD1GI1I2caseIIIap}) (left) and the trace distance between the ground state and the case IIIb double-particle bound states $D_1^\XXX(\r_{A,G},\r_{A,I_1I_2})$ (\ref{XXXD1GI1I2caseIIIbp}) (right) in the XXX chain.
  The horizonal axes $u$ is defined as $v=\f{u}{L}$.
  In the left and right panels, the dotted lines are the lower bound of the trace distance $D_1(\r_{A,G},\r_{A,k})=x$.
  In the right panel, the dashed lines are the upper bound $D_1^\bos(\r_{A,G},\r_{A,k^2})=x(1-x)$.}
  \label{FigureXXXD1}
\end{figure}

We summarize the trace distance (\ref{XXXD1GI1I2caseIIIa}) in the table~\ref{tableIIIa}.
We show the trace distance (\ref{XXXD1GI1I2caseIIIap}) in the left panel of figure~\ref{FigureXXXD1}.

\begin{table}[t]
  \centering
  \begin{tabular}{|c|c|c|c|c|c|}
  \hline
  \mr{2}{IIIa states}       & \mc{2}{c|}{$v=\f{u}{L}$}  & \mr{2}{$\cdots$} & \mr{2}{$v$ finite} & \mr{2}{$v\to+\inf$}  \\ \cline{2-3}
                            & $u$ finite & $u\to+\inf$  &                  &            &              \\ \hline
   \mr{2}{$D_1^\XXX(\r_{A,G},\r_{A,I_1I_2})$} & \mc{2}{c|}{(\ref{XXXD1GI1I2caseIIIap})} & \mc{3}{c|}{} \\ \cline{3-3}
   & \mc{1}{c|}{} & \mc{4}{c|}{$D_1(\r_{A,G},\r_{A,k})=x$}\\\hline
  \end{tabular}
  \caption{The subsystem trace distance between the ground state and the case IIIa double-magnon bound states in XXX chain.}
  \label{tableIIIa}
\end{table}

\subsubsection{Case IIIb states}

For case IIIb states there are
\be
p_1=\f{\pi I}{L}+\ii v, ~~
p_2=\f{\pi I}{L}-\ii v, ~~
\th=\ii L v,
\ee
with the possible values of the total Bethe number
\be
I=2,4,\cdots,\f{L}{2},\f{3L}{2},\cdots,2L-2,\ee
and the Bethe numbers of the two quasiparticles
\be
I_1=I_2=\f{I}{2}.
\ee

We get the analytical trace distance
\bea
&& \hspace{-9mm}
D_1^\XXX(\r_{A,G},\r_{A,I_1I_2}) = 1 \label{XXXD1GI1I2caseIIIb} \\
&& \hspace{-9mm} ~~~
 - \f{2(L-\ell)\sinh v\sinh[(L-1)v] - \cosh(Lv) + \cosh[(L-2\ell)v] + 2(L-\ell)(L-\ell-1)\sinh^2v}
     {2L [ \sinh((L-1)v) + (L-1)\sinh v ] \sinh v}. \nn
\eea
In the scaling limit, the parameter $v$ is in the range
\be
\f{2\pi^2}{L^2} \lesssim v \leq +\inf.
\ee
For $v=\f{u}{L}$ with fixed $u$ in the scaling limit, the trace distance (\ref{XXXD1GI1I2caseIIIb}) becomes
\be \label{XXXD1GI1I2caseIIIbp}
D_1^\XXX(\r_{A,G},\r_{A,I_1I_2}) = 1 - \f{2(1-x)u[\sinh u+(1-x)u]-\cosh u+\cosh[(1-2x)u]}{2u(\sinh u + u)}.
\ee

We summarize the trace distance (\ref{XXXD1GI1I2caseIIIb}) in table~\ref{tableIIIb}.
We show the trace distance (\ref{XXXD1GI1I2caseIIIbp}) in the left panel of figure~\ref{FigureXXXD1}.

\begin{table}[t]
  \centering
  \begin{tabular}{|c|c|c|c|c|c|c|c|c|}
  \hline
  \mr{2}{IIIb states}               & \mr{2}{$v=\f{w}{L^2}$} & \mr{2}{$\cdots$} & \mc{3}{c|}{$v=\f{u}{L}$}            & \mr{2}{$\cdots$} & \mr{2}{$v$ finite} & \mr{2}{$v\to+\inf$}  \\ \cline{4-6}
                                    &                        &                  & $u\to0$ & $u$ finite & $u\to+\inf$  &                  &            &              \\ \hline
  \mr{2}{$D_1^\XXX(\r_{A,G},\r_{A,I_1I_2})$} & \mc{2}{c|}{}                              & \mc{3}{c|}{(\ref{XXXD1GI1I2caseIIIbp})}                        & \mc{3}{c|}{} \\ \cline{4-4} \cline{6-6}
                                    & \mc{3}{c|}{$D_1^\bos(\r_{A,G},\r_{A,k^2})=x(2-x)$}                  &            & \mc{4}{c|}{$D_1(\r_{A,G},\r_{A,k})=x$}\\\hline
\end{tabular}
  \caption{The subsystem trace distance between the ground state and the case IIIb double-magnon bound state in XXX chain.}
  \label{tableIIIb}
\end{table}

\subsection{$\r_{A,I_1}$ VS $\r_{A,I_2}$}

We get from the local mode method the trace distance and fidelity
\be \label{XXXD1I1I2}
D_1^\XXX(\r_{A,I_1},\r_{A,I_2}) = \sr{x^2 - \Big| \f{\sin\f{\pi I_{12} \ell}{L}}{L \sin\f{\pi I_{12}}{L}} \Big|^2},
\ee
\be \label{XXXFI1I2}
F^\XXX(\r_{A,I_1},\r_{A,I_2}) = 1 - x + \Big| \f{\sin\f{\pi I_{12} \ell}{L}}{L \sin\f{\pi I_{12}}{L}} \Big|,
\ee
with the difference of the Bethe numbers $I_{12}\equiv I_1-I_2$. Note that for the single-magnon state, the Bethe number $I\in[0,L-1]$ is the same as the momentum $k$.
The trace distance (\ref{XXXD1I1I2}) is the same as the trace distances in the free fermionic and bosonic chains (\ref{FermionDnD1k1k2}) and (\ref{BosonDnD1k1k2}).
The fidelity (\ref{XXXFI1I2}) is the same as the fidelities in the free fermionic and bosonic chains (\ref{FermionFrAk1rAk2}) and (\ref{BosonFrAk1rAk2}).

\subsection{Check of three conjectures}

As the XXX chain we consider has a finite positive transverse field, the model is finitely gapped, and the excited magnons always have large energies in the scaling limit, i.e. the large energy condition is always satisfied.
We reformulate the three conjectures for the subsystem distances among the ferromagnetic ground state and low-lying magnon scattering states in the ferromagnetic XXX chain.
\begin{itemize}
  \item For state $|\cI\rag=|I_1\cdots I_m\rag$ with finite number of excited magnons $m$ in the scaling limit, we follow \cite{Zhang:2021bmy} and group the magnons into $\a$ clusters according to the scaled Bethe numbers
  \be \io_i = \lim_{L\to+\infty} \f{I_i}{L}, ~ i=1,2,\cdots,m. \ee
  The magnons with $\io_i=0$ and the magnons with $\io_i=1$ are grouped in the same cluster, and other magnons with the same $\io_i\in(0,1)$ are grouped in the same cluster. We have $\cI=\bigcup_{a=1}^\a \cI_a$ with $\cI_a=\{I_{ab}|b=1,2,\cdots,\b_a\}$ and $m=\sum_{a=1}^\a\b_a$. We also denote the same state by the momenta $|K\rag=|k_1\cdots k_m\rag$ with the momenta related to the Bethe numbers as (\ref{XXXkifromIi}). The momenta are grouped into $\a$ clusters in the same way for the Bethe numbers $K=\bigcup_{a=1}^\a K_a$. We conjecture the trace distance and fidelity
  \be \label{XXXConjectureID1}
  D_1^\XXX(\r_{A,G},\r_{A,K}) = 1 - \prod_{a=1}^\a \cF_{A,G,K_a}^\XXX,
  \ee
  \be  \label{XXXConjectureIF}
  F^\XXX(\r_{A,G},\r_{A,K}) = \prod_{a=1}^\a \sqrt{\cF_{A,G,K_a}^\XXX},
  \ee
  with
  \be
  \cF_{A,G,K_a}^\XXX \equiv \tr_A(\r_{A,G}\r_{A,K_a}).
  \ee
  For the cluster $K_a$ with $\io_a=0$ or $\io_a=1$, there is
  \be
  \cF_{A,G,K_a}^\XXX = \cF_{A,G,K_a}^\bos.
  \ee
  For a cluster $K_a$ with $\io_a\in(0,1)$, there is
  \be
  \cF_{A,G,K_a}^\XXX = \cF_{A,G,K_a}^\fer.
  \ee
  Here $\cF_{A,G,K_a}^\bos$ and $\cF_{A,G,K_a}^\fer$ are results in respectively the free fermionic and bosonic chains.
  \item For two states $|K'\rag$ and $|K\cup K'\rag$ denoted by the momenta satisfying the large momentum difference condition
  \be
  |k-k'|\to+\infty, ~ \forall k\in K, \forall k' \in K',
  \ee
  we conjecture the trace distance and fidelity
  \be \label{XXXConjectureIID1}
  D_1^\XXX(\r_{A,K'},\r_{A,K\cup K'}) = D_1^\XXX(\r_{A,G},\r_{A,K}),
  \ee
  \be  \label{XXXConjectureIIF}
  F^\XXX(\r_{A,K'},\r_{A,\cup K'}) = F^\XXX(\r_{A,G},\r_{A,K}).
  \ee
  The RHS of the conjecture (\ref{XXXConjectureIID1}) and (\ref{XXXConjectureIIF}) could be further simplified according the first conjecture  (\ref{XXXConjectureID1}) and (\ref{XXXConjectureIF}).
  \item For two states $|K_1\cup K'\rag$ and $|K_2\cup K'\rag$ denoted by the momenta satisfying the condition
  \be
  |k-k'|\to+\infty, ~ \forall k\in K_1\cup K_2, \forall k' \in K',
  \ee
  we conjecture the trace distance and fidelity
  \be \label{XXXConjectureIIID1}
  D_1^\XXX(\r_{A,K_1\cup K'},\r_{A,K_2\cup K'}) = D_1^\XXX(\r_{A,K_1},\r_{A,K_2}),
  \ee
  \be  \label{XXXConjectureIIIF}
  F^\XXX(\r_{A,K_1\cup K'},\r_{A,K_2\cup K'}) = F^\XXX(\r_{A,K_1},\r_{A,K_2}).
  \ee
\end{itemize}

\begin{figure}[p]
  \centering
  \includegraphics[height=\textwidth]{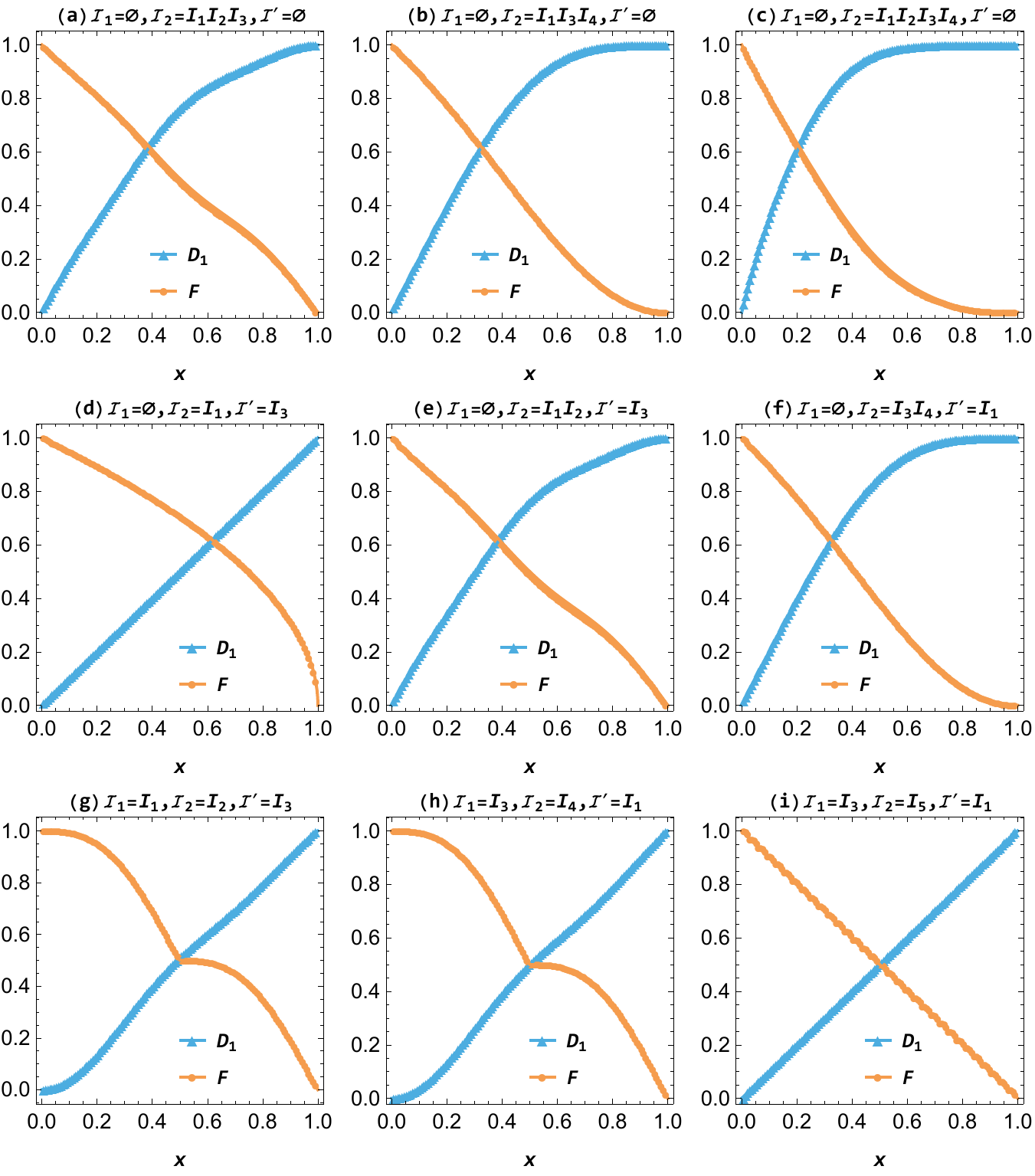}\\
  \caption{Checks of the first conjecture (\ref{XXXConjectureID1}) and (\ref{XXXConjectureIF}) (the first row), the second conjecture (\ref{XXXConjectureIID1}) and (\ref{XXXConjectureIIF}) (the second row), and the third conjecture (\ref{XXXConjectureIIID1}) and (\ref{XXXConjectureIIIF}) (the third row) in the ferromagnetic phase of the spin-1/2 XXX chain.
  The symbols in each panel are numerical results for the trace distance $D_1 \equiv D_1^\XXX(\r_{A,K_1\cup K'},\r_{A,K_2\cup K'})$ and fidelity $F \equiv F^\XXX(\r_{A,K_1\cup K'},\r_{A,K_2\cup K'})$, which are from the local mode method.
  The solid lines are the analytical conjectured results from the subsystem mode method in the free fermionic and bosonic chains.
  We have used the Bethe numbers of the excited magnons to denote the states.
  We have set the Bethe numbers $(I_1,I_2,I_3,I_4,I_5)=(1,3,\f{L}{4},\f{L}{4}+2,\f{L}{2})$.
  For the analytical results  we have set $L=+\infty$, and for numerical results we have set $L=128$.}
  \label{FigureXXXConjectures}
\end{figure}

We check the above three conjectures in the XXX chain in figure~\ref{FigureXXXConjectures}.

\section{Conclusion and discussion} \label{SectionConclusion}

We have calculated the subsystem Schatten distance, trace distance and fidelity in the quasiparticle excited states of free and coupled fermionic and bosonic chains and the ferromagnetic phase of the spin-1/2 XXX chain from various methods and found consistency for the results.
In the free fermionic and bosonic chains, we obtained the subsystem distances from the subsystem mode method, which are still valid in the coupled fermionic and bosonic chains and the XXX chain under certain limit.
We followed the universal R\'enyi and entanglement entropies in \cite{Castro-Alvaredo:2018dja,Castro-Alvaredo:2018bij,Castro-Alvaredo:2019irt,Castro-Alvaredo:2019lmj} and obtained the universal R\'enyi and entanglement entropies in the large energy and large momentum difference limit.
More generally, we followed the three conjectures for the R\'enyi and entanglement entropies in \cite{Zhang:2021bmy} and formulated three conjectures for subsystem distances and checked the conjectures in the coupled fermionic and bosonic chains and XXX chain.
The results in this paper support the scenario that quasiparticles with large energies decouple from the ground state and two sets of quasiparticles with large momentum differences decouple from each other. In particular, we think that the same kind of phenomena should be valid in other integrable models too. Most notably, following the ideas in \cite{Zhang:2021bmy} combined with the results of current paper, calculating the universal subsystem trace distances and their corrections in the XXZ chain is straightforward.

The trace distance is usually difficult to evaluate.
For the cases with a few quasiparticles excited in the free fermionic and bosonic chain, we could calculate the trace distance using the subsystem mode method.
To calculate the trace distance directly in the coupled fermionic chain, we need to construct the explicit RDMs and this method is unfortunately only applicable for a subsystem a very small number of sites.
It is worse in the coupled bosonic chain, and we do not have a direct way to calculate the trace distance, even for a small subsystem.
In the coupled bosonic chain, it is also difficult to calculate the fidelity.
We hope to come back to these problems in the future.

\section*{Acknowledgements}

We thank Olalla Castro-Alvaredo and Benjamin Doyon for reading a previous version of the draft and helpful comments and suggestions.
MAR thanks CNPq and FAPERJ (grant number 210.354/2018) for partial support.

\appendix

\section{Calculations for states in nonorthonormal basis} \label{appNOB}

In this appendix, we give an efficient procedure to calculate the Schatten and trace distances and fidelity for density matrices in a general nonorthonormal basis, similar to the calculations of the R\'enyi and entanglement entropies in \cite{Zhang:2021bmy}.

We consider the general density matrix
\be \label{rcP}
\r_\cP=\sum_{i,j} \cP_{ij} |\phi_i\rag\lag\phi_j|,
\ee
in a general nonorthonormal basis $|\phi_i\rag$ with the positive matrix
\be
\cQ_{ij}=\lag\phi_i|\phi_j\rag.
\ee
It is convenient to define another matrix
\be
\cR=\cP\cQ.
\ee
Here $\r_\cP$ could be the density matrix of the total system or the RDM of a subsystem.
For two density matrices $\r_\cP$, $\r_{\cP'}$, there are the Schatten and trace distances
\be \label{DnrcPrcPp}
D_n(\r_\cP,\r_{\cP'}) = \f{1}{2^{1/n}} ( \tr|\cR-\cR'|^n )^{1/n},
\ee
\be \label{D1rcPrcPp}
D_1(\r_\cP,\r_{\cP'}) = \f{1}{2} \tr|\cR-\cR'|.
\ee
When the matrices $\cR$ and $\cR'$ are of a finite dimension, we calculate the Schatten and trace distances directly, without resorting to the replica trick.

The positive matrix $\cQ$ could be written in terms of the unitary matrix $U$ and the positive diagonal matrix $\Lam$
\be
\cQ = U \Lam U^\dag,
\ee
with $U$ and $\Lam$ being constructed respectively by the eigenvectors and eigenvalues of $\cQ$.
We get the orthonormal basis
\be
|\psi_i\rag = \sum_j (U \Lam^{-1/2})_{ji} |\phi_j\rag,
\ee
satisfying $\lag\psi_i|\psi_j\rag=\d_{ij}$.
We further write the density matrix as
\be
\r_\cP=\sum_{i,j} \cS_{ij} |\psi_i\rag\lag\psi_j|,
\ee
with
\be
\cS = \Lam^{1/2} U^\dag \cP U \Lam^{1/2}.
\ee
Then we get the fidelity of two density matrices $\r_\cP$, $\r_{\cP'}$
\be
F(\r_\cP,\r_{\cP'})=\tr [ (\cS^{1/2}\cS'\cS^{1/2})^{1/2} ].
\ee
Noting that $\cS=\Lam^{1/2}U^\dag\cR U\Lam^{-1/2}$ and $\cS'=\Lam^{1/2}U^\dag\cR'U\Lam^{-1/2}$ we obtain the fidelity calculated as
\be
F(\r_\cP,\r_{\cP'})=\tr [ (\cR^{1/2}\cR'\cR^{1/2})^{1/2} ].
\ee

When the matrices $\cP$, $\cQ$, $\cR$, $\cS$ are block diagonal
\be
\cP = \bigoplus_b \cP_{(b)}, ~~
\cQ = \bigoplus_b \cQ_{(b)}, ~~
\cR = \bigoplus_b \cR_{(b)}, ~~
\cS = \bigoplus_b \cS_{(b)}.
\ee
We further write the Schatten and trace distances and fidelity as
\be
D_n(\r_\cP,\r_{\cP'}) = \Big( \f12\sum_b \tr|\cR_{(b)}-\cR'_{(b)}|^n \Big)^{1/n},
\ee
\be
D_1(\r_\cP,\r_{\cP'}) = \f{1}{2} \sum_b \tr|\cR_{(b)}-\cR'_{(b)}|,
\ee
\be
F(\r_\cP,\r_{\cP'}) 
                    = \sum_b \tr \big[ (\cR_{(b)}^{1/2} \cR'_{(b)} \cR_{(b)}^{1/2})^{1/2} \big].
\ee

\section{Derivation of the recursive formula (\ref{RecursiveC})} \label{appRec}

In this appendix, we give a derivation of the recursive formula (\ref{RecursiveC}), following the derivation of (\ref{RecursiveGamma}) in \cite{Fagotti:2010yr}.

For the interval $A=[1,\ell]$, the RDM $\r_C$ corresponding to the $\ell\times\ell$ correlation matrix $C$ is \cite{Cheong:2002ukf,Peschel:2002jhw}
\be
\r_C=\det(1-C)\ep^{-c^\dag H c},
\ee
with the relation $H=\log\f{1-C}{C}$ and the shorthand
\be
c^\dag H c \equiv \sum_{j_1,j_2=1}^\ell H_{j_1j_2}c^\dag_{j_1}c_{j_2}.
\ee
Note that the RDM has been properly normalized $\tr \r_C=1$.
From
\be
\r_{C_1}\r_{C_2}=\tr(\r_{C_1}\r_{C_2})\r_{C_3},
\ee
we get
\bea
&& \tr(\r_{C_1}\r_{C_2}) = \f{\det(1-C_1)\det(1-C_2)}{\det(1-C_3)}, \nn\\
&& \ep^{-c^\dag H_1 c}\ep^{-c^\dag H_2 c}=\ep^{-c^\dag H_3 c}.
\eea
Note that
\be
[c^\dag H_1 c,c^\dag H_2 c]=c^\dag [H_1,H_2] c,
\ee
we get from the Baker-Campbell-Hausdorff formula
\be
\ep^{-H_1}\ep^{-H_2}=\ep^{-H_3},
\ee
which is just
\be
\f{C_1}{1-C_1}\f{C_2}{1-C_2}=\f{C_3}{1-C_3}.
\ee
With some simple algebra, we get
\be
C_3=C_1(1-C_1-C_2+2C_2C_1)^{-1}C_2=1-(1-C_2)(1-C_1-C_2+2C_1C_2)^{-1}(1-C_1).
\ee
Then we obtain the trace
\be
\tr(\r_{C_1}\r_{C_2}) = \det(1-C_1-C_2+2C_1C_2).
\ee
Then the recursive formula (\ref{RecursiveC}) is derived.


\begin{thebibliography}{10}

\bibitem{Nielsen:2010oan}
M.~A. Nielsen and I.~L. Chuang, \textit{{Quantum Computation and Quantum
  Information}}.
\newblock Cambridge University Press, Cambridge, UK, 10th anniversary~ed.,
  2010,
  \href{http://dx.doi.org/10.1017/CBO9780511976667}{10.1017/CBO9780511976667}.

\bibitem{Watrous:2018rgz}
J.~Watrous, \textit{{The Theory of Quantum Information}}.
\newblock Cambridge University Press, Cambridge, UK, 2018,
  \href{http://dx.doi.org/10.1017/9781316848142}{10.1017/9781316848142}.

\bibitem{Coles:2019kdj}
P.~J. {Coles}, M.~{Cerezo} and L.~{Cincio}, \textit{{Strong bound between trace
  distance and Hilbert-Schmidt distance for low-rank states}},
  \href{http://dx.doi.org/10.1103/PhysRevA.100.022103}{Phys. Rev. A {\bfseries
  100}, 022103 (2019)}, [\href{https://arxiv.org/abs/1903.11738}{{\ttfamily
  arXiv:1903.11738}}].

\bibitem{Cerezo:2019tuq}
M.~{Cerezo}, A.~{Poremba}, L.~{Cincio} and P.~J. {Coles}, \textit{{Variational
  Quantum Fidelity Estimation}},
  \href{http://dx.doi.org/10.22331/q-2020-03-26-248}{Quantum {\bfseries 4}, 248
  (2020)}, [\href{https://arxiv.org/abs/1906.09253}{{\ttfamily
  arXiv:1906.09253}}].

\bibitem{Chen:2020zpo}
R.~Chen, Z.~Song, X.~Zhao and X.~Wang, \textit{{Variational quantum algorithms
  for trace distance and fidelity estimation}},
  \href{http://dx.doi.org/10.1088/2058-9565/ac38ba}{Quantum Sci. Technol.
  {\bfseries 7}, 015019 (2021)},
  [\href{https://arxiv.org/abs/2012.05768}{{\ttfamily arXiv:2012.05768}}].

\bibitem{Li:2021jiv}
S.-J. {Li}, J.-M. {Liang}, S.-Q. {Shen} and M.~{Li}, \textit{{Variational
  quantum algorithms for trace norms and their applications}},
  \href{http://dx.doi.org/10.1088/1572-9494/ac1938}{{Commun. Theor. Phys.}
  {\bfseries 73}, 105102 (2021)}.

\bibitem{Agarwal:2021yol}
R.~{Agarwal}, S.~{Rethinasamy}, K.~{Sharma} and M.~M. {Wilde},
  \textit{{Estimating distinguishability measures on quantum computers}},
  \href{https://arxiv.org/abs/2108.08406}{{\ttfamily arXiv:2108.08406}}.

\bibitem{Bombelli:1986rw}
L.~Bombelli, R.~K. Koul, J.~Lee and R.~D. Sorkin, \textit{{A quantum source of
  entropy for black holes}},
  \href{http://dx.doi.org/10.1103/PhysRevD.34.373}{Phys. Rev. D {\bfseries 34},
  373 (1986)}.

\bibitem{Srednicki:1993im}
M.~Srednicki, \textit{{Entropy and area}},
  \href{http://dx.doi.org/10.1103/PhysRevLett.71.666}{Phys. Rev. Lett.
  {\bfseries 71}, 666 (1993)},
  [\href{https://arxiv.org/abs/hep-th/9303048}{{\ttfamily
  arXiv:hep-th/9303048}}].

\bibitem{Callan:1994py}
C.~G. Callan~Jr. and F.~Wilczek, \textit{{On geometric entropy}},
  \href{http://dx.doi.org/10.1016/0370-2693(94)91007-3}{Phys. Lett. B
  {\bfseries 333}, 55 (1994)},
  [\href{https://arxiv.org/abs/hep-th/9401072}{{\ttfamily
  arXiv:hep-th/9401072}}].

\bibitem{Holzhey:1994we}
C.~Holzhey, F.~Larsen and F.~Wilczek, \textit{{Geometric and renormalized
  entropy in conformal field theory}},
  \href{http://dx.doi.org/10.1016/0550-3213(94)90402-2}{Nucl. Phys. B
  {\bfseries 424}, 443 (1994)},
  [\href{https://arxiv.org/abs/hep-th/9403108}{{\ttfamily
  arXiv:hep-th/9403108}}].

\bibitem{Peschel:1998ftd}
I.~Peschel, M.~Kaulke and {\"O}.~Legeza, \textit{Density-matrix spectra for
  integrable models},
  \href{http://dx.doi.org/10.1002/(SICI)1521-3889(199902)8:2<153::AID-ANDP153>3.0.CO;2-N}{Ann.
  der Phys. {\bfseries 8}, 153--164 (1999)},
  [\href{https://arxiv.org/abs/cond-mat/9810174}{{\ttfamily
  arXiv:cond-mat/9810174}}].

\bibitem{Peschel:1999pkr}
I.~{Peschel} and M.-C. {Chung}, \textit{{Density matrices for a chain of
  oscillators}}, \href{http://dx.doi.org/10.1088/0305-4470/32/48/305}{J. Phys.
  A: Math. Gen. {\bfseries 32}, 8419--8428 (1999)},
  [\href{https://arxiv.org/abs/cond-mat/9906224}{{\ttfamily
  arXiv:cond-mat/9906224}}].

\bibitem{Chung:2000tqg}
M.-C. {Chung} and I.~{Peschel}, \textit{{Density-matrix spectra for
  two-dimensional quantum systems}},
  \href{http://dx.doi.org/10.1103/PhysRevB.62.4191}{Phys. Rev. B {\bfseries
  62}, 4191--4193 (2000)},
  [\href{https://arxiv.org/abs/cond-mat/0004222}{{\ttfamily
  arXiv:cond-mat/0004222}}].

\bibitem{Chung:2001oyk}
M.-C. Chung and I.~Peschel, \textit{Density-matrix spectra of solvable
  fermionic systems}, \href{http://dx.doi.org/10.1103/PhysRevB.64.064412}{Phys.
  Rev. B {\bfseries 64}, 064412 (2001)},
  [\href{https://arxiv.org/abs/cond-mat/0103301}{{\ttfamily
  arXiv:cond-mat/0103301}}].

\bibitem{Cheong:2002ukf}
S.-A. Cheong and C.~L. Henley, \textit{{Many-body density matrices for free
  fermions}}, \href{http://dx.doi.org/10.1103/PhysRevB.69.075111}{Phys. Rev. B
  {\bfseries 69}, 075111 (2004)},
  [\href{https://arxiv.org/abs/cond-mat/0206196}{{\ttfamily
  arXiv:cond-mat/0206196}}].

\bibitem{Vidal:2002rm}
G.~Vidal, J.~I. Latorre, E.~Rico and A.~Kitaev, \textit{{Entanglement in
  Quantum Critical Phenomena}},
  \href{http://dx.doi.org/10.1103/PhysRevLett.90.227902}{Phys. Rev. Lett.
  {\bfseries 90}, 227902 (2003)},
  [\href{https://arxiv.org/abs/quant-ph/0211074}{{\ttfamily
  arXiv:quant-ph/0211074}}].

\bibitem{Peschel:2002jhw}
I.~Peschel, \textit{{Calculation of reduced density matrices from correlation
  functions}}, \href{http://dx.doi.org/10.1088/0305-4470/36/14/101}{J. Phys. A:
  Math. Gen. {\bfseries 36}, L205 (2003)},
  [\href{https://arxiv.org/abs/cond-mat/0212631}{{\ttfamily
  arXiv:cond-mat/0212631}}].

\bibitem{Latorre:2003kg}
J.~I. Latorre, E.~Rico and G.~Vidal, \textit{{Ground state entanglement in
  quantum spin chains}}, \href{http://dx.doi.org/10.26421/QIC4.1}{Quant. Inf.
  Comput. {\bfseries 4}, 48 (2004)},
  [\href{https://arxiv.org/abs/quant-ph/0304098}{{\ttfamily
  arXiv:quant-ph/0304098}}].

\bibitem{Jin:2003pgk}
B.-Q. Jin and V.~E. Korepin, \textit{{Quantum spin chain, Toeplitz determinants
  and the Fisher-Hartwig conjecture}},
  \href{http://dx.doi.org/10.1023/B:JOSS.0000037230.37166.42}{J. Stat. Phys.
  {\bfseries 116}, 79--95 (2004)},
  [\href{https://arxiv.org/abs/quant-ph/0304108}{{\ttfamily
  arXiv:quant-ph/0304108}}].

\bibitem{Korepin:2004zz}
V.~Korepin, \textit{{Universality of Entropy Scaling in One Dimensional Gapless
  Models}}, \href{http://dx.doi.org/10.1103/PhysRevLett.92.096402}{Phys. Rev.
  Lett. {\bfseries 92}, 096402 (2004)},
  [\href{https://arxiv.org/abs/cond-mat/0311056}{{\ttfamily
  arXiv:cond-mat/0311056}}].

\bibitem{Plenio:2004he}
M.~Plenio, J.~Eisert, J.~Dreissig and M.~Cramer, \textit{{Entropy,
  entanglement, and area: analytical results for harmonic lattice systems}},
  \href{http://dx.doi.org/10.1103/PhysRevLett.94.060503}{Phys. Rev. Lett.
  {\bfseries 94}, 060503 (2005)},
  [\href{https://arxiv.org/abs/quant-ph/0405142}{{\ttfamily
  arXiv:quant-ph/0405142}}].

\bibitem{Calabrese:2004eu}
P.~Calabrese and J.~L. Cardy, \textit{{Entanglement entropy and quantum field
  theory}}, \href{http://dx.doi.org/10.1088/1742-5468/2004/06/P06002}{J. Stat.
  Mech. (2004) P06002}, [\href{https://arxiv.org/abs/hep-th/0405152}{{\ttfamily
  arXiv:hep-th/0405152}}].

\bibitem{Cramer:2005mx}
M.~Cramer, J.~Eisert, M.~Plenio and J.~Dreissig, \textit{{An Entanglement-area
  law for general bosonic harmonic lattice systems}},
  \href{http://dx.doi.org/10.1103/PhysRevA.73.012309}{Phys. Rev. A {\bfseries
  73}, 012309 (2006)},
  [\href{https://arxiv.org/abs/quant-ph/0505092}{{\ttfamily
  arXiv:quant-ph/0505092}}].

\bibitem{Casini:2005rm}
H.~Casini, C.~Fosco and M.~Huerta, \textit{{Entanglement and alpha entropies
  for a massive Dirac field in two dimensions}},
  \href{http://dx.doi.org/10.1088/1742-5468/2005/07/P07007}{J. Stat. Mech.
  (2005) P07007}, [\href{https://arxiv.org/abs/cond-mat/0505563}{{\ttfamily
  arXiv:cond-mat/0505563}}].

\bibitem{Casini:2005zv}
H.~Casini and M.~Huerta, \textit{{Entanglement and alpha entropies for a
  massive scalar field in two dimensions}},
  \href{http://dx.doi.org/10.1088/1742-5468/2005/12/P12012}{J. Stat. Mech.
  (2005) P12012}, [\href{https://arxiv.org/abs/cond-mat/0511014}{{\ttfamily
  arXiv:cond-mat/0511014}}].

\bibitem{Casini:2009sr}
H.~Casini and M.~Huerta, \textit{{Entanglement entropy in free quantum field
  theory}}, \href{http://dx.doi.org/10.1088/1751-8113/42/50/504007}{J. Phys. A:
  Math. Gen. {\bfseries 42}, 504007 (2009)},
  [\href{https://arxiv.org/abs/0905.2562}{{\ttfamily arXiv:0905.2562}}].

\bibitem{Calabrese:2009qy}
P.~Calabrese and J.~Cardy, \textit{{Entanglement entropy and conformal field
  theory}}, \href{http://dx.doi.org/10.1088/1751-8113/42/50/504005}{J. Phys. A:
  Math. Gen. {\bfseries 42}, 504005 (2009)},
  [\href{https://arxiv.org/abs/0905.4013}{{\ttfamily arXiv:0905.4013}}].

\bibitem{Peschel:2009iuj}
I.~Peschel and V.~Eisler, \textit{Reduced density matrices and entanglement
  entropy in free lattice models},
  \href{http://dx.doi.org/10.1088/1751-8113/42/50/504003}{J. Phys. A: Math.
  Gen. {\bfseries 42}, 504003 (2009)},
  [\href{https://arxiv.org/abs/0906.1663}{{\ttfamily arXiv:0906.1663}}].

\bibitem{Peschel:2011jed}
I.~Peschel, \textit{{Special review: Entanglement in solvable many-particle
  models}}, \href{http://dx.doi.org/10.1007/s13538-012-0074-1}{Braz. J. Phys.
  {\bfseries 42}, 267--291 (2012)},
  [\href{https://arxiv.org/abs/1109.0159}{{\ttfamily arXiv:1109.0159}}].

\bibitem{Alba:2009th}
V.~Alba, M.~Fagotti and P.~Calabrese, \textit{{Entanglement entropy of excited
  states}}, \href{http://dx.doi.org/10.1088/1742-5468/2009/10/P10020}{J. Stat.
  Mech. (2009) P10020}, [\href{https://arxiv.org/abs/0909.1999}{{\ttfamily
  arXiv:0909.1999}}].

\bibitem{Alcaraz:2011tn}
F.~C. Alcaraz, M.~I. Berganza and G.~Sierra, \textit{{Entanglement of
  low-energy excitations in Conformal Field Theory}},
  \href{http://dx.doi.org/10.1103/PhysRevLett.106.201601}{Phys. Rev. Lett.
  {\bfseries 106}, 201601 (2011)},
  [\href{https://arxiv.org/abs/1101.2881}{{\ttfamily arXiv:1101.2881}}].

\bibitem{Berganza:2011mh}
M.~I. Berganza, F.~C. Alcaraz and G.~Sierra, \textit{{Entanglement of excited
  states in critical spin chians}},
  \href{http://dx.doi.org/10.1088/1742-5468/2012/01/P01016}{J. Stat. Mech.
  (2012) P01016}, [\href{https://arxiv.org/abs/1109.5673}{{\ttfamily
  arXiv:1109.5673}}].

\bibitem{Pizorn:2012aut}
I.~{Pizorn}, \textit{{Universality in entanglement of quasiparticle
  excitations}},  \href{https://arxiv.org/abs/1202.3336}{{\ttfamily
  arXiv:1202.3336}}.

\bibitem{Essler:2012rai}
F.~H.~L. {Essler}, A.~M. {L{\"a}uchli} and P.~{Calabrese},
  \textit{{Shell-Filling Effect in the Entanglement Entropies of Spinful
  Fermions}}, \href{http://dx.doi.org/10.1103/PhysRevLett.110.115701}{Phys.
  Rev. Lett. {\bfseries 110}, 115701 (2013)},
  [\href{https://arxiv.org/abs/1211.2474}{{\ttfamily arXiv:1211.2474}}].

\bibitem{Berkovits:2013mii}
R.~{Berkovits}, \textit{{Two-particle excited states entanglement entropy in a
  one-dimensional ring}},
  \href{http://dx.doi.org/10.1103/PhysRevB.87.075141}{Phys. Rev. B {\bfseries
  87}, 075141 (2013)}, [\href{https://arxiv.org/abs/1302.4031}{{\ttfamily
  arXiv:1302.4031}}].

\bibitem{Taddia:2013kxu}
L.~{Taddia}, J.~C. {Xavier}, F.~C. {Alcaraz} and G.~{Sierra},
  \textit{{Entanglement entropies in conformal systems with boundaries}},
  \href{http://dx.doi.org/10.1103/PhysRevB.88.075112}{{Phys. Rev. B} {\bfseries
  88}, 075112 (2013)}, [\href{https://arxiv.org/abs/1302.6222}{{\ttfamily
  arXiv:1302.6222}}].

\bibitem{Storms:2013wzf}
M.~{Storms} and R.~R.~P. {Singh}, \textit{{Entanglement in ground and excited
  states of gapped free-fermion systems and their relationship with Fermi
  surface and thermodynamic equilibrium properties}},
  \href{http://dx.doi.org/10.1103/PhysRevE.89.012125}{Phys. Rev. E {\bfseries
  89}, 012125 (2014)}, [\href{https://arxiv.org/abs/1308.6257}{{\ttfamily
  arXiv:1308.6257}}].

\bibitem{Palmai:2014jqa}
T.~P\'almai, \textit{{Excited state entanglement in one dimensional quantum
  critical systems: Extensivity and the role of microscopic details}},
  \href{http://dx.doi.org/10.1103/PhysRevB.90.161404}{Phys. Rev. B {\bfseries
  90}, 161404 (2014)}, [\href{https://arxiv.org/abs/1406.3182}{{\ttfamily
  arXiv:1406.3182}}].

\bibitem{Calabrese:2014ntv}
P.~Calabrese, F.~H.~L. Essler and A.~M. Lauchli, \textit{{Entanglement
  Entropies of the quarter filled Hubbard model}},
  \href{http://dx.doi.org/10.1088/1742-5468/2014/09/P09025}{J. Stat. Mech.
  (2014) P09025}, [\href{https://arxiv.org/abs/1406.7477}{{\ttfamily
  arXiv:1406.7477}}].

\bibitem{Molter:2014qsb}
J.~{M{\"o}lter}, T.~{Barthel}, U.~{Schollw{\"o}ck} and V.~{Alba},
  \textit{{Bound states and entanglement in the excited states of quantum spin
  chains}}, \href{http://dx.doi.org/10.1088/1742-5468/2014/10/P10029}{J. Stat.
  Mech. (2014) 10029}, [\href{https://arxiv.org/abs/1407.0066}{{\ttfamily
  arXiv:1407.0066}}].

\bibitem{Taddia:2016dbm}
L.~Taddia, F.~Ortolani and T.~P\'almai, \textit{{R\'enyi entanglement entropies
  of descendant states in critical systems with boundaries: conformal field
  theory and spin chains}},
  \href{http://dx.doi.org/10.1088/1742-5468/2016/09/093104}{J. Stat. Mech.
  (2016) 093104}, [\href{https://arxiv.org/abs/1606.02667}{{\ttfamily
  arXiv:1606.02667}}].

\bibitem{Castro-Alvaredo:2018dja}
O.~A. Castro-Alvaredo, C.~De~Fazio, B.~Doyon and I.~M. Sz\'ecs\'enyi,
  \textit{{Entanglement Content of Quasiparticle Excitations}},
  \href{http://dx.doi.org/10.1103/PhysRevLett.121.170602}{Phys. Rev. Lett.
  {\bfseries 121}, 170602 (2018)},
  [\href{https://arxiv.org/abs/1805.04948}{{\ttfamily arXiv:1805.04948}}].

\bibitem{Castro-Alvaredo:2018bij}
O.~A. Castro-Alvaredo, C.~De~Fazio, B.~Doyon and I.~M. Sz\'ecs\'enyi,
  \textit{{Entanglement content of quantum particle excitations. Part I. Free
  field theory}}, \href{http://dx.doi.org/10.1007/JHEP10(2018)039}{JHEP
  {\bfseries 10} (2018) 039},
  [\href{https://arxiv.org/abs/1806.03247}{{\ttfamily arXiv:1806.03247}}].

\bibitem{Murciano:2018cfp}
S.~Murciano, P.~Ruggiero and P.~Calabrese, \textit{{Entanglement and relative
  entropies for low-lying excited states in inhomogeneous one-dimensional
  quantum systems}}, \href{http://dx.doi.org/10.1088/1742-5468/ab00ec}{J. Stat.
  Mech. (2019) 034001}, [\href{https://arxiv.org/abs/1810.02287}{{\ttfamily
  arXiv:1810.02287}}].

\bibitem{Castro-Alvaredo:2019irt}
O.~A. Castro-Alvaredo, C.~De~Fazio, B.~Doyon and I.~M. Sz\'ecs\'enyi,
  \textit{{Entanglement content of quantum particle excitations. Part II.
  Disconnected regions and logarithmic negativity}},
  \href{http://dx.doi.org/10.1007/JHEP11(2019)058}{JHEP {\bfseries 11} (2019)
  058}, [\href{https://arxiv.org/abs/1904.01035}{{\ttfamily
  arXiv:1904.01035}}].

\bibitem{Castro-Alvaredo:2019lmj}
O.~A. Castro-Alvaredo, C.~De~Fazio, B.~Doyon and I.~M. Sz\'ecs\'enyi,
  \textit{{Entanglement Content of Quantum Particle Excitations III. Graph
  Partition Functions}}, \href{http://dx.doi.org/10.1063/1.5098892}{J. Math.
  Phys. {\bfseries 60}, 082301 (2019)},
  [\href{https://arxiv.org/abs/1904.02615}{{\ttfamily arXiv:1904.02615}}].

\bibitem{Miao:2019xpp}
Q.~Miao and T.~Barthel, \textit{{Eigenstate Entanglement: Crossover from the
  Ground State to Volume Laws}},
  \href{http://dx.doi.org/10.1103/PhysRevLett.127.040603}{Phys. Rev. Lett.
  {\bfseries 127}, 040603 (2021)},
  [\href{https://arxiv.org/abs/1905.07760}{{\ttfamily arXiv:1905.07760}}].

\bibitem{Jafarizadeh:2019xxc}
A.~Jafarizadeh and M.~Rajabpour, \textit{{Bipartite entanglement entropy of the
  excited states of free fermions and harmonic oscillators}},
  \href{http://dx.doi.org/10.1103/PhysRevB.100.165135}{Phys. Rev. B {\bfseries
  100}, 165135 (2019)}, [\href{https://arxiv.org/abs/1907.09806}{{\ttfamily
  arXiv:1907.09806}}].

\bibitem{Barthel:2019zor}
T.~Barthel and Q.~Miao, \textit{{Scaling functions for eigenstate entanglement
  crossovers in harmonic lattices}},
  \href{http://dx.doi.org/10.1103/PhysRevA.104.022414}{Phys. Rev. A {\bfseries
  104}, 022414 (2021)}, [\href{https://arxiv.org/abs/1912.10045}{{\ttfamily
  arXiv:1912.10045}}].

\bibitem{Capizzi:2020jed}
L.~Capizzi, P.~Ruggiero and P.~Calabrese, \textit{{Symmetry resolved
  entanglement entropy of excited states in a CFT}},
  \href{http://dx.doi.org/10.1088/1742-5468/ab96b6}{J. Stat. Mech. (2020)
  073101}, [\href{https://arxiv.org/abs/2003.04670}{{\ttfamily
  arXiv:2003.04670}}].

\bibitem{You:2020osa}
Y.~You, E.~Wybo, F.~Pollmann and S.~Sondhi, \textit{{Observing Quasiparticles
  through the Entanglement Lens}},
  \href{https://arxiv.org/abs/2007.04318}{{\ttfamily arXiv:2007.04318}}.

\bibitem{Haque:2020ewo}
M.~{Haque}, P.~A. {McClarty} and I.~M. {Khaymovich}, \textit{{Entanglement of
  mid-spectrum eigenstates of chaotic many-body systems -- deviation from
  random ensembles}},
  \href{http://dx.doi.org/10.1103/PhysRevE.105.014109}{Phys. Rev. E {\bfseries
  105}, 014109 (2022)}, [\href{https://arxiv.org/abs/2008.12782}{{\ttfamily
  arXiv:2008.12782}}].

\bibitem{Zhang:2020ouz}
J.~Zhang and M.~A. Rajabpour, \textit{{Excited state R\'enyi entropy and
  subsystem distance in two-dimensional non-compact bosonic theory. Part I.
  Single-particle states}},
  \href{http://dx.doi.org/10.1007/JHEP12(2020)160}{JHEP {\bfseries 12} (2020)
  160}, [\href{https://arxiv.org/abs/2009.00719}{{\ttfamily
  arXiv:2009.00719}}].

\bibitem{Miao:2020hkj}
Q.~Miao and T.~Barthel, \textit{{Eigenstate entanglement scaling for critical
  interacting spin chains}},
  \href{http://dx.doi.org/10.22331/q-2022-02-02-642}{Quantum {\bfseries 6}, 642
  (2022)}, [\href{https://arxiv.org/abs/2010.07265}{{\ttfamily
  arXiv:2010.07265}}].
  
\bibitem{Zhang:2020vtc}
J.~Zhang and M.~A. Rajabpour, \textit{{Universal R\'enyi entanglement entropy
  of quasiparticle excitations}},
  \href{http://dx.doi.org/10.1209/0295-5075/ac130e}{EPL {\bfseries 135}, 60001
  (2021)}, [\href{https://arxiv.org/abs/2010.13973}{{\ttfamily
  arXiv:2010.13973}}].

\bibitem{Wybo:2020fiz}
E.~Wybo, F.~Pollmann, S.~L. Sondhi and Y.~You, \textit{{Visualizing
  quasiparticles from quantum entanglement for general one-dimensional
  phases}}, \href{http://dx.doi.org/10.1103/PhysRevB.103.115120}{Phys. Rev. B
  {\bfseries 103}, 115120 (2021)},
  [\href{https://arxiv.org/abs/2010.15137}{{\ttfamily arXiv:2010.15137}}].

\bibitem{Zhang:2020dtd}
J.~Zhang and M.~A. Rajabpour, \textit{{Corrections to universal R\'enyi entropy
  in quasiparticle excited states of quantum chains}},
  \href{http://dx.doi.org/10.1088/1742-5468/ac1f28}{J. Stat. Mech. (2021)
  093101}, [\href{https://arxiv.org/abs/2010.16348}{{\ttfamily
  arXiv:2010.16348}}].

\bibitem{Zhang:2020txb}
J.~Zhang and M.~A. Rajabpour, \textit{{Excited state R\'enyi entropy and
  subsystem distance in two-dimensional non-compact bosonic theory. Part II.
  Multi-particle states}},
  \href{http://dx.doi.org/10.1007/JHEP08(2021)106}{JHEP {\bfseries 08} (2021)
  106}, [\href{https://arxiv.org/abs/2011.11006}{{\ttfamily
  arXiv:2011.11006}}].

\bibitem{Chowdhury:2021qja}
B.~G. Chowdhury and J.~R. David, \textit{{Entanglement in descendants}},
  \href{http://dx.doi.org/10.1007/JHEP02(2022)003}{JHEP {\bfseries 02} (2022)
  003}, [\href{https://arxiv.org/abs/2108.00898}{{\ttfamily
  arXiv:2108.00898}}].

\bibitem{Zhang:2021bmy}
J.~Zhang and M.~A. Rajabpour, \textit{{Entanglement of magnon excitations in
  spin chains}}, \href{http://dx.doi.org/10.1007/JHEP02(2022)072}{JHEP
  {\bfseries 02} (2022) 072},
  [\href{https://arxiv.org/abs/2109.12826}{{\ttfamily arXiv:2109.12826}}].

\bibitem{Mussardo:2021gws}
G.~Mussardo and J.~Viti, \textit{{The $\hbar \rightarrow 0$ limit of the
  entanglement entropy}},
  \href{http://dx.doi.org/10.1103/PhysRevA.105.032404}{Phys. Rev. A {\bfseries
  105}, 032404 (2022)}, [\href{https://arxiv.org/abs/2112.06840}{{\ttfamily
  arXiv:2112.06840}}].

\bibitem{Fagotti:2013jzu}
M.~Fagotti and F.~H. Essler, \textit{Reduced density matrix after a quantum
  quench}, \href{http://dx.doi.org/10.1103/PhysRevB.87.245107}{Phys. Rev. B
  {\bfseries 87}, 245107 (2013)},
  [\href{https://arxiv.org/abs/1302.6944}{{\ttfamily arXiv:1302.6944}}].

\bibitem{Cardy:2014rqa}
J.~Cardy, \textit{{Thermalization and Revivals after a Quantum Quench in
  Conformal Field Theory}},
  \href{http://dx.doi.org/10.1103/PhysRevLett.112.220401}{Phys. Rev. Lett.
  {\bfseries 112}, 220401 (2014)},
  [\href{https://arxiv.org/abs/1403.3040}{{\ttfamily arXiv:1403.3040}}].

\bibitem{Lashkari:2014yva}
N.~Lashkari, \textit{{Relative Entropies in Conformal Field Theory}},
  \href{http://dx.doi.org/10.1103/PhysRevLett.113.051602}{Phys. Rev. Lett.
  {\bfseries 113}, 051602 (2014)},
  [\href{https://arxiv.org/abs/1404.3216}{{\ttfamily arXiv:1404.3216}}].

\bibitem{Lashkari:2015dia}
N.~Lashkari, \textit{{Modular Hamiltonian for Excited States in Conformal Field
  Theory}}, \href{http://dx.doi.org/10.1103/PhysRevLett.117.041601}{Phys. Rev.
  Lett. {\bfseries 117}, 041601 (2016)},
  [\href{https://arxiv.org/abs/1508.03506}{{\ttfamily arXiv:1508.03506}}].

\bibitem{Arias:2016nip}
R.~Arias, D.~Blanco, H.~Casini and M.~Huerta, \textit{{Local temperatures and
  local terms in modular Hamiltonians}},
  \href{http://dx.doi.org/10.1103/PhysRevD.95.065005}{Phys. Rev. D {\bfseries
  95}, 065005 (2017)}, [\href{https://arxiv.org/abs/1611.08517}{{\ttfamily
  arXiv:1611.08517}}].

\bibitem{Sarosi:2016atx}
G.~S\'arosi and T.~Ugajin, \textit{{Relative entropy of excited states in
  conformal field theories of arbitrary dimensions}},
  \href{http://dx.doi.org/10.1007/JHEP02(2017)060}{JHEP {\bfseries 02} (2017)
  060}, [\href{https://arxiv.org/abs/1611.02959}{{\ttfamily
  arXiv:1611.02959}}].

\bibitem{Sarosi:2016oks}
G.~S\'arosi and T.~Ugajin, \textit{{Relative entropy of excited states in two
  dimensional conformal field theories}},
  \href{http://dx.doi.org/10.1007/JHEP07(2016)114}{JHEP {\bfseries 07} (2016)
  114}, [\href{https://arxiv.org/abs/1603.03057}{{\ttfamily
  arXiv:1603.03057}}].

\bibitem{He:2017vyf}
S.~He, F.-L. Lin and J.-j. Zhang, \textit{{Subsystem eigenstate thermalization
  hypothesis for entanglement entropy in CFT}},
  \href{http://dx.doi.org/10.1007/JHEP08(2017)126}{JHEP {\bfseries 08} (2017)
  126}, [\href{https://arxiv.org/abs/1703.08724}{{\ttfamily
  arXiv:1703.08724}}].

\bibitem{Basu:2017kzo}
P.~Basu, D.~Das, S.~Datta and S.~Pal, \textit{{Thermality of eigenstates in
  conformal field theories}},
  \href{http://dx.doi.org/10.1103/PhysRevE.96.022149}{Phys. Rev. E {\bfseries
  96}, 022149 (2017)}, [\href{https://arxiv.org/abs/1705.03001}{{\ttfamily
  arXiv:1705.03001}}].

\bibitem{Arias:2017dda}
R.~Arias, H.~Casini, M.~Huerta and D.~Pontello, \textit{{Anisotropic Unruh
  temperatures}}, \href{http://dx.doi.org/10.1103/PhysRevD.96.105019}{Phys.
  Rev. D {\bfseries 96}, 105019 (2017)},
  [\href{https://arxiv.org/abs/1707.05375}{{\ttfamily arXiv:1707.05375}}].

\bibitem{He:2017txy}
S.~He, F.-L. Lin and J.-j. Zhang, \textit{{Dissimilarities of reduced density
  matrices and eigenstate thermalization hypothesis}},
  \href{http://dx.doi.org/10.1007/JHEP12(2017)073}{JHEP {\bfseries 12} (2017)
  073}, [\href{https://arxiv.org/abs/1708.05090}{{\ttfamily
  arXiv:1708.05090}}].

\bibitem{Suzuki:2019xdq}
Y.~Suzuki, T.~Takayanagi and K.~Umemoto, \textit{{Entanglement Wedges from
  Information Metric in Conformal Field Theories}},
  \href{http://dx.doi.org/10.1103/PhysRevLett.123.221601}{Phys. Rev. Lett.
  {\bfseries 123}, 221601 (2019)},
  [\href{https://arxiv.org/abs/1908.09939}{{\ttfamily arXiv:1908.09939}}].

\bibitem{Zhang:2019kwu}
J.~Zhang and P.~Calabrese, \textit{{Subsystem distance after a local operator
  quench}}, \href{http://dx.doi.org/10.1007/JHEP02(2020)056}{JHEP {\bfseries
  02} (2020) 056}, [\href{https://arxiv.org/abs/1911.04797}{{\ttfamily
  arXiv:1911.04797}}].

\bibitem{Kusuki:2019hcg}
Y.~Kusuki, Y.~Suzuki, T.~Takayanagi and K.~Umemoto, \textit{{Looking at Shadows
  of Entanglement Wedges}}, \href{http://dx.doi.org/10.1093/ptep/ptaa152}{PTEP
  {\bfseries 2020}, 11B105 (2020)},
  [\href{https://arxiv.org/abs/1912.08423}{{\ttfamily arXiv:1912.08423}}].

\bibitem{Zhang:2019wqo}
J.~Zhang, P.~Ruggiero and P.~Calabrese, \textit{{Subsystem Trace Distance in
  Quantum Field Theory}},
  \href{http://dx.doi.org/10.1103/PhysRevLett.122.141602}{Phys. Rev. Lett.
  {\bfseries 122}, 141602 (2019)},
  [\href{https://arxiv.org/abs/1901.10993}{{\ttfamily arXiv:1901.10993}}].

\bibitem{Mendes-Santos:2019tmf}
T.~Mendes-Santos, G.~Giudici, M.~Dalmonte and M.~A. Rajabpour,
  \textit{{Entanglement Hamiltonian of quantum critical chains and conformal
  field theories}}, \href{http://dx.doi.org/10.1103/PhysRevB.100.155122}{Phys.
  Rev. B {\bfseries 100}, 155122 (2019)},
  [\href{https://arxiv.org/abs/1906.00471}{{\ttfamily arXiv:1906.00471}}].

\bibitem{Zhang:2019itb}
J.~Zhang, P.~Ruggiero and P.~Calabrese, \textit{{Subsystem trace distance in
  low-lying states of $(1+1)$-dimensional conformal field theories}},
  \href{http://dx.doi.org/10.1007/JHEP10(2019)181}{JHEP {\bfseries 10} (2019)
  181}, [\href{https://arxiv.org/abs/1907.04332}{{\ttfamily
  arXiv:1907.04332}}].

\bibitem{Zhang:2020mjv}
J.~Zhang, P.~Calabrese, M.~Dalmonte and M.~A. Rajabpour, \textit{{Lattice
  Bisognano-Wichmann modular Hamiltonian in critical quantum spin chains}},
  \href{http://dx.doi.org/10.21468/SciPostPhysCore.2.2.007}{SciPost Phys. Core
  {\bfseries 2}, 007 (2020)},
  [\href{https://arxiv.org/abs/2003.00315}{{\ttfamily arXiv:2003.00315}}].

\bibitem{Arias:2020sgz}
R.~Arias and J.~Zhang, \textit{{R\'enyi entropy and subsystem distances in
  finite size and thermal states in critical XY chains}},
  \href{http://dx.doi.org/10.1088/1742-5468/ababfd}{J. Stat. Mech. (2020)
  083112}, [\href{https://arxiv.org/abs/2004.13096}{{\ttfamily
  arXiv:2004.13096}}].

\bibitem{deBoer:2020snb}
J.~de~Boer, V.~Godet, J.~Kastikainen and E.~Keski-Vakkuri, \textit{{Quantum
  hypothesis testing in many-body systems}},
  \href{http://dx.doi.org/10.21468/SciPostPhysCore.4.2.019}{SciPost Phys. Core
  {\bfseries 4}, 019 (2021)},
  [\href{https://arxiv.org/abs/2007.11711}{{\ttfamily arXiv:2007.11711}}].

\bibitem{Kudler-Flam:2021rpr}
J.~Kudler-Flam, \textit{{Relative Entropy of Random States and Black Holes}},
  \href{http://dx.doi.org/10.1103/PhysRevLett.126.171603}{Phys. Rev. Lett.
  {\bfseries 126}, 171603 (2021)},
  [\href{https://arxiv.org/abs/2102.05053}{{\ttfamily arXiv:2102.05053}}].

\bibitem{Yang:2021enf}
R.-Q. Yang, \textit{{Gravity duals of quantum distances}},
  \href{http://dx.doi.org/10.1007/JHEP08(2021)156}{JHEP {\bfseries 08} (2021)
  156}, [\href{https://arxiv.org/abs/2102.01898}{{\ttfamily
  arXiv:2102.01898}}].

\bibitem{Chen:2021pls}
H.-H. Chen, \textit{{Symmetry decomposition of relative entropies in conformal
  field theory}}, \href{http://dx.doi.org/10.1007/JHEP07(2021)084}{JHEP
  {\bfseries 07} (2021) 084},
  [\href{https://arxiv.org/abs/2104.03102}{{\ttfamily arXiv:2104.03102}}].

\bibitem{Capizzi:2021zga}
L.~Capizzi and P.~Calabrese, \textit{{Symmetry resolved relative entropies and
  distances in conformal field theory}},
  \href{http://dx.doi.org/10.1007/JHEP10(2021)195}{JHEP {\bfseries 10} (2021)
  195}, [\href{https://arxiv.org/abs/2105.08596}{{\ttfamily
  arXiv:2105.08596}}].

\bibitem{Kudler-Flam:2021alo}
J.~Kudler-Flam, V.~Narovlansky and S.~Ryu, \textit{{Distinguishing Random and
  Black Hole Microstates}},
  \href{http://dx.doi.org/10.1103/PRXQuantum.2.040340}{PRX Quantum {\bfseries
  2}, 040340 (2021)}, [\href{https://arxiv.org/abs/2108.00011}{{\ttfamily
  arXiv:2108.00011}}].

\bibitem{Casini:2018cxg}
H.~Casini, R.~Medina, I.~Salazar~Landea and G.~Torroba, \textit{{R\'enyi
  relative entropies and renormalization group flows}},
  \href{http://dx.doi.org/10.1007/JHEP09(2018)166}{JHEP {\bfseries 09} (2018)
  166}, [\href{https://arxiv.org/abs/1807.03305}{{\ttfamily
  arXiv:1807.03305}}].

\bibitem{Lieb:1961fr}
E.~H. Lieb, T.~Schultz and D.~Mattis, \textit{{Two soluble models of an
  antiferromagnetic chain}},
  \href{http://dx.doi.org/10.1016/0003-4916(61)90115-4}{Annals Phys. {\bfseries
  16}, 407 (1961)}.

\bibitem{Katsura:1962hqz}
S.~Katsura, \textit{{Statistical mechanics of the anisotropic linear Heisenberg
  model}}, \href{http://dx.doi.org/10.1103/PhysRev.127.1508}{{Phys. Rev.}
  {\bfseries 127}, 1508 (1962)}.

\bibitem{Pfeuty:1970ayt}
P.~Pfeuty, \textit{{The one-dimensional Ising model with a transverse field}},
  \href{http://dx.doi.org/10.1016/0003-4916(70)90270-8}{Annals Phys. {\bfseries
  57}, 79 (1970)}.

\bibitem{Balian:1969tb}
R.~Balian and E.~Brezin, \textit{{Nonunitary bogoliubov transformations and
  extension of Wick's theorem}},
  \href{http://dx.doi.org/10.1007/BF02710281}{Nuovo Cim. B {\bfseries 64},
  37--55 (1969)}.

\bibitem{Fagotti:2010yr}
M.~Fagotti and P.~Calabrese, \textit{{Entanglement entropy of two disjoint
  blocks in XY chains}},
  \href{http://dx.doi.org/10.1088/1742-5468/2010/04/P04016}{J. Stat. Mech.
  (2010) P04016}, [\href{https://arxiv.org/abs/1003.1110}{{\ttfamily
  arXiv:1003.1110}}].

\bibitem{Banchi:2014uht}
L.~{Banchi}, P.~{Giorda} and P.~{Zanardi}, \textit{{Quantum
  information-geometry of dissipative quantum phase transitions}},
  \href{http://dx.doi.org/10.1103/PhysRevE.89.022102}{Phys. Rev. E {\bfseries
  89}, 022102 (2014)}, [\href{https://arxiv.org/abs/1305.4527}{{\ttfamily
  arXiv:1305.4527}}].

\bibitem{Becker:1973qv}
H.~G. Becker, \textit{{On the transformation of a complex skew-symmetric matrix
  into a real normal form and its application to a direct proof of the
  Bloch-Messiah theorem}}, \href{http://dx.doi.org/10.1007/BF02906230}{Lett.
  Nuovo Cimento {\bfseries 8}, 185--188 (1973)}.

\bibitem{Gaudin:1983kpuCaux:2014uuq}
M.~Gaudin, \textit{La Fonction d'Onde de Bethe}.
\newblock Masson, 1983.
[Translated by
J.-S. Caux, \textit{The Bethe Wavefunction}.
\newblock Cambridge University Press, 2014,
  \href{http://dx.doi.org/10.1017/CBO9781107053885}{10.1017/CBO9781107053885}.]

\bibitem{Karbach:1998abi}
M.~{Karbach} and G.~{Muller}, \textit{{Introduction to the Bethe ansatz I}},
  \href{http://dx.doi.org/10.1063/1.4822511}{Comput. Phys. {\bfseries 11}, 36
  (1997)}, [\href{https://arxiv.org/abs/cond-mat/9809162}{{\ttfamily
  arXiv:cond-mat/9809162}}].

\end{thebibliography}

\providecommand{\href}[2]{#2}\begingroup\raggedright\endgroup

\end{document}